\documentclass[aps,prd,10pt,showpacs,amsmath,amssymb, 
nofootinbib,nobibnotes,preprintnumbers,showkeys,longbibliography]{revtex4-1}
\pdfoutput=1 % if your are submitting a pdflatex (i.e. if you have images in pdf, png or jpg format)
\usepackage{float}
\usepackage{mathrsfs,amsfonts}
\usepackage{mathtools}
\usepackage{tensor}
\usepackage[T1]{fontenc} 
\usepackage{xcolor}
\usepackage{url}
\usepackage{subfig}
\usepackage[hyperindex,breaklinks]{hyperref}
\usepackage{graphicx}
\usepackage{tensor}
\usepackage{booktabs}
\usepackage{tikz}
\usetikzlibrary{arrows,decorations.pathmorphing,decorations.markings,patterns,decorations.pathreplacing}
\tikzset{
	graviton/.style={decorate,
		decoration={coil,amplitude=2pt, segment length=2pt}} 
}
\usepackage{nicefrac}
\usepackage[markup=red]{changes}
\begin{document}
	\allowdisplaybreaks[1]
	\title{Towards a unitary, renormalizable and\\
		ultraviolet-complete quantum theory of gravity}
	
	\author{Christian F. Steinwachs}
	\email{christian.steinwachs@physik.uni-freiburg.de}
	\affiliation{Physikalisches Institut, Albert-Ludwigs-Universit\"at Freiburg,\\
		Hermann-Herder-Str.~3, 79104 Freiburg, Germany}
	
\begin{abstract}
For any fundamental quantum field theory, unitarity, renormalizability, and
relativistic invariance are considered to be essential properties. Unitarity is inevitably connected to the probabilistic interpretation of the quantum theory, while renormalizability guarantees its completeness. Relativistic invariance, in turn, is a symmetry which derives from the structure of spacetime. So far, the perturbative attempt to formulate a fundamental local quantum field theory of gravity based on the metric field seems to be in conflict with at least one of these properties. 
In quantum Ho\v{r}ava gravity, a quantum Lifshitz field theory of gravity characterized by an anisotropic scaling between space and time, unitarity and renormalizability can be retained while Lorentz invariance is sacrificed at high energies and must emerge only as approximate symmetry at low energies.
I review various approaches to perturbative quantum gravity with a particular focus on recent progress in the quantization of Ho\v{r}ava gravity, supporting its theoretical status as a unitary, renormalizable and ultraviolet-complete quantum theory of gravity.
\end{abstract}

\maketitle

\section{Introduction}

The search for a consistent quantum theory of gravity might be dated back almost 90 years to the work of Rosenfeld \cite{Rosenfeld1930}. Since then, many different approaches have been suggested, each of them with its own assumptions, predictions (if any) and limitations, see \cite{Kiefer2007} for an overview.
Prominent roads to quantum gravity include canonical approaches such as Quantum Geometrodynamics \cite{DeWitt1967,Hartle:1983ai} and Loop quantum gravity \cite{Rovelli:1994ge,Rovelli:1997yv,Ashtekar:2004eh,Thiemann:2007zz}, discrete approaches such as Causal Dynamical Triangulations \cite{Ambjorn:2004qm,Ambjorn:2005db,Loll:2019rdj}, and unified approaches such as String Theory \cite{Lust:1989tj,Polchinski:1998rq,Polchinski:1998rr,Kiritsis:2007zza, Schomerus:2017lqg}.

In this review, I restrict the discussion to \textit{local} field theories, in which gravity is fundamentally described by the \textit{metric field}. For non-local (infinite derivative) theories of gravity, see e.~g.\cite{Krasnikov:1987yj,Gorbar:2002pw,Smilga:2005gb,Shapiro:2015uxa,Biswas:2011ar,Modesto:2011kw,Tomboulis:2015esa,Tomboulis:2015gfa,Edholm:2016hbt,Modesto:2017sdr} and for non-metric theories of gravity, see e.g.~\cite{Sezgin:1979zf,Sezgin:1981xs,Hehl:1994ue,Shapiro:2001rz,Pagani:2015ema,Percacci:2019hxn}.
In view of the tremendous success of perturbative quantum field theory in different areas of physics, including the Standard Model of particle physics,
it seems  natural to quantize gravity within this highly developed and strongly tested unified framework along with the fundamental interactions between the matter fields. For most of the content in this review, I focus on the \textit{covariant perturbative} approach to quantum gravity.
Much of the progress in this approach might be attributed to Bryce S.~DeWitt, who pioneered the field and set the standards for most of its developments in the following decades \cite{DeWitt1965,DeWitt1967b,DeWitt1967a}.

While the direct approach to quantize General Relativity perturbatively is considered to fail due to its non-renormalizability in the strict sense \cite{Hooft1974,Goroff1985}, the perturbative quantization and  renormalization can be consistently carried out when treating General Relativity as an Effective Field Theory \cite{Weinberg1979,Donoghue1994,Burgess2004}. However, by construction the effective description breaks down at a finite energy scale and therefore does not extend to arbitrarily high energies required for a fundamental theory of quantum gravity.
In this respect, the non-perturbative Asymptotic Safety program to quantum gravity might offer a solution in providing a consistent ultraviolet completion \cite{Reuter1998,Codello2009,Reuter2012,Litim2011}. 
A different strategy, which retains the perturbative treatment, is based on the quantization of modifications of General Relativity. Quadratic Gravity, the extension of the Einstein-Hilbert action by all quadratic curvature invariants, is a perturbatively renormalizable quantum theory of gravity \cite{Stelle1977}.
While the higher derivatives in Quadratic Gravity improve the ultraviolet behavior, relativistic invariance necessarily implies the inclusion of higher time derivatives, which in turn result in an enlarged particle spectrum, including a massive spin-two ghost. At the classical level, the presence of the ghost leads to runaway solutions known as Ostrogradsky instability \cite{Ostrogradsky1850}. At the quantum level, within the usual quantization prescription, the ghost was found to lead to a violation of unitarity \cite{Stelle1977}. Recent proposals, which involve different quantization prescriptions for the ghost, preserve unitarity but instead lead to a violating of micro-causality \cite{Anselmi2018,Donoghue2019a}.

In view of these problems, it has been suggested to explore the consequences of the assumption that Lorentz invariance is not a fundamental symmetry, but only emerges as an approximate symmetry at low energies. In this way, higher spatial derivatives can be introduced to tame the ultraviolet divergences, while retaining only second-order time derivatives to avoid the problems associated with the occurrence of higher-derivative ghosts. The breaking of relativistic invariance at a fundamental level is naturally realized in \textit{Lifshitz theories} by an anisotropic scaling between space and time \cite{Horava2009a,Horava2009}.

After a brief overview of various relativistic approaches to perturbative quantum gravity, I review several aspects of the Lifshitz theory of gravity, \textit{Ho\v{r}ava Gravity} \cite{Horava2009}, in ${D=2+1}$ and ${D=3+1}$ dimensions, including the consequences of the reduced invariance group of foliation-preserving diffeomorphisms, the geometrical formulation in terms of Arnowitt-Deser-Misner variables, the phenomenological implications of the additional propagating gravitational scalar degree of freedom and the current status of the experimental constraints. I discuss the quantization of projectable Ho\v{r}ava Gravity, a particular version of Ho\v{r}ava Gravity in which the lapse function is not a propagating degree of freedom. I sketch the proof that projectable Ho\v{r}ava Gravity is a perturbatively renormalizable quantum theory of gravity \cite{Barvinsky2016,Barvinsky2018} and report recent results on its renormalization group flow \cite{Barvinsky2017,Barvinsky2019a}.

The article is structured as follows.
In Sec. \ref{Sec:PQDG}, I introduce the general formalism for the perturbative quantization of local field theories.
In Sec. \ref{Sec:PQDGR}, I summarize the essential properties of General Relativity and the major drawback of its perturbative quantization -- non-renormalizability.
In Sec. \ref{Sec:EFT}, I briefly comment on the status of General Relativity as an Effective Field Theory. 
In Sec. \ref{Sec:AS}, I discuss several aspects of the Asymptotic Safety conjecture in the context of gravity and its status as a possible ultraviolet complete scenario for a quantum theory of gravity.
In Sec. \ref{Sec:HDG}, I review the perturbatively renormalizable theory of Quadratic Gravity and discuss the ghost problem.
In Sec. \ref{Sec:HG}, I present various aspects in the classical theory of Ho\v{r}ava Gravity in $D=2+1$ and $D=3+1$ dimensions.
In Sec. \ref{Sec:QHG}, I discuss the perturbative quantization of projectable Ho\v{r}ava Gravity, its perturbative renormalizability and its status as ultraviolet-complete theory.
Finally, I conclude in Sec. \ref{Sec:CAO} with a short summary and a brief outlook on important further steps towards a unitary, renormalizable and ultraviolet-complete quantum theory of gravity in $D=3+1$ dimensions.

\section{Perturbative Quantum Field Theory-- general formalism}
\label{Sec:PQDG}
Consider a \textit{local} field theory, which is defined by the action functional $S$,
\begin{align}
S[\phi]=\sum_{n}\int\mathrm{d}^DXc_{n}\mathcal{O}_{n}(\phi,\partial).\label{act}
\end{align}  
Locality means that the operators $\mathcal{O}_n(\phi,\partial)$ are functions of a \textit{finite} number of derivatives (including no derivative) of the generalized field(s) $\phi^{i}=\phi^{A}(x)$, evaluated \textit{at the same point} $x$. The operators $\mathcal{O}_n$ are restricted by the symmetries of $S$. The $c_{n}$ are the coupling constants characterizing the ``strength'' of the interaction associated with the operator $\mathcal{O}_n$.\footnote{I use the ultra-condensed DeWitt notation, in which the generalized index $i=\{A,X\}$ of a generalized field $\phi^{i}=\phi^{A}(X)$ encompasses the discrete bundle index $A$ and the continuous spacetime point $X$. Summation over $i$ implies summation over $A$ as well as integration over $X$, i.e. $\phi_{i}\phi^{i}=\int\mathrm{d}^DX\phi_{A}(X)\phi^{A}(X)$.} The main object in the Quantum Field Theory (QFT) is the \textit{quantum effective action} $\Gamma$.

\subsection{Perturbation Theory}
\label{sec:PerturbationTheory}
Starting point for the formal derivation of the Euclidean effective action is the partition function $Z$, which is defined by the functional integral over the field configurations $\phi^{i}$ and is a functional of the external source $J_{i}$, 
\begin{align}
Z[J]:=e^{-W[J]}=\int\mathcal{D}\phi\, e^{-\left(S[\phi]+J_i\phi^{i}\right)}.\label{PI}
\end{align}
The mean field $\varphi^{i}$ is defined as the quantum average in the presence of the source $J_{i}$,
\begin{align}
\varphi^{i}:=\langle \phi^{i}\rangle_{J}=\frac{\delta W[J]}{\delta J_{i}}.\label{MF}
\end{align}
The quantum effective action $\Gamma$ is defined as functional Legendre transformation of the Schwinger functional $W$,
\begin{align}
\Gamma[\varphi]:=W[J]-\varphi^{i}J_i.\label{QEA}
\end{align}
Combining \eqref{PI}-\eqref{MF}, leads to the functional integro-differential equation\footnote{Postfix notation with indices separated by a comma denote functional derivatives with respect to the argument, e.g. $\Gamma_{,i}=\delta\Gamma[\varphi]/\delta\varphi^i$. }
\begin{align}
e^{-\Gamma[\varphi]}=\int\mathcal{D}\phi\, e^{-\left\{S[\phi]-(\varphi^{i}-\phi^{i})\Gamma_{,i}[\varphi]\right\}}.\label{FIDE}
\end{align}
Equation \eqref{FIDE} provides the starting point for the perturbative expansion of $\Gamma$ (reinserting powers of $\hbar$),
\begin{align}
\Gamma[\varphi]=S[\varphi]+\hbar\Gamma_{1}[\varphi]+\hbar^2\Gamma_{2}[\varphi]+\mathcal{O}(\hbar^3).\label{EAE}
\end{align}
The diagrammatic representation of the expansion \eqref{EAE} is given in terms of vacuum diagrams in which the number of loops corresponds to the power of $\hbar$ in \eqref{EAE},
\begin{figure}[h!]
	\begin{center}
		\begin{tikzpicture}[scale=1.6]
		\tikzset{boson/.style={decorate, decoration=snake}}
		\tikzset{graviton/.style={decorate, decoration={coil,amplitude=3pt, segment length=2.75pt}}} 
		\draw[] (1.7,0) circle (0.5cm);
		\draw[] (3.2,0) circle (0.5cm);
		\draw[] (4.2,0) circle (0.5cm);
		\draw[] (5.8,0) circle (0.5cm);
		\draw[] (5.8,0.5)-- (5.8,-0.5);
		\draw[fill] (3.7,0) circle (0.04cm);
		\draw[fill] (5.8,0.5) circle (0.04cm);
		\draw[fill] (5.8,-0.5) circle (0.04cm);
		\node at (0.5,0){$\Gamma=S\;+$};
		\node at (1,0){$\frac{1}{2}$};
		\node at (2.45,0){$+\frac{1}{8}$};
		\node at (5,0){$+\frac{1}{12}$};
		\node at (6.7,0){$+\cdots$};
	\begin{scope}[xshift=3.6cm]
		\draw [thick,decoration={brace,mirror,raise=0.5cm},decorate](-2.7,-0.3)--(-1.5,-0.3);
		\draw [thick,decoration={brace,mirror,raise=0.5cm},decorate](-1.1,-0.3)--(3.1,-0.3);
		\node at (-2.1,-0.85) {$\hbar\Gamma_1$};
		\node at (1.1,-0.85) { $\hbar^2\Gamma_2$};
	\end{scope}
		\end{tikzpicture}
	\end{center}
	\caption{The diagrammatic expansion of the quantum effective action runs in powers of loops.}
	\label{Fig:EffAct}
\end{figure}

\noindent
In the background field method (BFM), $\phi^{i}$ is decomposed into a background field $\bar{\phi}^{i}$ and a linear perturbation $\delta\phi^{i}$,
\begin{align}
\phi^{i}=\bar{\phi}^{i}+\delta\phi^{i}.\label{BFM}
\end{align}
 The first two orders of the expansion \eqref{EAE} correspond to the vacuum diagrams shown in Fig. \ref{Fig:EffAct}, 
\begin{align}
\Gamma_{1}=\frac{1}{2}\mathrm{Tr}\ln F_{ij},\qquad \Gamma_{2}=\frac{1}{8}G^{ij}S_{,ijk\ell}G^{k\ell}+\frac{1}{12}S_{,ijk}G^{i\ell}G^{jm}G^{kn}S_{,\ell mn}\label{OneLoop}.
\end{align}
Here, $\mathrm{Tr}$ is the functional trace, $F_{ij}$ the fluctuation operator, and the Green's function $G^{ij}$ (propagator) its inverse
\begin{align}
F_{ij}G^{ik}=-\tensor{\delta}{_{i}^{k}}.\label{GreensFunc}
\end{align}
The operator $F_{ij}$, which propagates the linear perturbations $\delta\phi^{i}$ on the background $\bar{\phi}^{i}$ is defined as the Hessian of $S$,
\begin{align}
F_{ij}(\bar{\nabla}):=\left.S_{,ij}\right|_{\phi=\bar{\phi}}.\label{DefFlucOp}
\end{align}
The covariant derivative $\nabla_{\mu}$ defines the commutator (``bundle'') curvature
\begin{align}
\tensor{\mathscr{R}}{_{\mu\nu}^{i}_{j}}\phi^{j}:=[\nabla_{\mu},\nabla_{\nu}]\phi^{i}.
\end{align}  
The effective action is the generating functional of off-shell one-particle-irreducible (1PI) $n$-point correlation functions
\begin{align}
\langle\phi_{i_1},\ldots,\phi_{i_n}\rangle=\Gamma_{,i_1,\ldots,i_n}.\label{Corr}
\end{align}
In particular, for $J_i=0$, the mean field $\varphi^i=\langle\phi\rangle$ is the solution of the quantum effective equations of motion
\begin{align}
\Gamma_{,i}=0.
\end{align}
Physical observables which derive from the S-matrix of scattering amplitudes are calculated from the off-shell correlation functions \eqref{Corr} via the Lehmann-Symanzik-Zimmermann (LSZ) reduction formula \cite{Lehmann1955}.

\subsection{Gauge Theories}
\label{sec:GaugeTheories}
In gauge theories, different field configurations which correspond to the same physical state are related by a gauge transformation
\begin{align}
\phi^{i}_{\varepsilon}:=\delta_{\varepsilon}\phi^{i}=\tensor{R}{^{i}_{\alpha}}\varepsilon^{\alpha}.\label{GTP}
\end{align}
The $\tensor{R}{^{i}_{\alpha}}(\phi)$ are the generators of gauge transformations and the $\varepsilon^{\alpha}$ the infinitesimal gauge parameter.\footnote{The generalized DeWitt gauge index $\alpha=(a,X)$ is taken from the beginning of the Greek alphabet and not to be confused with indices $\mu,\nu,\ldots$ from the tangent bundle.} For linearly realized symmetries (considered here) $\tensor{R}{_{i}^{\alpha}_{,jk}}=0$. For gauge algebras which close off-shell, the generators satisfy 
\begin{align}
\tensor{R}{^{i}_{\alpha,j}}\tensor{R}{^{j}_{\beta}}-\tensor{R}{^{i}_{\beta,j}}\tensor{R}{^{j}_{\alpha}}=\tensor{R}{^{i}_{\gamma}}\tensor{C}{^{\gamma}_{\alpha\beta}}.
\end{align}
The $\tensor{C}{^{\gamma}_{\alpha\beta}}$ are the structure functions (here assumed to be field independent $\tensor{C}{^{\gamma}_{\alpha\beta}_{,i}}=0$) and satisfy the Jacobi identity
\begin{align}
 C^{\epsilon}_{\alpha\beta}C^{\delta}_{\epsilon\gamma}+ C^{\epsilon}_{\gamma\alpha}C^{\delta}_{\epsilon\beta}+ C^{\epsilon}_{\beta\gamma}C^{\delta}_{\epsilon\alpha}=0.\label{Jacobi}
\end{align}
Gauge invariance $\delta_{\varepsilon}S=0$ of the action \eqref{act} implies the Noether identity
\begin{align}
S_{,i}\tensor{R}{^{i}_{\alpha}}=0.\label{Noether}
\end{align}
Differentiation of \eqref{Noether} shows that the fluctuation operator \eqref{DefFlucOp} for gauge theories is \textit{degenerate} (on shell $S_{,i}=0$),
\begin{align}
F_{ij}R^{i}_{\alpha}=0.\label{DegF}
\end{align} 
The gauge degeneracy $\mathrm{Det}(F_{ij})=0$ prevents the construction of the inverse $\left(F^{-1}\right)^{ij}$ and the associated Green's function $G^{ij}$ does not exit. In order to break the gauge degeneracy, a gauge-breaking action must be added
\begin{align}
S_{\mathrm{gb}}=\chi^{\alpha}O_{\alpha\beta}(\bar{\nabla})\chi^{\beta}.\label{GBAct}
\end{align} 
The background covariant gauge condition $\chi^{\alpha}(\bar{\phi};\delta\phi)$ depends linearly on the difference $\delta\phi^{i}-\bar{\phi}^{i}$ between the ``quantum field'' $\delta\phi^{i}$, i.e. the variable which is integrated over in the path integral and the background field $\bar{\phi}^{i}$. But, like the operator $O_{\alpha\beta}(\bar{\phi};\bar{\nabla})$, it might have an arbitrary (non-linear) parametric dependence on the background field $\bar{\phi}^{i}$. In this way, invariance of the effective action under \textit{background gauge transformations} is realized. For the linear split \eqref{BFM}, an infinitesimal, linearly realized gauge transformation \eqref{GTP} can be distributed in different ways, in particular by
\begin{align}
\delta^{Q}_{\varepsilon}\bar{\varphi}^i={}0,\qquad\delta^{Q}_{\varepsilon}\delta\phi^i=R^{i}_{\alpha}(\bar{\phi}+\delta\phi)\varepsilon^{\alpha},\qquad\mathrm{or}\qquad
\delta^{B}_{\varepsilon}\bar{\varphi}^i={}R^{i}_{\alpha}(\bar{\phi})\varepsilon^{\alpha},\qquad\delta^{B}_{\varepsilon}\delta\phi^i={}R^{i}_{\alpha}(\delta\phi)\varepsilon^{\alpha}.
\end{align}
While the linearity of the generators ensures that in both cases $\delta_{\varepsilon}\phi^{i}=\delta_{\varepsilon}^{Q}\left(\bar{\phi}^{i}+\delta\phi^{i}\right)=\delta_{\varepsilon}^{B}\left(\bar{\phi}^{i}+\delta\phi^{i}\right)=R^{i}_{\alpha}(\phi)\varepsilon^{\alpha}$,
the ``quantum gauge transformation'' $\delta_{\varepsilon}^{\mathrm{Q}}$ does not affect the background field $\bar{\phi}^{i}$ but only the ``quantum'' field $\delta\varphi^{i}$, while for the background gauge transformations $\delta_{\varepsilon}^{\mathrm{B}}$, the transformation \eqref{GTP} is split between the background field and the quantum field according to \eqref{BFM}. 
The gauge-breaking action \eqref{GBAct} must be compensated by the ghost action 
\begin{align}
S_{\mathrm{gh}}=c^{*}_{\alpha}\tensor{Q}{^{\alpha}_{\beta}}c^{\beta}.\label{Ghact}
\end{align} 
The anticommuting independent ghost field $c^{\alpha}$ and anti-ghost field $c^{*}_{\alpha}$ have fermionic statistics. The ghost operator $\tensor{Q}{^{\alpha}_{\beta}}$ is defined as the variation of the gauge-transformed gauge condition
\begin{align}
\tensor{Q}{^{\alpha}_{\beta}}(\bar{\nabla}):=\frac{\delta \chi^{\alpha}[\phi^{i}_{\varepsilon}]}{\delta\varepsilon^{\beta}}.\label{Gop}
\end{align}
Summarizing, for gauge theories, the partition function \eqref{PI} generalizes to
\begin{align}
Z[J]=\mathrm{Det}\left(O_{\alpha\beta}\right)^{1/2}\int\mathcal{D}[\phi,c,c^{*}]e^{-\left\{S_{\mathrm{tot}}[\phi,c,c^*]+J_i\phi^{i}\right\}},
\end{align}
with the total action $S_{\mathrm{tot}}$ defined as the sum of \eqref{act}, \eqref{GBAct} and \eqref{Ghact},
\begin{align}
S_{\mathrm{tot}}[\phi,c,c^{*}]=S[\phi]+S_{\mathrm{gb}}[\bar{\phi};\phi]+S_{\mathrm{gh}}[\bar{\phi},c,c^{*}].\label{STot}
\end{align}
In particular, the gauge-fixed fluctuation operator is no longer degenerate and can be inverted
\begin{align}
F_{ij}^{\mathrm{gb}}(\bar{\nabla}):=\left.\left(S+S_{\mathrm{gb}}\right)_{ij}\right|_{\phi=\bar{\phi}}.\label{GaugeFixedOp}
\end{align}
The structure of the effective action and the proof of perturbative renormalizability of a local gauge theory are described in more general terms by exploiting the residual non-linearly realized Becchi-Rouet-Stora-Tyutin (BRST) symmetry of the gauge-fixed action \cite{Becchi1976,Tyutin1975}. For the application of these methods in the context of General Relativity and Yang-Mills theories, see \cite{Barnich1995}, for a generalization to non-relativistic theories see \cite{Barvinsky2018}.

\subsection{Functional traces and heat-kernel technique}
\label{sec:HeatKernel}
In addition to the abstract formalism presented in Sec. \ref{Sec:PQDG}, explicit calculations in the perturbative expansion \eqref{EAE} require to evaluate functional traces, for which the combination of the BFM with heat-kernel techniques provides a manifest covariant and efficient tool.\footnote{The heat-kernel is in particular very efficient for the extraction of the one-loop divergences. For calculations involving higher-loop orders, it is not so well developed, see however \cite{Barvinsky1987a}. } For the connection between the heat-kernel technique and position space Feynman diagrams in curved spacetime, see e.g.~\cite{Barvinsky1985, Vilkovisky:1992za}. For an introduction into the background field method, see \cite{Abbott1982,Abbott:1980hw,Abbott:1983zw}. For an overview of flat-space Feynman-diagrammatic calculations in momentum space, see e.g.~\cite{Ellis:2011cr} and \cite{Dixon:1996wi} for an introduction in modern on-shell methods. An explicit illustration of the connection between the different techniques is given Sec.~\ref{Sec:RG2D} in the context of the one-loop divergences for projectable Ho\v{r}ava gravity.

The heat-kernel technique, originally developed in mathematics in the context of asymptotic expansions, partial differential equations and the geometric analysis of the Laplace operator \cite{Hadamard1923,Minakshisundaram1949,Minakshisundaram1953,Atiyah1973, Atiyah1975,Gilkey1975}, turned out to be also a very fruitful tool in physics, and, in particular, in the context of renormalization in quantum field theory on a curved background \cite{DeWitt1965,Barvinsky1985,Avramidi2000,Vassilevich2003}.
Recalling the definition of the ultra-condensed DeWitt notation, the (gauge-fixed) fluctuation operator \eqref{GaugeFixedOp} acquires the general form $F_{ij}(\bar{\nabla})=F_{AB}(\bar{\nabla}_{X_{A}})\delta(X_{A},X_{B})$.
The operator with proper index positions $\tensor{F}{^{A}_{\;B}}$, acting on the fluctuation field $\delta\phi^{A}(X)$ is obtained from $F_{AB}$ by raising the bundle index $A$ with the (ultra-local) configuration space metric $C_{AB}$,\footnote{If the configuration space of fields $\mathcal{C}$ is viewed as differentiable manifold, the configuration space metric defines the invariant line element $\mathrm{d}S^2=\mathcal{C}_{ij}\mathrm{d}\phi^{i}\mathrm{d}\phi^{j}$. Ultralocality means that $\mathcal{C}_{ij}=C_{AB}\delta(X_{A},X_{B})$ with $\mathcal{C}_{AB}$ involving no derivatives. For $2k$th-order derivative theories, defined by an action functional \eqref{act}, the configuration space metric $C_{AB}$ might be defined by the coefficient of the (minimal part of the) highest derivative term in the fluctuation operator $F_{AB}=C_{AB}\Delta^k+\ldots$. The inverse is defined via $C^{AC}C_{CB}=\delta^{A}_{B}=\mathbf{1}$. The boldface notation is exclusively reserved for matrix-valued operators with proper index positions. Since the content of this section holds for general operators $\mathbf{F}$, no background tensors appear in what follows. }
\begin{align}
\mathbf{F}(\nabla):=\tensor{F}{^{A}_{B}}(\nabla)=C^{AC}F_{CB}(\nabla).\label{Op}
\end{align} 
Inverse powers and the logarithm of the operator \eqref{Op}, which appear in the perturbative expansion \eqref{EAE}, are conveniently expressed in terms of the Schwinger integral representation\footnote{The inverse $\mathbf{F}^{-1}$ of the operator $\mathbf{F}$ is denoted as $\mathbf{1}/F$. It is assumed that $\mathbf{F}$ is positive definite. In the integral relation for the logarithm \eqref{SchwingerInt}, an (infinite) constant has been neglected. The precise relation can be defined by a regularizing mass damping factor, i.e.~by defining $\mathbf{G}(m^2):=\int_{0}^{\infty}\mathrm{d}se^{-s m^2}e^{-s\mathbf{F}}$, the logarithm of $\mathbf{F}$ is obtained as limit $\ln\mathbf{F}=\underset{m^2\to\infty}{\lim}\left[\ln m^2\mathbf{1}-\int_{0}^{m^2}\mathrm{d}\mu^2\mathbf{G}(\mu^2)\right]$.} over ``proper time'' $s$,
\begin{align}
\frac{\mathbf{1}}{F^n}={}&\int_{0}^{\infty}\frac{{\rm d}s}{(n-1)!}s^{n-1}\,e^{-s\,\mathbf{F}},\qquad \ln\mathbf{F}={}-\int_{0}^{\infty}\frac{\mathrm{d}s}{s}e^{-s\mathbf{F}}.\label{SchwingerInt}
\end{align}
The \textit{heat-kernel} $\mathbf{K}_{F}(s|X,Y)$, associated with the operator $\mathbf{F}$, formally satisfies the heat equation
\begin{align}
\mathbf{K}_{F}(s|X,Y):=e^{ -s\,\mathbf{F}(\nabla_X)}\delta(X,Y),\qquad \left[\frac{\partial}{\partial s}+\mathbf{F}(\nabla)\right]\mathbf{K}_{F}(s|X,Y)=0.\label{HK}
\end{align} 
In terms of the heat-kernel \eqref{HK}, the one-loop contribution to the effective action \eqref{OneLoop} acquires the form
\begin{align}
\Gamma_{1}
={}&-\frac{1}{2}\int_{0}^{\infty}\frac{{\rm d}s}{s}\,\text{Tr}\,\mathbf{K}_F(s|X,Y)=-\frac{1}{2}\int_{0}^{\infty}\frac{{\rm d}s}{s}\,\int{\rm d}^dx\,\text{tr}\,\left[K_{F}\right]^{A}_{\;\;B}(s|,X,X).\label{FuncTrace}
\end{align}
The last equation might be viewed as the definition of the functional trace $\mathrm{Tr}$ and requires to evaluate the spacetime integral over the internal trace $\mathrm{tr}$ of the coincidence limit $y\to x$ of the matrix valued two-point kernel $\left[K_{F}\right]^{A}_{\;\;B}(s|,X,Y)$. UV divergences arise from the lower integration bound in \eqref{FuncTrace}, i.e. the $s\to0$  limit.

For a \textit{minimal second-order operator} with (positive definite) Laplacian $\mathbf{\Delta}=-g^{\mu\nu}\nabla_{\mu}\nabla_{\nu}\mathbf{1}$ and potential $\mathbf{P}$,
\begin{align}
\mathbf{F}(\nabla)=\mathbf{\Delta}+\mathbf{P},\label{secor}
\end{align}
there is an ansatz for the associated heat-kernel at \textit{non-coincident points}, introduced in \cite{DeWitt1965},
\begin{align}
\mathbf{K}(s|X,Y)=\frac{g^{1/2}(y)}{(4\,\pi\,s)^{d/2}}\,\mathfrak{D}^{1/2}(X,Y)\,e^{-\frac{\sigma(X,Y)}{2\,s}}\,\mathbf{\Omega}(s|X,Y).\label{ansatz}
\end{align}
Synge's world function $\sigma(X,Y)$ is a bi-scalar \cite{Synge1960}, which measures one-half of the geodesic distance squared between the points $X$ and $Y$, and $\mathfrak{D}(X,Y)$ is the de-densitized Van Vleck determinant, a bi-scalar defined as
\begin{align}
\mathfrak{D}(X,Y):=g^{-1/2}(X)\det\left(\frac{\partial^2\sigma(X,Y)}{\partial X^{\mu}\partial Y^{\nu}}\right)g^{-1/2}(Y).
\end{align}
The bi-tensor $\mathbf{\Omega}$ can be obtained in the form of an asymptotic expansion in proper time
\begin{align}
\mathbf{\Omega}(s|X,Y):=\sum_{n=0}^{\infty} \mathbf{a}_{n}(X,Y)\,s^n, \qquad \mathbf{a}_n(X,X)\propto \underbrace{\nabla_X...\nabla_X}_{2p}\,\underbrace{\mathfrak{R}....\mathfrak{R}}_{m},\quad n=p+m.
\end{align}
The \textit{Schwinger-DeWitt coefficients} (SDW) at coincidence points $\mathbf{a}_{n}(X,X)$ are local functions of the background fields and the generalized curvature $\mathfrak{R}$ encompasses three different types of \textit{background curvatures} $\mathfrak{R}=\{R_{\mu\nu\rho\sigma}\mathbf{1},\tensor{\pmb{\mathscr{R}}}{_{\mu\nu}},\mathbf{P}\}$.

For the minimal second-order operators \eqref{secor}, a closed-form algorithm for the calculation of the one-loop divergences $\Gamma^{\mathrm{div}}_1$ is available.  In general, dimensional regularization annihilates all power-law divergences and is only sensitive to logarithmic divergences, which are isolated as poles in dimension ${\epsilon^{-1}=2/(4-D)}$. In $D=4$, the logarithmically UV-divergent part of the one-loop contributions to the effective action \eqref{FuncTrace} for the minimal second-order operator \eqref{secor} are determined by the coincidence limit of  $\mathbf{a}_{2}(x,x)$ \cite{DeWitt1965},
\begin{align}
\Gamma_{1}^{\rm div}={}&-\frac{1}{\epsilon}\frac{1}{32\,\pi^2}\int{\rm d}^4X\,g^{1/2}\,\text{tr}\,\mathbf{a}_{2}(X,X).\label{1Ldiv4}
\end{align}
The coincidence limits of the Schwinger-DeWitt coefficients $\mathbf{a}_{n}(x,x)$ can be calculated iteratively
by inserting the ansatz \eqref{ansatz} into the heat equation \eqref{HK}, leading to the recurrence relation (for $n\geq0$),
\begin{align}
\left[(n+1)+\sigma^{\mu}\nabla_{\mu}\right]\mathbf{a}_{n+1}={}&\mathfrak{D}^{-1/2}\mathbf{F}(\nabla)\left(\mathfrak{D}^{1/2}\mathbf{a}_{n}\right)=0.
\end{align}
In order to obtain $\mathbf{a}_{2}(X,X)$ in this way, the coincidence limits of $\sigma$, $\mathfrak{D}$, $\mathbf{a}_0$, $\mathbf{a}_1$ and derivatives thereof must be calculated. The successive pattern of this calculation is illustrated in Table \ref{Table1}. 
\begin{table}[h!]
	\begin{center}
		\begin{tabular}{lccccc}
			\toprule
			$\mathfrak{R}\quad$&$\mathfrak{R}^{0}\quad$&$\mathfrak{R}^{\nicefrac{1}{2}}\quad$&$\mathfrak{R}\quad$&$\mathfrak{R}^{\nicefrac{3}{2}}\quad$&$\mathfrak{R}^2$\\
			\midrule
			\addlinespace
			$\sigma$&$\nabla^2\sigma$&$\nabla^3\sigma$&$\nabla^4\sigma$&$\nabla^5\sigma$&$\nabla^6\sigma$\\
			\midrule
			\addlinespace
			$\mathfrak{D}$&$\mathfrak{D}$&$\nabla\mathfrak{D}$&$\nabla^2\mathfrak{D}$&$\nabla^3\mathfrak{D}$&$\nabla^4\mathfrak{D}$\\
			\midrule
			\addlinespace
			$\mathbf{a}_0$&$\mathbf{a}_0$&$\nabla\mathbf{a}_0$&$\nabla^2\mathbf{a}_0$&$\nabla^3\mathbf{a}_0$&$\nabla^4\mathbf{a}_0$\\
			\midrule
			\addlinespace
			$\mathbf{a}_1$&{}&{}&$\mathbf{a}_{1}$&$\nabla\mathbf{a}_1$&$\nabla^2\mathbf{a}_1$\\
			\midrule
			\addlinespace
			$\mathbf{a}_2$&{}&{}&{}&{}&$\mathbf{a}_2$\\
			\bottomrule
		\end{tabular}
	\end{center}
	\caption{Coincidence limits required for the calculation of $\mathbf{a}_{2}(X,X)$. }
	\label{Table1}
\end{table}

\noindent The coincidence limits of $\sigma$, $\mathfrak{D}$, $\mathbf{a}_0$ and their derivatives can be obtained by successive differentiation of the ``defining equations'' for $\sigma$, $\mathfrak{D}$ and $\mathbf{a}_{0}$,
\begin{align}
\sigma^{\mu}\sigma_{\mu}=2\sigma,\qquad\mathfrak{D}^{-1}\nabla^{\mu}(\mathfrak{D}\,\sigma_{\mu})={}d,\qquad \sigma^{\mu}\,\nabla_{\mu}\mathbf{a}_0=0,
\end{align}
provided with the ``initial conditions'' $\sigma|_{y=x}=0$, $\mathfrak{D}\big|_{y=x}={}1$ and $\mathbf{a}_0|_{y=x}=\mathbf{1}$.
In this way, the coincidence limit of $\mathbf{a}_{2}(X,Y)$ is found as \cite{DeWitt1965,Barvinsky1985},
\begin{align}
\mathbf{a}_{2}(X,X)={}&\frac{1}{180}\left(R_{\mu\nu\rho\sigma}R^{\mu\nu\rho\sigma}-R_{\mu\nu}R^{\mu\nu}-6\Delta R\right)\mathbf{1}+\frac{1}{2}\left(\mathbf{P}^2-\frac{1}{6}R\mathbf{1}\right)^2+\frac{1}{12}\pmb{\mathscr{R}}_{\mu\nu}\pmb{\mathscr{R}}^{\mu\nu}+\frac{1}{6}\Delta\,\mathbf{P}.\label{a2}
\end{align}

For higher-order and non-minimal operators there is no closed expression for the one-loop divergences \eqref{1Ldiv4} in terms of a single SDW coefficient as for the minimal second-order operator \eqref{secor}. Nevertheless, in \cite{Barvinsky1985} a closed algorithm has been developed, which allows to reduce the calculation of the one-loop divergences for higher-order and non-minimal operators to the heat-kernel of the second-order minimal operator \eqref{ansatz} and a few \textit{universal functional traces}
\begin{align}
\pmb{\mathscr{U}}_{\mu_1\ldots\mu_p}^{(p,n)}:=\left.\nabla_{\mu_1}\ldots\nabla_{\mu_p}\frac{\mathbf{1}}{\Delta^n}\right|^{\mathrm{div}}_{Y=X}.\label{uft}
\end{align} 
The perturbative algorithm underlying the generalized Schwinger-DeWitt technique relies on the non-degeneracy of the \textit{principal symbol} $\mathbf{D}$ of the operator $\mathbf{F}$. There are, however, important physical theories, for which the principal symbol of the fluctuation operator is degenerate and the (generalized) Schwinger-DeWitt algorithm is not directly applicable. In such cases, more general methods are required; see \cite{Ruf2018,Ruf2018b,Ruf2018a} for heat-kernel calculations involving operators with degenerate principal part and \cite{Heisenberg2019a,Heisenberg2019} for operators with Laplacians constructed from an effective (background field dependent) metric. In the context of Lifshitz theories, the development of heat-kernel technique for anisotropic operators has recently been initiated \cite{Nesterov2011,DOdorico2014,Barvinsky2017a}.

\section{Perturbative quantum General Relativity}
\label{Sec:PQDGR}
\subsection{Classical General Relativity}
\label{sec:ClassicalGR}
In the theory of General Relativity (GR), the gravitational interaction manifests itself \textit{geometrically} as curvature of spacetime and couples \textit{universally} to all fields, which, when combined with the attractive nature of gravity, implies that it cannot be shielded. In Einstein's theory, the \textit{dynamical} character of the spacetime geometry is encoded in the dynamics of the metric field $g_{\mu\nu}(X)$. 
The action functional of GR is the Einstein-Hilbert action,
\begin{align}
S_{\mathrm{EH}}=\frac{M_{\mathrm{P}}^2}{2}\int\mathrm{d}^DX\sqrt{-g}\left(R-2\Lambda\right).\label{EHact}
\end{align}
The action \eqref{EHact} involves the invariant volume element with determinant $g=\mathrm{det}(g_{\mu\nu})$, the Ricci scalar $R=g^{\mu\nu}R_{\mu\nu}$ as well as the cosmological constant $\Lambda$.\footnote{I work on a $D-$dimensional (pseudo)-Riemannian manifold $\mathcal{M}$ with local coordinates $X^{\mu}$, $\mu=0,1,2,3$, a metric structure $g_{\mu\nu}$ with inverse $g^{\mu\nu}$ defined via $g^{\mu\rho}g_{\rho\nu}=\delta^{\mu}_{\nu}$ and the torsion-free metric-compatible Christoffel connection ${\Gamma^{\rho}_{\mu\nu}=g^{\rho\sigma}\left(\partial_{\mu}g_{\sigma\nu}+\partial_{\nu}g_{\mu\sigma}-\partial_{\sigma}g_{\mu\nu}\right)/2}$, which defines the covariant derivative $\nabla_{\mu}$. I use the following conventions for the Lorentzian signature $\mathrm{sig}(g)=\mathrm{diag}(-1,1,1,\ldots,1)$, the Riemann curvature tensor $R^{\rho}_{\;\mu\sigma\nu}=\partial_{\sigma}\Gamma^{\rho}_{\mu\nu}- \partial_{\nu}\Gamma^{\rho}_{\mu\sigma}+\Gamma^{\lambda}_{\mu\nu}\Gamma^{\rho}_{\lambda\sigma}-\Gamma^{\lambda}_{\mu\sigma}\Gamma^{\rho}_{\lambda\nu}$, and the Ricci tensor $R_{\mu\nu}=R^{\rho}_{\;\mu\rho\nu}$. I use natural units in which the speed of light $c$ and Planck's constant $\hbar$ are set to one $c=\hbar=1$ and Newton's constant $G_{\mathrm{N}}$ can be expressed in terms of the the reduced Planck mass $M_{\mathrm{P}}:=1/\sqrt{8\pi G_{\mathrm{N}}}$ . } 
The dynamics of $g_{\mu\nu}$ is determined by Einstein's field equations, obtained from extremizing the total action ${S[g,\Psi]=S_{\mathrm{EH}}[g]+S_{\mathrm{M}}[g,\Psi]}$ with respect to $g_{\mu\nu}$,
\begin{align}
R_{\mu\nu}-\frac{1}{2}g_{\mu\nu}R+\Lambda g_{\mu\nu}=M_{\mathrm{P}}^{-2}T_{\mu\nu}.\label{Eeq}
\end{align}
The energy momentum tensor $T_{\mu\nu}$ derives from the ``matter'' action $S_{\mathrm{M}}[\Psi]$, with all non-geometrical ``matter'' fields collectively denoted by $\Psi$,
\begin{align}
T_{\mu\nu}=-\frac{2}{\sqrt{-g}}\frac{\delta S_{\mathrm{M}}[\Psi,g]}{\delta g^{\mu\nu}}.
\end{align}
Infinitesimal spacetime distances $\mathrm{d}s$ measured by the metric field $g_{\mu\nu}$ are defined by the line element 
\begin{align}
\mathrm{d}s^2=g_{\mu\nu}(X)\mathrm{d}X^{\mu}\mathrm{d}X^{\nu}.\label{LineElement}
\end{align}
Denoting the \textit{mass dimension} by $[\ldots]_{\mathrm{M}}$ and assigning coordinates $X^{\mu}$ the dimension of a length ${[X]_{\mathrm{M}}=-1}$, implies 
\begin{align}
[\partial_{\mu}]_{\mathrm{M}}={}1,\qquad [g_{\mu\nu}]_{\mathrm{M}}=0,\qquad [R_{\mu\nu\rho\sigma}]_{\mathrm{M}}=2,\qquad [G_{\mathrm{N}}]_{\mathrm{M}}={}-(D-2),\qquad[\Lambda]_{\mathrm{M}}=2.\label{MDimCouplings}
\end{align}
The Ricci scalar $R$ is the only curvature invariant involving exactly two spacetime derivatives. Except for the cosmological constant, all other curvature invariants necessarily contain higher derivatives. In $D=4$, these are the only two classically relevant local curvature operators.\footnote{I call an operator $O$ classically relevant if $[\mathcal{O}]_{\mathrm{M}}<D$, classically marginal if $[\mathcal{O}]_{\mathrm{M}}=D$ and classically irrelevant if $[\mathcal{O}]_{\mathrm{M}}>D$.}

The metric field transforms as a rank $(0,2)$ tensor under $D$-dimensional coordinate transformations ${X^{\mu}\to\tilde{X}^{\mu}(X)}$, 
\begin{align}
g_{\mu\nu}(X)\mapsto\tilde{g}_{\mu\nu}(\tilde{X})=g_{\alpha\beta}(X)\frac{\partial X^{\alpha}}{\partial \tilde{X}^{\mu}}\frac{\partial X^{\beta}}{\partial \tilde{X}^{\nu}}.
\end{align}
The invariance group of GR are the $D$-dimensional diffeomorphisms $\mathrm{Diff}(\mathcal{M})$. The change of the metric field under an infinitesimal diffeomorphism $\delta_{\xi}$ generated by the vector field $\xi^{\mu}$ is given by the Lie derivative of $g_{\mu\nu}$ along $\xi^{\mu}$,
\begin{align}
\delta_{\xi}g_{\mu\nu}=\left(\mathcal{L}_{\xi}g\right)_{\mu\nu}=\xi^{\rho}\partial_{\rho}g_{\mu\nu}+2g_{\rho(\nu}\partial_{\mu)}\xi^{\rho}=2\nabla_{(\mu}\xi_{\nu)}.\label{Diffeog}
\end{align}
Round brackets in \eqref{Diffeog} denote symmetrization among the enclosed indices with unit weight and $\xi_{\mu}=g_{\mu\rho}\xi^{\rho}$.
Since the gravitational field equations \eqref{Eeq} relate geometry with matter, consistency requires that $S_{\mathrm{M}}[g,\Psi]$ must as well be invariant under $\mathrm{Diff}(\mathcal{M})$, which implies the ``on-shell'' covariant conservation of the energy momentum tensor ${\nabla^{\mu}T_{\mu\nu}=0}$.

\subsection{Quantum GR}
\label{Sec:QuantumGR}
In order to establish contact with the general formalism of perturbative QFT reviewed in Sec. \ref{Sec:PQDG}, the generalized field $\phi^{i}$ in GR is to be identified with the metric field $\phi^{i}\mapsto g_{\mu\nu}(X)$. Comparison of \eqref{act} with the Einstein-Hilbert action \eqref{EHact} implies that the operators $\mathcal{O}_i(g,\partial)$ and the coupling constants $c_i$ are to be identified as follows
\begin{align}
\mathcal{O}_{1}(g)\mapsto\sqrt{-g},\qquad  c_1\mapsto -M_{\mathrm{P}}^2\Lambda,\qquad
\mathcal{O}_{2}(g,\partial)\mapsto\sqrt{-g}R,\qquad c_2\mapsto \frac{M_{\mathrm{P}}^2}{2}.\label{RelOp}
\end{align}
The particle spectrum of GR is derived by expanding the action \eqref{EHact} to quadratic order in the linear perturbations,
\begin{align}
h_{\mu\nu}=g_{\mu\nu}-\bar{g}_{\mu\nu}
\end{align} 
around a flat background $\bar{g}_{\mu\nu}=\eta_{\mu\nu}$.\footnote{The particle spectrum of a QFT is usually derived by expanding the action up to quadratic order in the linear perturbation around the vacuum. In relativistic QFTs, the natural vacuum is Minkowski space, which, even in the presence of gravity, might be justified locally by the equivalence principle. Minkowski space is a maximal symmetric space whose isometries are generated by the ${D(D+1)/2}$ linearly independent Killing vectors, which correspond to the generators of infinitesimal transformations of the Poincar\'e group. In this way, the Minkowski vacuum is connected to the representation theory of the Poincar\'e group ultimately giving rise to Wigner's classification \cite{Wigner1939}, in which particles are classified according to their mass and their spin, i.e. the eigenvalues of the Casimir operators of the Poincar\'e group. A positive cosmological constant $\Lambda>0$ suggests however that the global vacuum is De Sitter space rather than Minkowski space. De Sitter space is also a maximally symmetric space whose Killing vectors are the generators of the De Sitter group. More generally, this also suggests that for an arbitrary spacetime without any symmetry, the very concept of a particle is not really well defined.} Absorbing a factor of $M_{\mathrm{P}}/2$ in the definition of $h_{\mu\nu}$, i.e. $h_{\mu\nu}\mapsto 2h_{\mu\nu}/M_{\mathrm{P}}$, defining $h=\eta^{\mu\nu}h_{\mu\nu}$ and $\partial^2:=\eta^{\mu\nu}\partial_{\mu}\partial_{\nu}$, upon integration by parts the result reads
\begin{align}
S_{\mathrm{EH}}^{(2)}|_{\bar{g}=\eta}=\int\mathrm{d}^DX\left[h^{\mu\nu} \partial^2h_{\mu\nu} -  h \partial^2h - 2 h^{\mu\nu} \partial_{\nu}\partial_{\rho}h_{\mu}{}^{\rho}+2h^{\mu\nu} \partial_{\nu}\partial_{\mu}h\right].
\end{align}
After Fourier transformation $\partial_{\mu}\mapsto iP_{\mu}$ with four momentum $P^{\mu}$ and square $P^2=\eta_{\mu\nu}P^{\mu}P^{\nu}$, the fluctuation operator \eqref{DefFlucOp} in momentum space might be expressed in terms of spin-projection operators
\begin{align}
F^{\mu\nu,\rho\sigma}(-P^2)=\left[\tensor{\Pi}{^{(2)}^{\mu\nu\rho\sigma}}-(D-2)\tensor{\Pi}{^{(0,ss)}^{\mu\nu\rho\sigma}}\right]\left(-P^2\right).\label{FopMom}
\end{align}
The spin-projection operators acting on the symmetric rank-two tensor $h_{\mu\nu}$ read
\begin{align}
\tensor{\Pi}{^{(2)}_{\mu\nu}^{\rho\sigma}}={}&\frac{1}{2}\left(\tensor{\Pi}{^{(\mathrm{T})}_{\mu}^{\rho}}\tensor{\Pi}{^{(\mathrm{T})}_{\nu}^{\sigma}}+\tensor{\Pi}{^{(\mathrm{T})}_{\mu}^{\sigma}}\tensor{\Pi}{^{(\mathrm{T})}_{\nu}^{\rho}}\right)-\frac{1}{D-1}\tensor{\Pi}{^{(\mathrm{T})}_{\mu\nu}}\tensor{\Pi}{^{(\mathrm{T})}^{\rho\sigma}},\label{PS2}\\
\tensor{\Pi}{^{(1)}_{\mu\nu}^{\rho\sigma}}={}&\frac{1}{2}\left(\tensor{\Pi}{^{(\mathrm{T})}_{\mu}^{\rho}}\tensor{\Pi}{^{(\mathrm{L})}_{\nu}^{\sigma}}+\tensor{\Pi}{^{(\mathrm{T})}_{\mu}^{\sigma}}\tensor{\Pi}{^{(\mathrm{L})}_{\nu}^{\rho}}+\tensor{\Pi}{^{(\mathrm{T})}_{\nu}^{\rho}}\tensor{\Pi}{^{(\mathrm{L})}_{\mu}^{\sigma}}+\tensor{\Pi}{^{(\mathrm{T})}_{\nu}^{\sigma}}\tensor{\Pi}{^{(\mathrm{L})}_{\mu}^{\rho}}\right),\\
\tensor{\Pi}{^{(0,ss)}_{\mu\nu}^{\rho\sigma}}={}&\frac{1}{D-1}\tensor{\Pi}{^{(\mathrm{T})}_{\mu\nu}}\,\tensor{\Pi}{^{(\mathrm{T})}^{\rho\sigma}}\label{PScal1},\\
\tensor{\Pi}{^{(0,ww)}_{\mu\nu}^{\rho\sigma}}={}&\tensor{\Pi}{^{(\mathrm{L})}_{\mu\nu}}\;\tensor{\Pi}{^{(\mathrm{L})}^{\rho\sigma}},\label{PScal2}\\
\tensor{\Pi}{^{(0,sw)}_{\mu\nu}^{\rho\sigma}}={}&\frac{1}{\sqrt{D-1}}\tensor{\Pi}{^{(\mathrm{T})}_{\mu\nu}}\,\tensor{\Pi}{^{(\mathrm{L})}^{\rho\sigma}},\\
\tensor{\Pi}{^{(0,ws)}_{\mu\nu}^{\rho\sigma}}={}&\frac{1}{\sqrt{D-1}}\tensor{\Pi}{^{(\mathrm{L})}^{\rho\sigma}}\,\tensor{\Pi}{^{(\mathrm{T})}_{\mu\nu}},\label{PScal4}
\end{align}
Here, $\tensor{\Pi}{^{(\mathrm{T})}}$ and $\tensor{\Pi}{^{(\mathrm{L})}}$ are the transversal and longitudinal vector field projectors
\begin{align}
\tensor{\Pi}{^{(\mathrm{T})}_{\mu}^{\nu}}=\delta_{\mu}^{\nu}-\frac{P_{\mu}P^{\nu}}{P^2},\qquad \tensor{\Pi}{^{(\mathrm{L})}_{\mu}^{\nu}}=\frac{P_{\mu}P^{\nu}}{P^2}.
\end{align}
Note that the scalar sector \eqref{PScal1}-\eqref{PScal4} is non-diagonal, such that aside from the diagonal projection operators $P^{(0,ss)}$ and  $P^{(0,ww)}$ there are the two intertwining operators $\Pi^{(0,sw)}$ and $\Pi^{(0,ws)}$ which connect the two spin-$0$ representations $s$ and $w$. The operators satisfy the algebra (orthogonality and idempotency relations)
\begin{align}
\tensor{\Pi}{^{(I,ij)}_{\mu\nu}^{\alpha\beta}}\tensor{\Pi}{^{(J,kl)}_{\alpha\beta}^{\rho\sigma}}=\delta^{IJ}\delta^{ik}\tensor{\Pi}{^{(J,jl)}_{\mu\nu}^{\rho\sigma}},
\end{align}
with $J=2,1,0$ labeling the spin of the representation and $i,j,k,l=s,w$ labeling the different spin-$0$ operators. In addition, the diagonal operators \eqref{PS2}-\eqref{PScal2} satisfy the completeness relation
\begin{align}
\tensor{\Pi}{^{(2)}_{\mu\nu}^{\rho\sigma}}+\tensor{\Pi}{^{(1)}_{\mu\nu}^{\rho\sigma}}+\tensor{\Pi}{^{(0,ss)}_{\mu\nu}^{\rho\sigma}}+\tensor{\Pi}{^{(0,ww)}_{\mu\nu}^{\rho\sigma}}=\delta_{\mu\nu}^{\rho\sigma},\label{Completness}
\end{align} 
with $\delta_{\mu\nu}^{\rho\sigma}=(\delta_{\mu}^{\rho}\delta_{\nu}^{\sigma}+\delta_{\nu}^{\rho}\delta_{\mu}^{\sigma})/2$ denoting the identity in the space of symmetric rank-two tensors. Finally, the traces of the operators \eqref{PS2}-\eqref{PScal2} yield the dimensions of the invariant subspaces, which, according to \eqref{Completness}, add up to the $D(D+1)/2$ components of a symmetric rank-two tensor $h_{\mu\nu}$,
\begin{align}
\mathrm{tr}\,\tensor{\Pi}{^{(2)}}={}\frac{1}{2}\left(D+1\right)\left(D-2\right),\qquad\mathrm{tr}\,\tensor{\Pi}{^{(1)}}={}D-1,\qquad\mathrm{tr}\,\tensor{\Pi}{^{(0,ss)}}={}1,\qquad \mathrm{tr}\,\tensor{\Pi}{^{(0,ww)}}={}1.
\end{align}
Despite the appearance of the spin-$0$ projector in \eqref{FopMom}, the spectrum of propagating particles in GR in $D$ dimensions only encompasses the massless spin-$2$ graviton -- the scalar mode can be eliminated by a residual gauge transformation and is not a physical degree of freedom. As explained in \eqref{DegF}, the operator \eqref{FopMom} is degenerate and a gauge-fixing is required for its inversion. Choosing $O_{\mu\nu}=-\eta_{\mu\nu}\delta(x-y)$ for the operator in \eqref{GBAct} and the De Donder gauge condition on a flat background 
\begin{align}
\chi^{\mu}[\eta,g]=\left(\eta^{\mu\rho}\eta^{\nu\sigma}-\frac{1}{2}\eta^{\rho\sigma}\eta^{\mu\nu}\right)\partial_{\nu}h_{\rho\sigma},\label{GCGRFlat}
\end{align}
the flat gauge-fixed fluctuation operator \eqref{GaugeFixedOp} of GR in momentum space reads 
\begin{align}
F^{\mu\nu,\rho\sigma}_{gf}(-P^2)=\frac{1}{2}\left[\eta^{\mu\rho}\eta^{\nu\sigma}+\eta^{\mu\sigma}\eta^{\nu\rho}-\eta^{\mu\nu}\eta^{\rho\sigma}\right]\left(-P^2\right).\label{FgfopMom}
\end{align}
Inversion of \eqref{FgfopMom} leads to the spin-$2$ propagator on a flat background\footnote{I reserve the symbol $G$ for the general Green's function in position space defined in \eqref{GreensFunc}, and use $\mathcal{P}$ instead for the flat space Green's function in momentum space.},
\begin{align}
\mathcal{P}_{\mu\nu,\rho\sigma}(-P^2)
=\frac{1}{2}\left(\eta_{\mu\rho}\eta_{\nu\sigma}+\eta_{\mu\rho}\eta_{\nu\sigma}-\frac{2}{D-2}\eta_{\mu\nu}\eta_{\rho\sigma}\right)\frac{1}{(-P^2)}.
\end{align}
The propagator $\mathcal{P}_{\mu\nu,\rho\sigma}$ defines the free theory and hence the particle spectrum in perturbation theory.
The massless graviton in $D$ dimensions has $D(D-3)/2$ polarization states, following from subtracting the $2D$ components of the independent ghost fields in \eqref{Ghact} from the $D(D+1)/2$ independent components of the symmetric rank-two tensor $h_{\mu\nu}$.

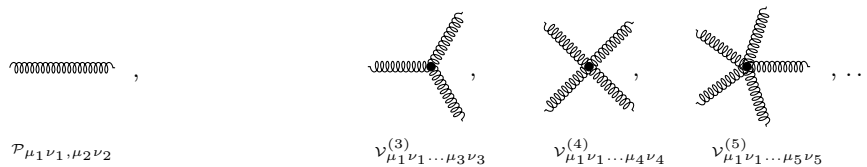
\begin{figure}[h!]
 \begin{center}
	\begin{tikzpicture}[scale=1.4]
	\begin{scope}[xshift=0cm]
	\draw [black,graviton](0,0)--(1,0);
	\node at (1.2,-0.1) {,};
	\node at (0.5,-0.8) {\tiny $\mathcal{P}_{\mu_1\nu_1,\mu_2\nu_2}$};
	\end{scope}
	\begin{scope}
	\begin{scope}[xshift=4cm,yshift=0.3]
	\filldraw[black] (0:0) circle (0.4mm);
	\draw[black,graviton] (60:0) -- (60:0.6);
	\draw[black,graviton] (180:0) -- (180:0.6);
	\draw[black,graviton] (300:0) -- (300:0.6);
	\node at (0.4,-0.1) {,};
	\node at (0,-0.8) {\tiny $\mathcal{V}^{(3)}_{\mu_1\nu_1\ldots\mu_3\nu_3}$};
	\end{scope}
	\begin{scope}[xshift=5.5cm,yshift=0.3]
	\filldraw[black] (0:0) circle (0.4mm);
	\draw[black,graviton] (45:0) -- (45:0.6);
	\draw[black,graviton] (135:0) -- (135:0.6);
	\draw[black,graviton] (225:0) -- (225:0.6);
	\draw[black,graviton] (315:0) -- (315:0.6);
	\node at (0.45,-0.1) {,};
	\node at (0.2,-0.8) {\tiny $\mathcal{V}^{(4)}_{\mu_1\nu_1\ldots\mu_4\nu_4}$};
	\end{scope}
	\begin{scope}[xshift=7cm,yshift=0.3]
	\filldraw[black] (0:0) circle (0.4mm);
	\draw[black,graviton] (72:0) -- (72:0.6);
	\draw[black,graviton] (72+72:0) -- (72+72:0.6);
	\draw[black,graviton] (72+72+72:0) -- (72+72+72:0.6);
	\draw[black,graviton] (72+72+72+72:0) -- (72+72+72+72:0.6);
	\draw[black,graviton] (72+72+72+72+72:0) -- (72+72+72+72+72:0.6);
	\node at (0.2,-0.8) {\tiny $\mathcal{V}^{(5)}_{\mu_1\nu_1\ldots\mu_{5}\nu_5}$};
	\node at (1,-0.1) {, $\ldots$};
	\end{scope}
	\end{scope}
	\end{tikzpicture}
	\caption{Diagrammatic representation of the propagator and the interaction vertices in GR.}
	\label{FeynmanDiagrasGR}
\end{center}
\end{figure}

The interactions in momentum space are defined by the higher $n$-point functions $\mathcal{V}^{(n)}_{\mu_1\nu_1\cdots\mu_{n}\nu_n}(P_1,\ldots,P_n)$, which derive from the Fourier transforms of the $n$th functional derivative of the action 
\begin{align}
\mathcal{V}^{(n)}_{\mu_1\nu_1\cdots\nu_n\mu_{n}}(X_1,\ldots,X_n):= \frac{\delta^n S_{\mathrm{EH}}[g]}{\delta g_{\mu_1\nu_1}(X_1)\ldots\delta g_{\mu_{n}\nu_{n}}(X_n)},\label{Vertex} \quad n>2.
\end{align}
The essential non-linearity of GR (i.e. the non-polynomial dependence of \eqref{EHact} on $g_{\mu\nu}$) is the origin for the infinite tower of interaction vertices \eqref{Vertex} with an increasing number of legs $n$.\footnote{This can be seen also as follows: Starting from a spin-$2$ particle freely propagating in flat spacetime with a linear field equation, locality and diffeomorphism invariance require to iteratively add non-linear self-interactions in a consistent way, which, when summed, recover the full non-linear theory of GR, see \cite{Deser1970}. The explicit expressions for the vertices in momentum space are rather lengthy and not very illuminating. The expression for the three-point and four-point vertices can e.g. be found in \cite{DeWitt1967a}. For these kind of calculation computer-algebra programs such as \texttt{FORM}, or the \texttt{Mathematica} based \texttt{xAct} bundle (in particular, the core package \texttt{xTensor} and the extension packages \texttt{xPert} and \texttt{xTras} packages) are indispensable \cite{FORM,xAct,xTensor,xPert,xTras}.} The diagrammatic representation of the propagator and the interaction vertices in GR are shown in Fig. \ref{FeynmanDiagrasGR}.

\noindent  The fact that the Einstein-Hilbert action is linear in the scalar curvature, implies that GR is a second-order derivative theory, such that (suppressing the index structure) the propagators have a momentum scaling $\mathcal{P}\propto P^{-2}$, while all $n$-point vertices in momentum-space scale as $\mathcal{V}^{(n)}\propto P^2$. 
Feynman diagrams with loops, such as in Fig. \eqref{Fig:EffAct}, correspond to a momentum space integral $\mathcal{I}$ which might diverge in the ultraviolet (UV).
A generic Feynman integral $\mathcal{I}$ in GR with $L$-loops, $I$ internal propagators and $V$ vertices has momentum scaling
\begin{align}
\mathcal{I}\propto\int \left(\mathrm{d}^D\,P\right)^L\frac{1}{\left(P^2\right)^I}\left(P^2\right)^{V}.
\end{align}
The superficial degree of divergence $D^{\mathrm{div}}(\mathcal{I})$ provides a simple way to estimate the leading divergence of $\mathcal{I}$ by power counting. Scaling each loop momentum by a constant factor $b$, taking the limit $b\to\infty$ and counting powers of $b$ defines $D^{\mathrm{div}}(\mathcal{I})$. If $D^{\mathrm{div}}(\mathcal{I})<0$, the associate diagram is superficially finite (i.e. finite modulo subdivergences) and if  if $D^{\mathrm{div}}(\mathcal{I})\geq0$, it is divergent.
Using the topological relation $ I-V=L-1$, valid on an abstract graph level ( i.e. independent of the underlying physical theory), the superficial degree of divergences of quantum GR reads
\begin{align}
D^{\mathrm{div}}_{\mathrm{GR}}=D\,L-2(L-1).\label{DoDGR}
\end{align}
The last equality shows that in $D=4$, the degree of divergence grows with the number of loops $L$ as $D^{\mathrm{div}}_{\mathrm{GR}}=2\,(L+1)$ and signals the perturbatively non-renormalizable character of GR, which in $D=4$ is directly connected to the negative mass dimension \eqref{MDimCouplings} of the gravitational coupling constant  $G_{N}=M_{\mathrm{P}}^{-2}$.

In addition to this simple power counting argument, the UV divergences of GR and its coupling to matter fields have been calculated in various approximations: for GR with and without a scalar field, the one-loop divergences were first derived in \cite{Hooft1974}. In subsequent works, the one-loop divergences were extended, including GR coupled to abelian and non-abelian gauge-fields \cite{Deser1974a,Deser1974}, GR coupled to fermions \cite{Deser1974b}, GR with a cosmological constant \cite{Gibbons1978a,Christensen1980},
GR with non-minimal gauges \cite{Barvinsky1985} and GR coupled non-minimally to a scalar field \cite{Barvinsky1993,Shapiro1995,Steinwachs2011}. At the two-loop order, the calculations of the UV divergences for pure gravity have first been performed in \cite{Goroff1985,Goroff1986} and later confirmed in \cite{Ven1992}, see also \cite{Barvinsky1987a}. 

In order to establish contact with the general formalism outlined in Sec. \ref{Sec:PQDG}, I briefly illustrate the calculation of the one-loop divergences for the Euclidean version of the Einstein-Hilbert action \eqref{EHact} in $D=4$,
\begin{align}
S_{\mathrm{EH}}[g]=-\frac{M_{\mathrm{P}}^2}{2}\int\mathrm{d}^4X\sqrt{g}\left(R-2\Lambda\right).\label{EHActEuclid}
\end{align}
The gauge-breaking action \eqref{GBAct} for the second-order theory \eqref{EHActEuclid} is given by
\begin{align}
S_{\mathrm{gb}}[\bar{g}_{\mu\nu};h_{\mu\nu}]=-\frac{1}{2}\int\mathrm{d}^4X\,\chi^{\mu}g_{\mu\nu}\chi^{\nu},\label{GBActGR}
\end{align}
with the ultra-local operator $O_{\alpha\beta}$ and De Donder gauge condition $\chi^{\alpha}$,
\begin{align}
O_{\alpha\beta}=-\frac{\sqrt{\bar{g}}}{2} \bar{g}_{\mu\nu}\delta^{(4)}(X,Y),\qquad \chi^{\mu}[\bar{g}_{\mu\nu};h_{\mu\nu}]=\left(\bar{g}^{\mu\rho}\bar{g}^{\nu\sigma}-\frac{1}{2}\bar{g}^{\rho\sigma}\bar{g}^{\mu\nu}\right)\bar{\nabla}_{\nu}h_{\rho\sigma}.\label{GCGR}
\end{align} 
Adding \eqref{GBActGR} to \eqref{EHActEuclid} results in a gauge-fixed fluctuation operator \eqref{GaugeFixedOp}, which is of the minimal second-order type \eqref{secor}, 
\begin{align}
F^{\mu\nu,\rho\sigma}=\bar{\mathcal{G}}^{\mu\nu,\tau\lambda}\tensor{F}{_{\tau\lambda}^{\rho\sigma}}=\bar{\mathcal{G}}^{\mu\nu,\tau\lambda}\left(\bar{\Delta}\delta_{\tau\lambda}^{\rho\sigma}+\tensor{\bar{P}}{_{\tau\lambda}^{\rho\sigma}}\right),\label{OpFGR}
\end{align}
with the positive definite background Laplacian $\bar{\Delta}=-\bar{g}^{\mu\nu}\bar{\nabla}_{\mu}\bar{\nabla}_{\nu}$ and the background values of the DeWitt metric $\mathcal{G}^{\mu\nu,\rho\sigma}$ and the potential $\tensor{P}{_{\tau\lambda}^{\rho\sigma}}$ defined as
\begin{align} \mathcal{G}^{\mu\nu,\rho\sigma}:={}&\frac{g^{1/2}}{4}\left(g^{\mu\rho}g^{\nu\sigma}+g^{\mu\sigma}g^{\nu\rho}-g^{\mu\nu}g^{\rho\sigma}\right),\label{GDeWitt}\\
\tensor{P}{_{\mu\nu}^{\rho\sigma}}:={}&-2\tensor{R}{^{\rho}_{(\mu}^\sigma_{\nu)}}-2\tensor*{\delta}{^{(\rho}_{(\mu}}\tensor*{R}{^{\sigma)}_{\nu)}}+\tensor{g}{_\mu_\nu}\tensor*{R}{^\rho^\sigma}+\tensor{g}{^\rho^\sigma}\tensor*{R}{_\mu_\nu}-\frac{1}{2}\tensor{g}{_\mu_\nu}\tensor{g}{^\rho^\sigma}R+(R-2\Lambda)\tensor*{\delta}{^{\rho\sigma}_{\mu\nu}}.
\end{align}
According to \eqref{Gop}, the ghost operator derives from \eqref{GCGR} and reads 
\begin{align}
\tensor{Q}{_{\mu}^{\nu}}
=\tensor*{\delta}{_{\mu}^{\nu}}\bar{\Delta}-\tensor*{\bar{R}}{_{\mu}^{\nu}}.
\label{GhostOPGR}
\end{align} 
The divergent part of the one-loop approximation \eqref{OneLoop} reduces to the evaluation of the two functional traces
\begin{align}
\Gamma_{1}^{\mathrm{div}}=\frac{1}{2}\left.\mathrm{Tr}\ln\left(\tensor{F}{^{\mu\nu}_{\rho\sigma}}\right)\right|^{\mathrm{div}}-\mathrm{Tr}\left.\ln\left(\tensor{Q}{_{\mu}^{\nu}}\right)\right|^{\mathrm{div}}.\label{OneLoopDivGRFuncTrace}
\end{align}
Terms proportional to $\delta^{(4)}(0)$ which arise from $\mathrm{Tr}\ln\left(\mathcal{G}_{\mu\nu\rho\sigma}\right)$ are zero in dimensional regularization.
The divergent parts of the functional traces \eqref{OneLoopDivGRFuncTrace} are most efficiently evaluated by the heat-kernel techniques presented in Sec. \ref{sec:HeatKernel}.
The operators \eqref{OpFGR} and \eqref{GhostOPGR} in \eqref{OneLoopDivGRFuncTrace} are both of the form \eqref{secor}, for which the divergent part is given by \eqref{1Ldiv4}. The final result for the one-loop divergences \eqref{OneLoopDivGRFuncTrace} reads
\begin{align}
\Gamma_1^{\mathrm{div}}
=&\frac{1}{16\pi^2\varepsilon}\int\mathrm{d}^4X\, \sqrt{\bar{g}}\left[-\frac{53}{90}\bar{\mathfrak{G}}-\frac{7}{20}\tensor*{\bar{R}}{_\mu_\nu}\tensor*{\bar{R}}{^\mu^\nu}-\frac{1}{120}\bar{R}^2+\frac{13}{6}\Lambda \bar{R}-\frac{5}{2}\Lambda^2\right].\label{OneLoopDivGrav}
\end{align}
The Euler characteristic $\chi(\mathcal{M})$ is a topological invariant, defined in terms of the quadratic Gauss-Bonnet invariant $\mathfrak{G}$,
\begin{align} \chi(\mathcal{M}):={}&\frac{1}{32\pi^2}\int_{\mathcal{M}}\mathrm{d}^4X\sqrt{-g}\,\mathfrak{G},\qquad \mathfrak{G}:=R_{\mu\nu\rho\sigma}R^{\mu\nu\rho\sigma}-4R_{\mu\nu}R^{\mu\nu}+R^2.\label{Euler}
\end{align}
It allows to eliminate squares of the Riemann tensor in \eqref{OneLoopDivGrav} in favor of squares of the Ricci tensor and squares of the Ricci scalar.
For gravity with a cosmological constant in vacuum, the field equations \eqref{Eeq} imply $R_{\mu\nu}=\Lambda g_{\mu\nu}$. Therefore, on-shell, quantum Einstein gravity with a cosmological constant at one-loop can be expressed in terms of the Euler characteristic \eqref{Euler} and the volume $\mathcal{V}(\mathcal{M}):=\int\mathrm{d}^DX\sqrt{\bar{g}}$,
\begin{align}
\Gamma_{1,\mathrm{on-shell}}^{\mathrm{div}}={}\frac{1}{\varepsilon}\left[-\frac{53}{45}\chi(\mathcal{ M})+\frac{87}{20}\frac{\Lambda^2}{12\pi^2}\mathcal{V}(\mathcal{M})\right].\label{OnShellEG}
\end{align}
As discussed in \cite{Christensen1980}, the result \eqref{OnShellEG} shows that, within the one-loop approximation, pure Einstein gravity in $D=4$ is on-shell renormalizable, as the divergences in \eqref{OnShellEG} can be absorbed by adding the topological term $\chi(\mathrm{M})$ (which does not affect the field equations) with some coefficient to the action \eqref{EHActEuclid} and by renormalizing this coefficient as well as the cosmological constant $\Lambda$. For the case of a vanishing cosmological constant, the fact that Einstein gravity is on-shell one-loop finite was first found in \cite{Hooft1974}.
However, as soon as matter fields are coupled, the one-loop divergences remain even on-shell \cite{Hooft1974}. For example, the one-loop divergences of GR with a minimally coupled scalar field $\varphi$ with quartic self-interaction induces a non-minimal coupling to gravity proportional to $R\varphi^2$ -- an operator not present in the original action \cite{Hooft1974,Barvinsky1993,Shapiro1995,Steinwachs2011}. At the two-loop order, even for a vanishing cosmological constant $\Lambda=0$, a divergent contribution of a single operator among the cubic curvature invariants survives the on-shell reduction \cite{Goroff1985, Goroff1986,Ven1992}, 
\begin{align}
\Gamma_{2,\mathrm{on-shell}}^{\mathrm{div}}={}&\frac{1}{\varepsilon}\frac{1}{(16\pi^2)^2}\frac{209}{1470}\frac{1}{M^2_{\mathrm{P}}}\int\mathrm{d}^4X\sqrt{-g}\, \bar{C}_{\mu\nu}^{\;\;\;\;\rho\sigma}\bar{C}_{\rho\sigma}^{\;\;\;\;\alpha\beta}\bar{C}_{\alpha\beta}^{\;\;\;\;\mu\nu},\label{TwoLoops}
\end{align}
thereby showing explicitly that GR is perturbatively non-renormalizable.\footnote{In a recent calculation of the two-loop divergences with modern on-shell methods, it was found that, using dimensional regularization, evanescence operators (such as the Gauss-Bonnet term) in divergent subdiagrams can alter the coefficient of the pole term \cite{Bern2015}.} 
In \eqref{TwoLoops}, the cubic Riemann curvature invariant is expressed in terms of the Weyl tensor $C_{\mu\nu\rho\sigma}$, which on-shell coincides with the Riemann tensor $R_{\mu\nu\rho\sigma}$ in view of the vacuum on-shell identity $R_{\mu\nu}=0$,
\begin{align}
C_{\mu\nu\rho\sigma}=R_{\mu\nu\rho\sigma}&-\frac{2}{D-2}\left(R_{\mu\rho}g_{\nu\sigma}+R_{\nu\rho}g_{\mu\sigma}+R_{\mu\sigma}g_{\nu\rho}+R_{\nu\sigma}g_{\nu\rho}\right)-\frac{R}{(D-1)(D-2)}\left(g_{\mu\rho}g_{\nu\sigma}-g_{\nu\rho}g_{\mu\sigma}\right).\label{Weyl}
\end{align}
In a perturbatively renormalizable quantum field theory, a finite number of free parameters (fields, masses and coupling constants) are sufficient to absorb all UV divergences to all orders in the perturbative expansion. As demonstrated in \eqref{DoDGR} based on power counting arguments and in \eqref{TwoLoops} based on explicit calculations, GR is not of that form. New higher-dimensional operators with divergent coefficients are induced at every loop order and have to be renormalized by the introduction of the corresponding counterterms, each of which introduces a new coupling constant which has a finite part that needs to be determined by a measurement. In this way, more and more free parameters are introduced at each order in the perturbative expansion and the theory ultimately looses its predictive power.

\section{Effective field theory of gravity}
\label{Sec:EFT}
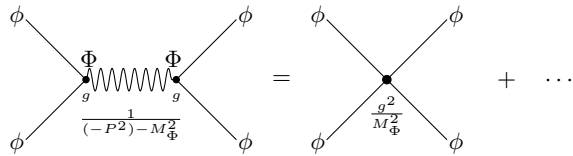
\begin{figure}[h!]
	\begin{center}
		\begin{tikzpicture}[scale=0.4]
		\begin{scope}
		\draw[black,thin](0,2)--(2,0);
		\node at (-0.3,2.1){$\phi$};
		\node at (-0.3,-2.1){$\phi$};
		\draw[black,thin](0,-2)--(2,0);
		\filldraw[black] (2,0) circle (3pt);
		\draw[black,thin,decorate, decoration={snake, segment length=1.5mm, amplitude=1.5mm}](2,0)--(5,0);
		\node at (2.1,0.7){$\Phi$};
		\node at (4.9,0.7){$\Phi$};
		\draw[black,thin](5,0)--(7,2);
		\draw[black,thin](5,0)--(7,-2);
		\filldraw[black] (5,0) circle (3pt);
		\node at (8.5,0){$=$};
		\node at (2,-0.6) {\tiny $g$};
		\node at (5,-0.6) {\tiny $g$};
		\node at (3.5,-1.5) {\tiny $\frac{1}{(-P^2)-M_{\Phi}^2}$};
		\node at (7.3,2.1){$\phi$};
		\node at (7.3,-2.1){$\phi$};
		\end{scope}
		\begin{scope}[xshift=10cm]
		\draw[black,thin](0,2)--(2,0);
		\draw[black,thin](0,-2)--(2,0);
		\filldraw[black] (2,0) circle (4pt);
		\draw[black,thin](2,0)--(4,2);
		\draw[black,thin](2,0)--(4,-2);
		\node at (2,-1.2) {\tiny $\frac{g^2}{M^2_{\Phi}}$};
		\node at (7,0) {$+\quad\cdots$};
		\node at (-0.3,2.1){$\phi$};
		\node at (-0.3,-2.1){$\phi$};
			\node at (4.3,2.1){$\phi$};
		\node at (4.3,-2.1){$\phi$};
		\end{scope}
		\end{tikzpicture} 
	\end{center}
	\caption{In the diagrammatic representation, to first order in the expansion \eqref{Propexp}, the $\Phi$-propagator is shrunk to a point, leading to an effective four-point contact interaction among the $\phi$-fields.}
	\label{Fig2}
\end{figure}

For many physical systems, an effective \textit{coarse grained} description is sufficient to accurately describe phenomena at low energies by the relevant degrees of freedom \cite{Weinberg1979}. Such an effective description might arise in two complementary ways often termed \textit{top-down} and \textit{bottom-up} approach. In case a (more) fundamental theory is known at high energy scales, a top-down approach leads to an effective low-energy theory by ``integrating out'' the heavy degrees of freedom.\footnote{Only in case the more fundamental theory is valid up to arbitrarily high energy scales, it qualifies as UV-complete theory. Instead of integrating out certain heavy particles, in the Wilsonian approach the effective action is defined at a given energy scale $E$ by integrating out \textit{all} particles with momenta $P^2$ greater than $E$.}
Denoting the heavy degrees of freedom collectively by $\Phi$ with characteristic mass scale $M_\Phi$ and the light degrees of freedom by $\phi$ with characteristic mass scale $M_\phi$, in a ``top-down''  scenario, there is a natural mass hierarchy $M_\Phi\gg M_\phi$. Integrating out the $\Phi$-fields from the combined action $S[\Phi,\phi]$ in the path integral defines the effective action $S_{\mathrm{eff}}[\phi]$ for the $\phi$-fields,
\begin{align*}
\int{\mathcal D}[\phi]e^{-S_{\rm eff}[\phi]}:=\int{\mathcal D}[\Phi,\phi]\,e^{-S[\phi,\Phi]}.
\end{align*}
In general, the process of integrating out $\Phi$-fields results in a non-local effective action $S_{\mathrm{eff}}[\phi]$. Within an energy expansion $E/M_{\Phi}\ll1$, it can be expanded in terms of \textit{local} operators $\mathcal{O}_n(\phi,\partial)$ for the $\phi$-fields,
\begin{align}
S_{\mathrm{eff}}[\phi]=S[\phi]+\sum_{n}\int\mathrm{d}^DXw_n\frac{\mathcal{O}_{n}(\phi,\partial)}{M_\Phi^{n-D}},\qquad \left[\mathcal{O}_n(\phi,\partial)\right]_{M}=n, \qquad [w_n]_{\mathrm{M}}=0.\label{Seff}
\end{align}
The higher-dimensional local operators $\mathcal{O}_n(\phi,\partial)$ parametrize the impact of the heavy degrees of freedom $\Phi$ on the effective low-energy theory for the light degrees of freedom $\phi$, and their interacting strength is characterized by the dimensionless \textit{Wilson coefficients} $w_n$. 
In terms of momentum space Feynman integrals, this expansion is associated to an expansion of the $\Phi$ propagators in inverse powers of the heavy mass scale $M_{\Phi}$,
\begin{align}
\frac{1}{(-P^2)-M_\Phi^2}=-\frac{1}{M_\Phi^2}-\frac{1}{(-M_\Phi)}(-P^2)\frac{1}{(-M^2_\Phi)}+\dots\label{Propexp}
\end{align}
For example, in this way, a $\phi\phi-\phi\phi$ interaction from a trivalent vertex $\propto g \Phi\phi^2 $ in $S[\Phi,\phi]$, leads to an effective quartic contact interaction among the $\phi$ fields $\propto (g^2/M_\Phi^2)\phi^4$ in $S_{\mathrm{eff}}[\phi]$ as diagrammatically illustrated in Fig. \eqref{Fig2}.

\noindent Since in the top-down approach calculations can be performed both ways, i.e. in the more fundamental theory as well as in the effective theory, scattering amplitudes can be compared at some scale below (but usually close to) $M_\Phi$ in order to fix the Wilson coefficients in terms of the parameters of the more fundamental theory -- a procedure called \textit{matching}.
Assuming $w_n=\mathcal{O}(1)$, the accuracy of the effective description is only limited by the ratio  $E/M_{\Phi}$, which controls the energy expansion, and completely breaks down for energies $E\approx M_{\Phi}$, where the propagation of the $\Phi$ particles is no longer suppressed.

\begin{figure}[h!]
	\begin{center}
		\begin{tikzpicture}[scale=1.3]
		\begin{scope}
		\draw[<->](-1,0)--(5,0);
		\node at (5.3,0) {UV};
		\draw (4,-0.1)--(4,0.1);
		\node at (4,0.4){$M_{\mathrm{P}}$};
		\node at (4.2,-0.4){$10^{18}$ GeV};
		\draw (-0.5,-0.1)--(-0.5,0.1);
		\node at (-0.5,0.4){$E_{\mathrm{SM}}$};
		\node at (-0.3,-0.4){$10^{2}$ GeV};	
		\node at (-1.3,0){IR};
		\node at (2,-0.4){Big desert or new physics?};
		\end{scope}
		\end{tikzpicture}	
	\end{center}
	\caption{Different energy scales. Is there new physics beyond the EW scale and the Planck scale or a ``big desert''?}
	\label{Fig3a}
\end{figure}
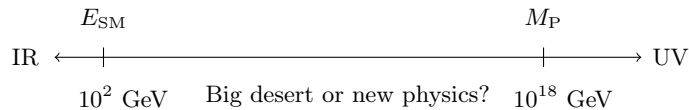

Importantly, the effective field theory description is still applicable, even if no (more) fundamental theory in the UV is known. This is the situation for GR, i.e. the effective field theory approach to gravity is necessarily a bottom up one \cite{Donoghue1994,Burgess2004}. In this case, the cutoff scale $M$ which limits the range of validity of the effective description is not known a priori. Assuming no new physics at scales in between the electroweak scale (EW) of the Standard Model (SM) of particle physics and the scale at which gravity becomes comparable to the other interactions (see Fig. \ref{Fig3a}), the Planck scale might be the natural cutoff scale $M=M_{\mathrm{P}}$.\footnote{This naive estimate might be modified in the presence of matter, see e.g. the discussion in the context of scalar-tensor theories with a strong non-minimal coupling such as in the model of Higgs inflation \cite{Burgess2009,Barbon2009,Burgess2010, Bezrukov2011, Barvinsky2012,Steinwachs2019a}.} 

\noindent It might be considered as a particular strength of the bottom-up approach that it is agnostic about the gravitational degrees of freedom in the UV -- the low-energy limit of the effective field theory (EFT) defines the field variables, symmetries and the particle spectrum. In the case of GR, these are the metric field, the diffeomorphisms and the massless spin-$2$ graviton. The ignorance about a more fundamental theory in the UV is parametrized by the systematic inclusion of higher-dimensional operators, which are compatible with the symmetries of the defining low-energy theory and suppressed by inverse powers of the cutoff scale. In the case of gravity, diffeomorphism invariance requires that the higher-dimensional purely gravitational operators $\mathcal{O}(g,\partial)$ have the form of curvature invariants proportional to $g^{\nicefrac{1}{2}}\nabla^{2n}R^m/M^{2(n+m)-D}$. For energy scales well below the cutoff $\nabla/M\ll1$, ${R}/M^2\ll1$, these higher-dimensional operators are strongly suppressed and the expansion can be truncated at a finite order determined by the required accuracy of the EFT.
In contrast to a fundamental theory, the higher-dimensional operators in an EFT are only viewed as correction terms, i.e. they lead to additional interaction vertices but do not modify the propagators of the theory and hence do not affect the particle spectrum, which is defined by the relevant operators at low energy.\footnote{Note however that a summation of operators with a fixed number of external fields but arbitrary number of derivatives results in non-local form factors which lead to IR modifications of the propagator. For a discussion of these non-local form factors in the context of gravity and the heat-kernel, see e.g.~\cite{Barvinsky:1990up,Barvinsky:1994ic}.}  
While the higher-dimensional operators in an EFT are included in a controlled way, the precise way of how such an expansion scheme is realized can differ. Depending on the requirements of the underlying physical model, such an expansion might be realized as derivative expansion, as vertex expansion, as the aforementioned combined ``energy expansion'', or according to a different scheme. 

In principle, the presence of the infinite tower of operators $g^{\nicefrac{1}{2}}\nabla^{2n}R^m/M^{2(n+m)-D}$ is required in an EFT to absorb all UV divergences by renormalizing the $w_i$.
However, according to the GR power counting \eqref{DoDGR}, the $L$th loop correction in $D=4$ induces divergent operators of the form $g^{\nicefrac{1}{2}}\nabla^{2n}R^m/M^{2(n+m)-D}$ with $n+m=L+1$. Thus, within a \textit{finite truncation}, the EFT of gravity can be perturbatively renormalized in the standard way and only finitely many renormalized parameters $w_i$ have to be measured, ultimately rendering the EFT predictive.\footnote{An important technical requirement for the consistent renormalization is that the counterterms have the same structure as the operators in the EFT expansion. Since the latter are restricted by symmetry, this requires the process of renormalization to preserve this symmetry, see e.g. the discussion in \cite{Gomis1996}. To show this property is non-trivial and has been proven for GR and Yang-Mills theory in \cite{Barnich1995}. Recently, this proof was extended to effective and non-relativistic theories by combining the BRST cohomology with the background field method \cite{Barvinsky2018}.} 

However, ultimately absorbing the UV divergences within a finite truncation rather provides a consistency condition than a prediction. In contrast to the local but unphyscial UV divergences, true predictions of the quantum theory are connected with IR effects which arise from long-range interactions dominated by massless particles. These contributions are connected to the non-analytic parts in scattering amplitudes. 
The most prominent example of how such IR effects can be extracted from QFT scattering amplitudes within the EFT of GR are the corrections to the Newtonian potential for two point masses $M_1$ and $M_2$, which after Fourier transformation reads \cite{Bjerrum-Bohr2003},
\begin{align}
V(r)=-\frac{G_{\mathrm{N}}M_1M_2}{r}\left[1+3\frac{G_{\mathrm{N}}(M_1+M_2)}{rc^2}+\frac{41}{10\pi}\frac{G_{\mathrm{N}}\hbar}{r^2 c^3}+\ldots\right].\label{CorrPot}
\end{align}
The second term is a purely classical relativistic correction related to the $\sqrt{P^2}$ part, while the third term is of genuine quantum origin and related to the $P^2\log(P^2)$ part of the one-loop contribution \cite{Bjerrum-Bohr2003}. Both contributions correspond to those parts of the scattering amplitude which have a non-analytic momentum dependence. They are independent of the higher curvature terms in the EFT expansion and therefore do not depend on a UV completion. While the general structure of the correction terms in \eqref{CorrPot} follows from dimensional analysis, the coefficients (in particular the sign) have to be calculated and provide a true prediction of quantum gravity. 

While the \textit{quantum} gravitational corrections are accompanied by powers of $G_{\mathrm{N}}\hbar$ and therefore very hard to measure, classical Post-Minkowskian (PM) corrections run in powers of $G_{\mathrm{N}}$. High-order PM corrections have been calculated by classical techniques \cite{Jaranowski1998,Buonanno1999,Damour2016,Schaefer2018}. Since the advent of gravitational wave astronomy, there is an increasing effort to extract the classical PM corrections within an EFT framework from QFT scattering amplitudes, which, in turn, can be efficiently calculated by modern on-shell techniques, see e.g. \cite{Goldberger2006,Bjerrum-Bohr2014,Porto2016,Bern2019,Bern2020,Blumlein:2020pog,Blumlein:2020znm}.
 
The EFT of GR is a powerful and universal approach which leads to universal quantum gravitational predictions from long-range effect of massless particles, but its range of applicability is limited by construction. Therefore, certain questions cannot be addressed within this framework but require a fundamental quantum theory of gravity.

\section{Asymptotic safety}
\label{Sec:AS}
While the question about a fundamental theory of gravity cannot be addressed in the framework of the perturbative EFT approach, the Asymptotic Safety (AS) program, initiated in \cite{Weinberg1976,Weinberg1980}, might offer a UV complete theory of quantum gravity.
The basic underlying idea is that the renormalization group (RG) flow drives the (dimensionless) \textit{essential} couplings $g_{n}$ of a theory towards a UV fixed point $g_n^{*}$.\footnote{In this section, I denote the coupling constants by $g_{n}$ to contrast with the $c_n$ in \eqref{act} and the $\omega_n$ in \eqref{Seff}, although when put in the right context they are all the same objects. The RG flow $g_n(k)$ is defined as the solution of the RG system $k\partial_kg_{n}=\beta_{g_n}$, with the abstract RG scale $k$ and the beta functions $\beta_{g_n}$. A fixed point $g^{*}_{n}$ is defined by the condition $\beta_{g_n}(g_{m}^{*})=0$, $\forall n$. Couplings $\tilde{g}_{n}$, which carry a canonical physical dimension $[\tilde{g}_{n}]_{\mathrm{M}}=\alpha_n$ are made dimensionless by a rescaling with the appropriate power of the RG scale $[k]_{\mathrm{M}}=1$, i.e.~$g_{n}=\tilde{g}_{n}k^{-\alpha_n}$, such that $[g_{n}]_{\mathrm{M}}=0$. Moreover, since only essential couplings enter physical observables, only they are required to acquire finite values in the UV. In contrast, inessential couplings, which can be changed by a field redefinition, do not enter physical observables and therefore might diverge in the UV.}
In this way, the AS scenario prevents the couplings form running into divergences at finite energy scales (Landau pole) and allows to extrapolate the RG flow to arbitrary energy scales $k\to\infty$.
However, in contrast to the asymptotic freedom scenario corresponding to a free (i.e.~non-interacting or ``Gaussian'') UV fixed point $g_{n}^{*}=0$, the AS scenario only requires the weaker condition $g_{n}^{*}=\text{const.}$, which includes the possibility of an interacting fixed point for $g_{n}^{*}\neq0$ \cite{Weinberg1976}. In particular, the couplings $g_{n}$ are not required to remain within the perturbative regime $g_{n}\ll1$ and consequently allow for a strongly interacting UV fixed point at which (at least some of) the couplings $g_{n}^{*}\gg1$. Clearly such a strongly interacting UV fixed point cannot be found within a perturbative approach. Thus, the AS scenario is an inherently \textit{non-perturbative} approach, which can be addressed within the Wilsonian approach to the RG \cite{Wilson1974}.

The main object is the \textit{averaged effective action} $\Gamma_k$ which defines the full quantum theory at a given RG scale $k$. The sliding scale $k$ interpolates between the bare action $\Gamma_{\infty}=S$ in the UV, corresponding to $k=\infty$, and the full effective action $\Gamma_{0}=\Gamma$ in the IR, corresponding to $k=0$.
Once the propagating degrees of freedom $\phi^{i}$ and their symmetries are identified, $\Gamma_k$ might be expressed in terms of symmetry-compatible operators $\mathcal{O}_{n}(\phi,\partial)$ with coupling strengths $g_n(k)$,
\begin{align}
\Gamma_{k}=\sum_{i=1}^{\infty}\int\mathrm{d}^DX\,g_{i}(k)\mathcal{O}_{n}(\phi,\partial).\label{aea}
\end{align}
The space of all coupling constants $g_i$ is called \textit{theory space}.
A suitable tool for a non-perturbative analysis is the Wetterich equation \cite{Wetterich1993,Morris1994,Reuter1994}, which describes the exact functional renormalization group (RG) flow of the averaged effective action $\Gamma_{k}$,
\begin{align}
k\partial_k\Gamma_k=\frac{1}{2}\mathrm{Tr}\left(\frac{k\partial_k\mathcal{R}_k}{\Gamma_k^{(2)}+\mathcal{R}_k}\right).\label{Wetterich}
\end{align}
Here, $\mathrm{Tr}$ is the functional trace, $\mathcal{R}_k$ a scale-dependent regulator and $\Gamma^{(2)}_k$ the Hessian of the averaged effective action $\Gamma_k$.  The Wetterich equation \eqref{Wetterich} has a similar structure as the one-loop approximation \eqref{OneLoop}, but involves the scale dependent regulator function $\mathcal{R}_k$ defined such that it acts as an effective mass term of the full propagator for quantum fluctuations with momenta $P^2\leq k^2$ and vanishes for momenta $P^2\gg k^2$. Together with the factor $\partial_k\mathcal{R}$ in \eqref{Wetterich}, which cuts off fluctuations with momenta $P^2\geq k^2$, the presence of the regulator ensures that only fluctuation with momenta peaked around $P^2\approx k^2$ contribute to the trace in \eqref{Wetterich}, thereby realizing the Wilsonian ``shell-by-shell'' integration.\footnote{In particular, once a cutoff is introduced, it does not matter whether the underlying theory is perturbatively renormalizable in the strict sense or not. All operators compatible with the symmetries of the theory have to be considered. This is similar as for the EFT, but in contrast to the EFT treatment, the particle content and the symmetries are not necessarily defined by the relevant operators of the low-energy approximation, but defined along with the averaged effective action \eqref{aea}. In general the theory space is infinite, but if the symmetry restriction is so strong that it only allows for a finite number of operators, the theory space might be finite.}  
Due to the presence of the regulator no divergences occur. In general, the Wetterich equation cannot be solved exactly. Instead of a semiclassical expansion in powers of loops such as in \eqref{EAE}, a finite truncation of the (in general infinite) set of operators included in $\Gamma_k$ is performed
\begin{align}
\Gamma_{k}=\sum_{n=1}^N\int\mathrm{d}^DX\,g_{n}(k)\mathcal{O}_{n}(\phi,\partial).\label{aeatrunc}
\end{align}
According to which criteria such a truncation is chosen practically might depend on the underlying physical problem. In most applications the operators are organized in terms of an energy expansion, i.e.~ordered by increasing canonical mass dimension. There are however also cases where a derivative expansion or a vertex expansion is more appropriate. In the case of gravity, diffeomorphism invariance requires that the $\mathcal{O}(g,\partial)$ are curvature invariants, schematically $\mathcal{O}(g,\partial)=\sqrt{g}(\nabla)^{2p}R^{m}$.
Substituting the ansatz \eqref{aeatrunc} into \eqref{Wetterich}, choosing a regulator $\mathcal{R}_{k}$ and evaluating the functional trace on the right-hand-side of \eqref{Wetterich}, the RG flow of the couplings $g_n(k)$ can be extracted by ``projecting'' to the operator basis $\mathcal{O}_{n}(\phi,\partial)$. Contributions of operators which are induced by the flow and lead out of the truncation \eqref{aeatrunc} are neglected.\footnote{This is a consistency requirement of the truncation. In case no operators which lead out of the truncation are induced, the flow closes and \eqref{Wetterich} is really an exact equation.}

For a successful realization of the AS scenario, the existence of a UV-fixed point $g_i^{*}$ is only a necessary condition, not a sufficient one. In addition, an appropriate fixed point must have a \textit{finite-dimensional} UV-critical surface.\footnote{The UV-critical surface might be thought as a subspace of the tangent space at $g_{i}^{*}$, consisting of those RG trajectories which are attracted towards the fixed-point. In general,there can be more than just one fixed-point and the RG flow might also allow for more exotic phenomena such as limit cycles. It might also happen that some of the fixed-points are dismissed on physical grounds. }
The finiteness of the UV-critical surface lies at the very heart of the AS scenario, as it implies that only a finite subset of the (in general infinitely many) coupling constants have to be measured, rendering the theory predictive. It is this feature which might qualify the AS scenario in providing a UV-complete quantum theory of gravity.\footnote{Compare this to the perturbative quantization of GR, discussed in Sec. \ref{Sec:PQDGR}. The perturbatively non-renormalizable character requires the measurements of an infinite number of couplings thereby leading to a loss of predictive power. Compare this also to the EFT approach to GR, discussed in Sec. \ref{Sec:EFT}. While only a finite number of couplings have to be measured within a finite truncation, the EFT cannot be extrapolated beyond a certain energy scale and therefore does not quality as a UV-complete theory.} Thus, in principle, if all UV relevant couplings would have been measured (and in this way select a particular RG trajectory emanating from the UV fixed point) all other UV irrelevant couplings are fixed. They therefore constitute predictions which could be falsified by additional measurements of these couplings. In practice, calculations are however limited to finite truncations and one must ensure that the properties of the fixed-point (and therefore any prediction derived from it) remain stable under an enlargement of the truncation. In principle, if a reliable measure of the quality of a given truncation would exist one could try to ultimately prove convergence, but since so far no such measure exists this is hard to realize in practice and one has to rely on systematic step-by-step enlargements of finite truncations. Nevertheless, as for the perturbative approach (fundamental or EFT), a particular strength of the AS approach to quantum gravity is its universality, i.e.~gravity and matter fields are treated within one and the same formalism. This not only allows for a unification, but also allows to test the techniques used in the context of quantum gravity in more controlled environments, in which also experimental data is available.      

The functional RG flow in the context of gravity \cite{Reuter1998,Niedermaier2006,Codello2009,Liberati2012,Reuter2012} has been studied in various truncations, starting with the Einstein-Hilbert truncation \cite{Reuter2002}, including higher curvature invariants \cite{Lauscher2002b,Lauscher2002,Codello2006,Benedetti2009,Falls2016b,Gies2016} and matter fields \cite{Eichhorn2012,Dona2014,Dona2016,Eichhorn2017,Eichhorn2018,Christiansen2018}, as well as closed flow equations for $f(R)$ gravity \cite{Machado2008,Codello2008}, and general scalar-tensor theories \cite{Narain2010,Narain2010a}. A pattern which emerges in most of these truncations is that an interacting UV fixed point can be found and that the dimension of the associated UV critical surface does not grow upon enlarging the truncation beyond the classically marginal operators. Since this program has been pushed to high orders in various truncations, it might give some confidence that the observed pattern is a generic feature and not an artefact of the truncation.

Despite these interesting results, there are a number of open questions associated with this program, see e.g. \cite{Donoghue2019}.
In general, the off-shell flow defined by $\Gamma_k$ suffers from a number of ambiguities connected to the choice of the regulator as well as to the gauge dependence and field parametrization dependence of the beta functions. Since different regulator choices, different gauges and different field parametrizations can even affect qualitative features such as the existence of a fixed-point, a satisfying resolution of these ambiguities seems to be crucial for the reliability of the predictions following from the AS conjecture.

In connection with the gauge and parameter dependence, a unique off-shell extension of the averaged effective action along the lines of the construction proposed in \cite{Vilkovisky1984} might offer an interesting option, but even without such a construction, the gauge and parametrization dependence should be absent in an on-shell scheme, see e.g.~\cite{Benedetti:2011ct}. However, making use of the equations of motion, in general leads to degeneracies among different operators in a given truncation and therefore does not allow to resolve and disentangle the individual RG flow of the couplings for these on-shell degenerated operators.\footnote{A similar problem occurs when working on special (in general highly symmetric) backgrounds, even if they do not correspond to on-shell configurations.} Nevertheless, extracting e.g.~physical observables from the S-matrix will anyway involve an on-shell reduction. By definition only \textit{essential} couplings span the theory space. In this sense, the ``on-shellness'' is already built into the formalism of the AS conjecture from the very beginning. However, especially in the context of gravity, the situation is more complicated, as e.g.~the question of whether Newton's constant is an essential or inessential coupling is not so clear and leads to conceptional intricacies, see e.g.~the discussion in \cite{Percacci:2004sb}. 
   
In any case, the starting point for the derivation of observables should be the effective action at $k=0$, which is independent of the regulator and formally obtained by integrating out all quantum fluctuations, i.e. by integrating the functional flow all the way down to the IR. One might be tempted to extract information from the averaged effective action $\Gamma_k$ at non-zero $k$ by performing a ``RG-improvement'' based on a heuristic identification of the abstract coarse graining RG scale $k$ with some characteristic physical scale. However, beside the fact that such an identification is typically only possible in highly symmetric backgrounds where a single scale is present, such as e.g.~the radius in the context of spherically symmetric black hole backgrounds, the Hubble parameter in the context of an isotropic and homogeneous cosmological Friedmann-Lema\^itre-Robertson-Walker background, the value of the scalar field in the Coleman-Weinberg-like radiatively induced symmetry breaking in a classically scale-invariant theory, or the momentum transfer in the context of scattering amplitudes, etc., it does not seem that such a naive identification can be based on a more general solid theoretical ground.
However, even when working with the effective action at $k=0$, another problem arises: The effective action is non-local (and non-analytic), and therefore not appropriately described by the finite number of local operators in a given truncation which do not capture essential IR contributions. In this context, the introduction of form factors in the context of the AS program, provide a more promising route. Including form factors in the truncation goes beyond a finite derivative expansion as it captures the full momentum dependence of propagators and vertices, which can either be studied by a flat-space vertex expansion \cite{Christiansen:2014raa,Christiansen:2015rva,Eichhorn:2018akn}, or in a general background by an expansion of the effective action in powers of external fields (curvatures in the context of gravity) \cite{Bosma:2019aiu,Knorr:2019atm}. The manifest covariant calculation of these non-local form factors are technically challenging and require heat-kernel-based methods developed in \cite{Barvinsky:1990up,Barvinsky:1990uq,Barvinsky:1994hw,Barvinsky:1994ic,Codello:2012kq}.

The analysis of form factors in the AS program might also shed some light on the status of the particle content -- a problem also shared by higher-derivative theories of gravity, discussed in Sec. \ref{Sec:HDG}. Any truncation based on a finite derivative expansion will in general lead to additional propagating degrees of freedom in the particle spectrum (defined by the quadratic action expanded around a flat background) and will almost always include higher-derivative ghosts among them. Having access to the pole structure of the propagators, including the full momentum dependence carried by the form factors, might ultimately reveal the status of the ghost degrees of freedom as an artifact of the finite truncation (realized, e.g. if the full propagators only have a single pole with positive residue). Technically, this program is closely related with the (ghost-free) non-local approach to quantum gravity, see e.g.~ \cite{Gorbar:2002pw,Shapiro:2015uxa,Biswas:2011ar,Modesto:2011kw,Edholm:2016hbt,Modesto:2017sdr}.

\section{Higher derivative gravity}
\label{Sec:HDG}
Before giving up on finding a fundamental theory of quantum gravity or abandoning the framework of perturbative QFT, yet another obvious approach is to modify the underlying classical theory of gravity and investigate the impact of these modifications on the resulting quantum theory. Adding higher-dimensional curvature invariants to the action might be the most natural generalization of GR. 
In contrast to the EFT treatment, when treating the modified theory as fundamental, the higher-dimensional operators are no longer considered as perturbations and correspondingly not only modify the interaction vertices but also the propagators. Ultimately, this leads to new additional propagating degrees of freedom.
There are many ways to modify GR. A simple and phenomenologically important extension of GR is $f(R)$ gravity, allowing for an arbitrary function $f$ of the Ricci scalar $R$,
\begin{align}
S_f[g]=\int\mathrm{d}^4X \sqrt{g}\,f(R).\label{ActfR}
\end{align}
In particular, \eqref{ActfR} encompasses the Starobinsky model \cite{Starobinsky1980}, which is highly relevant for inflationary cosmology
\begin{align}
f_{\mathrm{Star}}=\frac{M_{\mathrm{P}}^2}{2}\left[R+\frac{1}{6M_{0}^2}R^2\right].\label{ActStar}
\end{align}
In fact, \eqref{ActStar} was the first model of inflation and is strongly favored by the latest Planck data \cite{Akrami2018}. The one-loop divergences for $f(R)$ gravity \eqref{ActfR} have recently been calculated on an arbitrary background \cite{Ruf2018}, thereby essentially generalizing previous calculations obtained for spaces of constant curvature \cite{Cognola2005, Codello2008, Machado2008},
\begin{align}
\Gamma_1^{\mathrm{div}}=\frac{1}{32\pi^2\varepsilon}\int{\rm d}^4X\, \sqrt{g}&\left[
-\frac{71}{60}\mathfrak{G}
-\frac{259}{180}\tensor*{R}{_\mu_\nu}\tensor*{R}{^\mu^\nu}
-\frac{9}{2}\left(\frac{f}{f_1}\right)^2
-\frac{1}{18}\left(\frac{f_1}{f_2}\right)^2+\frac{9}{2}\frac{f}{f_1}R
+\frac{1}{3}\frac{f}{f_2}
-\frac{59}{360}R^2\right.\nonumber\\
&\;\,\left.+\frac{21}{2}\frac{f}{f_1}\left(\tensor{\Upsilon}{_\mu^{;\mu}}\right)
-\frac{33}{4}R\left(\tensor{\Upsilon}{_\mu^{;\mu}}\right)-\frac{371}{72}R\left(\tensor{\Upsilon}{_\mu}\tensor{\Upsilon}{^\mu}\right)+\frac{27}{4}\frac{f}{f_1}\left(\tensor{\Upsilon}{_\mu}\tensor{\Upsilon}{^\mu}\right)
+\frac{20}{9}\tensor{R}{_\mu_\nu}\tensor{\Upsilon}{^\mu}\tensor{\Upsilon}{^\nu}
\right.\nonumber\\
&\;\,\left.
-\frac{137}{24}\left(\tensor{\Upsilon}{_\mu^{;\mu}}\right)^2
-\frac{9}{8}\left(\tensor{\Upsilon}{_\mu}\tensor{\Upsilon}{^\mu}\right)\left(\tensor{\Upsilon}{_\nu^{;\nu}}\right)
-\frac{769}{96}\left(\tensor{\Upsilon}{_\mu}\tensor{\Upsilon}{^\mu}\right)^2
\right].\label{Gamma1fR}
\end{align}
The derivatives of the function $f$ are defined by $ f_n:=\partial^nf/\partial R^n$ and the vector $\tensor{\Upsilon}{_\mu}$ is defined as $\tensor{\Upsilon}{_\mu}:=R_{;\mu}f_2/f_1$.
Even for a general function $f$, the result \eqref{Gamma1fR} shows that $f(R)$ gravity is perturbatively non-renormalizable on a general background.
Although divergences accompanied by arbitrary functions of $R$ might be absorbed by renormalizing $f(R)$, due to the absence of the derivative structures $\tensor{\Upsilon}{_\mu}$ and the quadratic curvature structure $R_{\mu\mu}R^{\mu\nu}$ in \eqref{ActfR}, the associated divergences cannot be absorbed.\footnote{Even on-shell, there remain divergences associated with operators involving derivatives of the Ricci scalar, which are not total derivatives and cannot be absorbed in the function $f(R)$ \cite{Ruf2018}. On a constant curvature background $g_{\mu\nu}^{0}$, for which $R_{\mu\nu\rho\sigma}^{0}=R_0(g^{0}_{\mu\rho}g^{0}_{\nu\sigma}-g^{0}_{\mu\sigma}g^{0}_{\nu\rho})$, $\Upsilon_{\mu}=0$, $\int\mathrm{d}^4X\sqrt{g^{0}}=384\pi^2/R_0^2$ and the equations of motion reduce to the algebraic equation $2f-R_0f_1=0$, the one-loop divergences ${\Gamma_1^{\mathrm{div}}|_{0}^{\mathrm{on-shell}}=(1/\varepsilon)[\frac{97}{20}+4f/R_0^2f_2-8f^2/3(R_0f_2)^2]}$ can be absorbed by a renormalization of $f(R_0)$. } The higher derivatives in \eqref{ActfR} lead to a fourth-order fluctuation operator and imply the presence of an additional propagating scalar degree of freedom, the scalaron. In the context of the cosmological model \eqref{ActStar}, the scalaron drives the accelerated expansion of the early universe and its mass $M_0\approx10^{-5}M_{\mathrm{P}}$ is fixed by the observed anisotropy spectrum in the Cosmic Microwave Background radiation \cite{Akrami2018}.

What is the required extension of GR which qualifies as a candidate for a perturbatively renormalizable quantum theory of gravity? The power counting performed in \eqref{DoDGR} for GR can easily be generalized to higher derivative theories of gravity (HDG). Diffeomorphism invariance requires that all higher curvature invariants have a schematic structure (suppressing indices) $\sqrt{-g}\nabla^{2n}R^m$ with a total number of derivatives $p=2(n+m)$. The natural candidate HDG theory is the one which includes all classically relevant and marginal operators, i.e. in $D=4$ all operators with $p\leq4$. Aside from the relevant operators \eqref{RelOp} present already in the Einstein-Hilbert action \eqref{EHact}, the marginal operators with $p=4$ have either $m=2$ and $n=0$ or $m=1$ and $n=1$. For the latter case, there is only one scalar invariant $\mathcal{O}_3(g,\partial)=\sqrt{-g}\nabla_{\mu}\nabla^{\mu}R$, which is a total derivative. For the former case there are three possible scalar invariants quadratic in the curvature
\begin{align}
\mathcal{O}_4(g,\partial)=\sqrt{-g}R_{\mu\nu\rho\sigma}R^{\mu\nu\rho\sigma},\qquad \mathcal{O}_5(g,\partial)=\sqrt{-g}R_{\mu\nu}R^{\mu\nu},\qquad\mathcal{O}_6(g,\partial)=\sqrt{-g}R^2. \label{CurvInv}
\end{align}
The three curvature invariants \eqref{CurvInv} might be more conveniently parametrized in a different basis of quadratic curvature invariants involving the Gauss-Bonnet term and the Weyl tensor and the Ricci scalar, as the latter two are more directly related to the particle content,
\begin{align}
S_{\mathrm{QDG}}[g]=S_{\mathrm{EH}}[g]+\int\mathrm{d}^4X\sqrt{-g}\left[c_1\mathfrak{G}+c_2C_{\mu\nu\rho\sigma}C^{\mu\nu\rho\sigma}+c_3R^2\right].\label{ActQDG}
\end{align}     
The power counting in the UV is dominated by the marginal quadratic curvature operators and the momentum scaling of propagator is $\mathcal{P}\propto P^{-4}$, while that of the vertices is $\mathcal{V}^{(n)}\propto P^4$. Consequently, the superficial degree of divergences in Quadratic Gravity (QDG) in $D=4$ is
\begin{align}
D^{\mathrm{div}}_{\mathrm{QDG}}=4L-4(L-1)=4.\label{SDoDHDG}
\end{align}
Hence, in $D=4$, QDG is power counting renormalizable and indeed suggests that QDG is the required extension of GR.    
Going beyond this simple power counting argument requires more advanced methods; a strict proof that QDG \eqref{ActQDG} is a perturbatively renormalizable quantum theory of gravity has been given in \cite{Stelle1977}.

However, even if the perturbative renormalizability of QDG has been established, it remains to show that QDG is UV complete, i.e. whether the theory can be extended to arbitrary energy scale. To answer this question requires to study the RG flow determined by the divergence structure of the theory. In particular, for an UV-complete theory the absence of Landau poles, where couplings diverges at finite energies, must be assured. The one-loop divergences of QDG were first calculated in \cite{Fradkin1982} and later corrected in \cite{Avramidi1985}. The authors of \cite{Avramidi1985}  considered the Euclidean version of \eqref{ActQDG} with a different parametrization and basis for the quadratic curvature invariants
\begin{align}
S_{\mathrm{QDG}}[g]=\int\mathrm{d}^4X\sqrt{g}\left[\frac{2}{k^4}\lambda-\frac{1}{k^2}R+\frac{1}{\nu^2}\mathfrak{G}+\frac{1}{f^2}\left(R_{\mu\nu}R^{\mu\nu}-\frac{1}{3}R^2\right)-\frac{\omega}{3f^2}R^2\right],\label{ActQDGAB}
\end{align}     
with $1/k^2=M_{\mathrm{P}}^2/2$ and the dimensionless cosmological constant $\lambda=2\Lambda/M_{\mathrm{P}}^2$. The beta functions can directly be read off from the one-loop divergences and determine the running of the coupling constants with the logarithmic parameter $t:=1/(4\pi^2)\ln(\mu/\mu_0)$. Here $\mu$ is the sliding scale and $\mu_0$ an arbitrary renormalization point. Within the standard framework with the ``ordinary'' definition of the effective action as in \eqref{FIDE}, it was found in \cite{Avramidi1985} that the \textit{essential} couplings $1/\nu^2(t)$, $1/f^2(t)$, $\omega/f^2(t)$ are asymptotically free, provided that $1/\nu^2>0$, $1/f^2>0$, $\omega/f^2<0$, while $\lambda$ grows in the UV limit $t\to\infty$. Note, however, it was found in \cite{Salvio:2014soa} that $\omega/f^2>0$ is required in the Lorentzian regime to avoid a tachyonic instability of the scalaron. Fixing the correct sign, the running is no longer asymptotically free.

Newton's constant, or $k^2$ in terms of the parametrization in \eqref{ActQDGAB}, is an inessential coupling and does not run. In order to access the running of all couplings separately, including the running of $k^2$, an off-shell extension is required, which renders the effective action gauge independent and parametrization invariant.\footnote{See also \cite{Kamenshchik2015, Ruf2018c,Ohta2018a,Falls2019,Finn2019} for a discussion of the quantum parametrization dependence of the effective action in cosmology.} Such an off-shell extension was proposed in \cite{Vilkovisky1984} by a geometrically defined (field-covariant) ``unique'' effective action. At the one-loop level, the difference between the ``ordinary'' definition of the effective action and the ``unique'' effective action is a correction term proportional to the equations of motion. The ``unique'' off-shell one-loop beta functions for \eqref{ActQDGAB} have been calculated in \cite{Avramidi1985} and the running of $1/k^2(t)$ was extracted, with the result that $\lim_{t\to\infty}1/k^2(t)=0$ and $\lim_{t\to\infty}\Lambda(t)=0$. Thus, the UV limit $t\to\infty$ found in this way corresponds to the induced gravity scenario $M_{\mathrm{P}}^2\to0$ (i.e. $G_{\mathrm{N}}\to\infty$) with vanishing (dimensional) cosmological constant $\Lambda\to0$.\footnote{Since Newton's constant $G_{\mathrm{N}}(t)\sim1/k^2(t)$ exceeds the perturbative regime, a perturbative treatment does not seem reliable in the asymptotic limit $t\to\infty$. However, since Newton's coupling is an inessential coupling in the ordinary perturbative approach (even if its runs in the covariant Vilkovisky off-shell extension), it should never enter an on-shell observable in an isolated way, but only via a dimensionless combination with other couplings (including $\Lambda(t)$), whose beta function is gauge-independent. Thus, independently of whether $G_{\mathrm{N}}$ itself grows beyond perturbative control in the limit $t\to\infty$, the question should then rather be whether the RG running of this dimensionless combination stays under perturbative control.}

While the above quoted results support the status of QDG in $D=4$ as a perturbative renormalizable theory of quantum gravity, the reason why QDG is usually not regarded as consistent theory of quantum gravity is connected to its problem with the additional propagating spin-two ghost degrees of freedom. 
In analogy to \eqref{FopMom}, the momentum space fluctuation operator of QDG defined in the parametrization \eqref{ActQDG} for arbitrary $D$ on a flat background can be expressed in terms of the projectors \eqref{PS2} and \eqref{PScal2} and reads \cite{Stelle1978},
\begin{align}
F^{\mu\nu,\rho\sigma}(-P^2)={}&\frac{(-P^2)}{2}\left[1+8c_2\frac{D-3}{D-2}\frac{(-P^2)}{M_{\mathrm{P}}^2}\right]\tensor{P}{^{(2)}^{\mu\nu\rho\sigma}}-(D-2)\frac{(-P^2)}{2}\left[1-8c_3\frac{D-1}{D-2}\frac{(-P^2)}{M_{\mathrm{P}}^2}\right]\tensor{P}{^{(0,ss)}^{\mu\nu\rho\sigma}}.\label{FopMomQDG}
\end{align} 
Clearly, this reduces to \eqref{FopMom} for $c_2=c_3=0$. Moreover, due to the topological nature of the GB term $\mathfrak{G}$, $c_1$ does not enter \eqref{FopMomQDG}. Just as in GR, the diffeomorphism invariance of QDG renders the fluctuation operator \eqref{FopMomQDG} degenerate and a gauge-fixing is required to obtain the propagators. Nevertheless, the tree-level particle spectrum of QDG can already be analyzed on the basis of the pole structure in \eqref{FopMomQDG}. Defining the two effective masses for $D>3$,
\begin{align}
M_{2}^2:=-\frac{1}{8c_2}\frac{D-2}{D-3}M_{\mathrm{P}}^2,\label{M2}\qquad
M_{0}^2:=\frac{1}{8c_3}\frac{D-2}{D-1}M_{\mathrm{P}}^2,
\end{align} 
the pole structure of the propagators in the spin-$2$ and spin-$0$ sectors becomes more transparent \cite{Stelle1977},
\begin{align}
\mathcal{P}_{(2)}\propto{}&\frac{-M_{2}^2}{(-P^2)\left[(-P^2)-M_{2}^2\right]}=\frac{1}{(-P^2)}-\frac{1}{(-P^2)-M_{2}^2},\label{Prop2}\\
\mathcal{P}_{(0)}\propto{}&\frac{M_0^2}{(-P^2)\left[(-P^2)-M_{0}^2\right]}=-\frac{1}{(-P^2)}+\frac{1}{(-P^2)-M_{0}^2}.\label{Prop0}
\end{align}
The partial fraction in the second equality reveals that, compared to GR, in QDG there are two additional propagating particles with masses $M_{2}$ and $M_{0}$.
The first term in \eqref{Prop2} corresponds to a massless spin-$2$ particle and, just as in GR, combines with the first term in \eqref{Prop0} to the massless graviton. The second term of \eqref{Prop2} indicates the presence of a propagating massive spin-$2$ particle originating from the $C_{\mu\nu\rho\sigma}C^{\mu\nu\rho\sigma}$ term in \eqref{ActQDG}, while the second term in \eqref{Prop0} indicates the presence of a massive spin-$0$ particle originating from the $R^2$ term in \eqref{ActQDG}.
Excluding tachyons requires $M_{2}^2>0$ ($c_2<0$) and $M_{0}^2>0$ ($c_3>0$). The massive spin-$0$ particle, which can be identified with the scalaron in the model \eqref{ActStar}, is ``healthy'' (neither a ghost nor a tachyon), while the overall minus sign in the the second term of \eqref{Prop2} shows that the massive spin-$2$ particle is a higher-derivative ghost. The presence of ghosts corresponds to states of negative norm, leading to a violation of unitarity \cite{Stelle1977}, see also \cite{Pais1950,Barth1983,Hawking1985,Woodard2007}. 

Within an effective low energy treatment $P^2/M_2^2\ll1$, the propagation of the massive spin-$2$ ghost is strongly suppressed. Whether such an EFT, which still includes the scalaron as propagating degree of freedom (since the $R^2$ would not be treated as perturbation compared to the $R$ term) can be realized, strongly depends on the characteristic mass scales $M_2$ and $M_0$, i.e. the values of $c_2$ and $c_3$, respectively. It requires that $M_{2}^2$ is sufficiently large such that the effective description is valid up to energy scales at which the additional propagating scalaron has interesting phenomenology such as in the inflationary model \eqref{ActStar}, but at the same time, $M_0^2\ll1$ must be sufficiently small such that the scalaron can be considered as propagating degree of freedom, see e.g. \cite{Gundhi2018} for a discussion of such a scenario in the context of the scalaron-Higgs model. Solar system based experimental constraints on both $c_2$ and $c_3$ are extremely weak. However, while $c_2$ is practically unconstrained, a large $c_3=M_{\mathrm{P}}^2/(12M_0^2)\approx10^{9}$ is required in \eqref{ActStar} if the scalaron is supposed to drive inflation.
But even if the problem with the spin-$2$ ghost can effectively be neglected at sufficiently ``low'' energies, without a mechanism which prevents the occurrence of the higher derivative ghost at arbitrarily high energy scales, QDG cannot be considered as a fundamental theory.

Recently, the negative conclusion about the ghost-related loss of unitarity in QDG at the fundamental level have been questioned. They are related to early proposals about different quantization prescriptions, which modify the pole structure of the propagators in higher-derivative theories \cite{Lee:1969fy,Tomboulis:1977jk}. In \cite{Anselmi2018a,Anselmi2018} a new quantization prescription is proposed which turns higher-derivative ghosts into ``fakeons'' at the expense of a loss of micro-causality.
Another resolution of the unitarity problem was suggested in \cite{Donoghue2019a,Donoghue2019c}. A key point in this proposal is that the coupling of light matter particles to gravity render the heavy spin-two ghost unstable, such that the ghost is not part of the asymptotic particle spectrum. Extending the conclusion that unstable particles must be excluded in the sum of the unitarity relation \cite{Veltman:1963th} to the case of unstable ghost particles (which are nevertheless identified as such by the free-particle spectrum), it is concluded in \cite{Donoghue2019a} that there is no violation of unitarity in QDG. Nevertheless, in \cite{Donoghue2019a,Donoghue2019c} it is also found that the ghosts ``propagate backwards in time'' which leads to a violation of micro-causality. While this effect can in principle be tested experimentally, it becomes unobservably small for sufficiently heavy ghost masses, such as e.g.~in QDG if $M_2\approx M_{\mathrm{P}}$.       

Summarizing, in both proposals \cite{Anselmi2018a,Anselmi2018} and \cite{Donoghue2019a,Donoghue2019c} about the correct treatment of higher-derivative ghost particles, it is concluded that unitarity violation is avoided at the expense of violating mirco-causality, but it seems that a conclusive agreement on this controversially debated issue has not yet been reached. For related work on higher-derivative ghosts, see also \cite{Coleman:1969xz,Cutkosky:1969fq,Salam:1978fd,Tomboulis:1980bs,Boulware:1983vw,Hawking:2001yt,Mannheim:2006rd,Bender:2007wu,Grinstein:2008bg,Denner:2014zga,Salvio:2015gsi,Accioly:2016qeb,Mannheim:2018ljq}.
For a discussion of the ghost problem in the context of the non-perturbative AS program to quantum gravity, see e.~g.~\cite{Floreanini1995,Benedetti2009,Niedermaier2009, Becker:2017tcx,Narain2017a,Narain2018}.
For the non-local approach to a ghost-free quantum theory of gravity, see \cite{Krasnikov:1987yj,Gorbar:2002pw,Smilga:2005gb,Shapiro:2015uxa,Biswas:2011ar,Modesto:2011kw,Tomboulis:2015esa,Tomboulis:2015gfa,Edholm:2016hbt,Modesto:2017sdr}.

\section{Ho\v{r}ava-gravity}
\label{Sec:HG}
The picture which emerged from the previous described approaches in providing a consistent fundamental local quantum theory of gravity suggests that the basic principles of relativistic invariance, renormalizability and unitarity are incompatible in the context of the perturbative quantization of the gravitational interaction: quantum GR is a relativistic and unitary but perturbatively non-renormalizable QFT, while quantum QDG is a relativistic and perturbatively renormalizable but non-unitary QFT. Therefore, in \cite{Horava2009a,Horava2009}, Petr Ho\v{r}ava suggested to explore the consequences of abandoning relativistic invariance, while trying to preserve unitarity and perturbative renormalizability.

One of the key motivations for this proposal follows from the discussion of QDG. While the higher derivatives help to improve the UV behavior of the theory, the higher \textit{time} derivatives are responsible for the occurrence of the additional higher derivative ghost degrees of freedom and the associated problems with unitarity. The desire to keep the UV-improving effect of the higher derivatives, but, at the same time, to avoid the ghost problem, leads to the idea of allowing for \textit{higher spatial derivatives} but restrict to \textit{second order time derivatives}.
Obviously, such a proposal is not compatible with relativistic invariance. It is clear that ``sacrosanct'' principles such as relativistic invariance are not recklessly sacrificed -- not only because this changes the fundamental structure of spacetime, but also since there are highly strong experimental constraints on Lorentz violating effects.

With this proviso, I first review how this idea can be formalized by the notion of an anisotropic Lifshitz scaling between space and time and how it can be incorporated in a consistent mathematical framework by formulating the resulting anisotropic theory of gravity in terms of the geometric Arnowitt-Deser-Misner (ADM) variables, giving rise to the Lifshitz theory of gravity, Ho\v{r}ava Gravity (HG).
Within the ADM formulation, the main difference between GR and HG is the weaker invariance group underlying HG, the \textit{foliation-preserving diffeomorphisms} $\mathrm{Diff}_{\mathcal{F}}$, which form a subgroup of the full diffeomorphisms.

An important consequence of the anisotropic scaling and the less restrictive invariance group in HG are the modified dispersion relations and the presence of an additional propagating gravitational scalar degree of freedom. After a brief discussion of their phenomenological consequences in $D=2+1$ and $D=3+1$ dimensions, I review the quantum properties of HG. I first discuss the gauge and propagator structure of the theory and then review the essential steps in the proof of perturbative renormalizability of the projectable version of HG.

Finally, I discuss the UV properties of quantum HG based on the RG flow of the projectable theory in $D=2+1$ dimensions, which requires to explicitly calculate the one-loop divergences within a Lifshitz theory of gravity \cite{Barvinsky2017}. I close with a brief summary and an outlook on future perspectives of quantum HG. For earlier reviews on HG with a different focus, especially on the phenomenological constraints and the cosmological applications, see \cite{Mukohyama2010,Sotiriou2011,Wang2017,Blas2018}.

\subsection{Anisotropic scaling and modified propagators}
As briefly outlined before, the basic idea of Ho\v{r}ava gravity is to allow for higher spatial derivatives but restrict to second order time derivatives. Obviously, such a proposal implies that relativistic invariance is lost at the fundamental level. How precisely Lorentz invariance is broken in a way compatible with this proposal can be made concrete by introducing the anisotropic Lifshitz scaling between time and space \cite{Lifshitz1941,Horava2009a,Horava2009},
\begin{align}
t\to b^{-z}\,t,\qquad\, x^{i}\to b^{-1}\,x^{i}.\label{AS}
\end{align}
Here, $b$ is a constant scaling parameter and $z$ a dynamical scaling exponent. In analogy to the mass dimension $[\ldots]_{\mathrm{M}}$, introduced in Sec. \ref{sec:ClassicalGR}, the \textit{anisotropic scaling dimension} is defined by $[\ldots]_{\mathrm{S}}$. According to the anisotropic scaling law \eqref{AS}, the scaling dimensions of time and space are $[t]_{\mathrm{S}}=-z$ and $[x]_{\mathrm{S}}=-1$. This implies the scaling relations 
\begin{align}
 [\partial_{t}]_{\mathrm{S}}=z,\qquad [\partial_{i}]_{\mathrm{S}}=1,\qquad  [\omega]_{\mathrm{S}}=z,\qquad  [k_i]_{\mathrm{S}}=1.\label{ScalingDer}
\end{align}
Here, $\omega$ and $k_{i}$ are the frequency and spatial momentum, Fourier conjugate to $\partial_t$ and $\partial_i$.
The dynamical scaling exponent $z$ might be thought of measuring the degree of anisotropy between space and time, with $z=1$ restoring relativistic invariance. 
In view of \eqref{ScalingDer}, the (Euclidean) anisotropic propagator acquires the form
\begin{align}
\label{Aprop}
{\mathcal P}\propto\frac{1}{\omega^2+k^2+\ldots+G\,(k^{2})^z}
\simeq
\begin{cases}
\text{IR:  }\;\frac{1}{\omega^2+k^2}=\frac{1}{p^2}\\[2mm]
\text{UV:  }\;\frac{1}{\omega^2+G\,(k^{2})^z},\\
\end{cases}
\end{align}
with some coupling constant $[G]_{\mathrm{M}}=-2(z-1)$, $[G]_{\mathrm{S}}=0$.
This propagator illustrates the basic idea that Lorentz invariance is completely broken by the anisotropic scaling exponent $z$ for $G(k^2)^z\gg k^2$ in the UV-limit and effectively restored in a natural way for $k^2\gg G(k^2)^z$ in the IR-limit  \cite{Horava2009}. \footnote{In general, relevant deformations also lead to different coupling constants in front of different powers of $k^2$ in the propagator \eqref{Aprop}, which, as discussed in the context of HG in Sec.~\ref{Sec:ParticleSpectrumHG}, might prevent a direct restoration of Lorentz invariance in the IR.}

\subsection{Geometrical formulation in terms of ADM variables}
The anisotropic Lifshitz theory of gravity can be consistently formulated within a geometrical framework when described in terms of ADM variables.  
Following the presentation in \cite{Steinwachs:2017ihd}, I briefly review the ADM formulation in the context of GR, and highlight the differences in HG when the full diffeomorphism invariance $\mathrm{Diff}(\mathcal{M})$ is reduced to the foliation-preserving diffeomorphism $\mathrm{Diff}_{\mathcal{F}}(\mathcal{M})$.

\subsubsection{ADM variables and GR}
\begin{figure}[h!]
	\begin{center}
		\begin{tikzpicture}[scale=0.65]
		\begin{scope}
		\draw[black,thin,smooth](0,0) .. controls (1,0.8) .. (1.8,1.8);
		\draw[black,thin,smooth](8,0) .. controls (9,0.8) .. (9.8,1.8);
		
		\draw[black,thin,smooth](0,0) .. controls (1.5,0.5) .. (4,0);
		\draw[black,thin,smooth](4,0) .. controls (6.5,-0.2) .. (8,0);
		
		\draw[black,thin,smooth](1.8,1.8) .. controls (3.3,2.3) .. (5.8,1.8);
		\draw[black,thin,smooth](5.8,1.8) .. controls (7.3,1.6) .. (9.8,1.8);
		\end{scope}
		
		\begin{scope}[xshift=0.5cm,yshift=3cm]
		\draw[black,thin,smooth](0,0) .. controls (1,0.8) .. (1.8,1.8);
		\draw[black,thin,smooth](8,0) .. controls (9,0.8) .. (9.8,1.8);
		
		\draw[black,thin,smooth](0,0) .. controls (1.5,0.5) .. (4,0);
		\draw[black,thin,smooth](4,0) .. controls (6.5,-0.2) .. (8,0);
		
		\draw[black,thin,smooth](1.8,1.8) .. controls (3.3,2.3) .. (5.8,1.8);
		\draw[black,thin,smooth](5.8,1.8) .. controls (7.3,1.6) .. (9.8,1.8);
		\end{scope}

		\draw[black,thin](6,0.8)--(6,2.7);
		\draw[black,thin,dashed,->](6,2.7)--(6,3.7);
		\draw[black,thin,->](3.1,0.8)--(6,0.8);
		\fill[black] (6,3.8) circle (1.5pt);
		\draw[black,thin](3.1,0.8)--(5.2,3);
		\draw[black,thin,dashed,->](5.2,3)--(5.9,3.7);
		\fill[black] (3,0.8) circle (1.5pt);
		\node(a) at(2.9,1.1){\tiny$x^{i}(t)$};
		\node(a) at(6.4,4){\tiny$x^{i}(t+{\rm d}t)$};
		\node(a) at(10,1.3){\tiny$\Sigma_{t}$};
		\node(a) at(10.7,4.3){\tiny$\Sigma_{t+{\rm d}t}$};
		\node(a) at(4.5,0.6){\tiny$N^{\mu}{\rm d}t$};
		\node(a) at(6.6,2.3){\tiny$N\,n^{\mu}{\rm dt}$};
		\node(a) at(4.2,2.5){\tiny$t^{\mu}{\rm d}t$};
		\end{tikzpicture}
	\end{center}
	\caption{Foliation of $D$ dimensional spacetime into $d=D-1$ dimensional hypersurfaces of constant time $t$.}
\label{Figure1}
\end{figure}
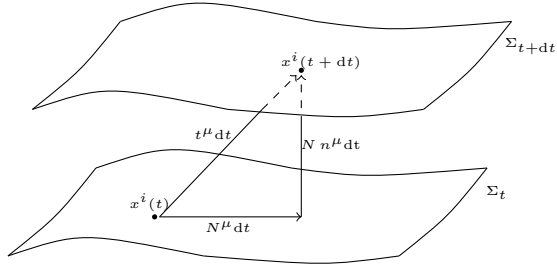

A point $X\in\mathcal{M}$ in the $D$-dimensional ambient spacetime $\mathcal{M}$ can be described by local coordinates $X^{\mu}$. For a globally hyperbolic ambient space, $\mathcal{M}$ can be foliated by a one-parameter 
family of $d=D-1$-dimensional spatial hypersurfaces $\Sigma_t$ of constant time $t$. 
The hypersurfaces $\Sigma_t$ might be thought of as level surfaces of a time field $t$. The gradient of $t$ defines a natural unit covector field
\begin{align}
n_\mu:=-\frac{\nabla_\mu t}{\sqrt{-g^{\mu\nu}\nabla_\mu t\nabla_\nu t}},\qquad n^{\mu}=g^{\mu\nu}n_{\nu},\qquad n^\mu n_\mu=-1.
\end{align}
By construction, at each point, the normal vector field $n^{\mu}(\mathbf{x},t)$ is orthogonal to $\Sigma_t$ and therefore allows for an orthogonal decomposition of tensor fields with respect to $n^{\mu}$. In particular, the ambient metric decomposes as 
\begin{align}
g_{\mu\nu}=\gamma_{\mu\nu}-n_\mu n_\nu.\label{Decg}
\end{align}
Here, $\gamma_{\mu\nu}$ is the tangential part of $g_{\mu\nu}$, 
that is $\gamma_{\mu\nu}n^{\mu}=0$. The hypersurfaces $\Sigma_t$
can be considered as the embeddings of an intrinsically 
$d$-dimensional manifold $\tilde{\Sigma}_t$ into the ambient 
space $\mathcal{M}$. A point $x\in\tilde{\Sigma}_t$ can be described
by the local coordinates $x^i$, $i=1,\ldots,d$. The $D$-dimensional 
coordinates $X^{\mu}=X^{\mu}(t,\mathbf{x})$ can be parametrized in terms of the time field $t$ and the spatial coordinates $x^{i}$. The change of $X^{\mu}$ with 
respect to $t$ and $x^{i}$ is given by the coordinate one-form 
\begin{align}
\mathrm{d} X^\mu
=t^{\mu}\mathrm{d} t+\tensor{e}{^{\mu}_{i}}\mathrm{d}x^{i}.\label{COF}
\end{align}
The time vector field $t^{\mu}$ 
and the soldering form $\tensor{e}{^{\mu}_{i}}$ appearing in \eqref{COF} are defined as
\begin{align}
t^{\mu}:=\frac{\partial X^\mu(t,\mathbf{x})}{\partial t},
\qquad 
\tensor{e}{^{\mu}_{i}}:=\frac{\partial X^\mu(t,\mathbf{x})}{\partial x^i}.
\end{align}
As illustrated in Fig. \ref{Figure1}, the lapse function $N(t,\mathbf{x})$ and the shift vector $N^{\mu}(t,\mathbf{x})$ are defined as the coefficients of the orthogonal decomposition of ${t^{\mu}:=N\,n^{\mu}+N^{\mu}}$ in the direction normal and tangential to $\Sigma_{t}$, respectively.

\noindent The soldering form $\tensor{e}{^{\mu}_{i}}$ transforms like a $D$-dimensional tangential vector w.r.t. the $\mu$ index, i.e. $\tensor{e}{^{\mu}_{i}}n_{\mu}=0$ and a $d$-dimensional
vector w.r.t. the $i$ index. It defines the pull-back of tangential
tensors in $\mathcal{M}$ to tensors in $\tilde{\Sigma}_t$,
\begin{align}
\tensor{e}{^{\mu}_{i}}\tensor{e}{_{\nu}^{i}}=\delta^{\mu}_{\nu},\qquad \tensor{e}{^{i}_{\mu}}\tensor{e}{^{\mu}_{j}}=\delta^{i}_{j}.
\end{align}
The pullback of $\gamma_{\mu\nu}$ and $N^{\mu}$ defines the spatial metric $\gamma_{ij}$ and the spatial shift-vector $N^{i}$,
\begin{align}
\gamma_{ij}:=\tensor{e}{^{\mu}_{i}}\tensor{e}{^{\nu}_j}\gamma_{\mu\nu},\qquad N^{i}:=\tensor{e}{_{\mu}^{i}}N^{\mu}.
\end{align}
In terms of $\mathrm{d}t$ and $\mathrm{d}x^i$, the ambient space coordinate one-form is expressed as
\begin{align}
\mathrm{d} X^\mu=Nn^\mu\mathrm{d} t+\tensor{e}{_{i}^{\mu}}\left(N^i\mathrm{d} t+\mathrm{d}x^i\right).\label{DiffCo}
\end{align}
Inserting this into \eqref{LineElement}, the ambient space line element acquires the familiar Arnowitt-Deser-Misner form \cite{Arnowitt1960},
\begin{align}
\mathrm{d}s^2=-N^2\mathrm{d} t^2
+\gamma_{ij}\left(N^i\mathrm{d} t+\mathrm{d}x^i\right)\left(N^j\mathrm{d} t+\mathrm{d}x^j\right).\label{LEADM}
\end{align} 
On $\tilde{\Sigma}_t$, the commutator of the (torsion-free and metric compatible $\nabla_k\gamma_{ij}=0$) spatial covariant derivative $\nabla_i$ defines the $d$-dimensional spatial curvature tensor by its action on a spatial vector field $v^{k}$,
\begin{align}
[\nabla_i,\,\nabla_j]v^{k}=\tensor{R}{^k_l_i_j}(\gamma)v^{l}.
\end{align} 
The relation between the scalar curvature of the $D$-dimensional ambient space $R(g)$ and the scalar curvature $R(\gamma)$ of the $d$-dimensional embedded space is given by the Gauss-Codazzi relation, see e.g. \cite{Kuchar1976},
\begin{align}
R(g)=R(\gamma)-\left(K^2-K_{ij}K^{ij}\right)-2\left(\nabla_i+ a_i\right)a^i+2\left(D_t+K\right)K.\label{GC}
\end{align}
Here, $K:=\gamma^{ij}K_{ij}$ is the trace of the extrinsic curvature $K_{ij}$, defined by the covariant time derivative $D_t$,
\begin{align}
K_{ij}:=\frac{1}{2}D_t\gamma_{ij}=\frac{1}{2N}\left(\partial_t\gamma_{ij}-\nabla_{i}N_{j}-\nabla_{i}N_{j}\right),\qquad D_t:=\frac{1}{N}\left(\partial_t-\mathcal{L}_{\mathbf{N}}\right),\label{CovDt}
\end{align}
with $\mathcal{L}_{\mathbf{N}}$ the Lie derivative along the spatial shift vector $N^i$. The acceleration vector $a^{i}$ in \eqref{GC} is defined as
\begin{align}
a_i:=\partial_i\ln{N}.\label{accV}
\end{align}
Note that the $D$-dimensional diffeomorphisms $\mathrm{Diff}(\mathcal{M})$ completely fix the structure and the numerical coefficients of the individual terms in \eqref{GC}.
In terms of the ADM variables \eqref{LEADM}, the volume element of $\mathcal{M}$ reads ${\sqrt{-g}=N\sqrt{\gamma}}$, and, modulo surface terms, the Einstein-Hilbert action \eqref{EHact} acquires the ADM form 
\begin{align}
S_{\mathrm{EH}}=\frac{M_{\rm P}^2}{2}\int{\rm d}t\,\mathrm{ d}^{3}x\,N\,\sqrt{\gamma}\left[K_{ij}K^{ij}-K^2+R(\gamma)\right].\label{EHADM}
\end{align}
It is natural to consider the first two terms in \eqref{EHADM}, which involve the square of the ``velocities'' $\partial_{t}\gamma_{ij}$, as the ``kinetic term'' for $\gamma_{ij}$, and to consider $R(\gamma)$ as the ``potential'', which only involves spatial derivatives $\partial_{k}\gamma_{ij}$. In particular, the invariance of the action \eqref{EHact} under $\mathrm{Diff}(\mathcal{M})$ implies that only the very specific combination of ADM operators in \eqref{EHADM} is $\mathrm{Diff}(\mathcal{M})$ invariant.  This illustrates how strongly the underlying $\mathrm{Diff}(\mathcal{M})$ invariance in GR restricts possible operators allowed in the EH action when expressed in terms of ADM variables.  

\subsubsection{Symmetry in GR and HG}
In GR, the ADM variables derive from the decomposition of the $D$-dimensional ambient space metric $g_{\mu\nu}$. Consequently, in this case, the symmetry group acting on the ADM variables are the full $D$-dimensional spacetime diffeomorphisms $\mathrm{Diff}(\mathcal{M})$, or general coordinate transformations,
\begin{align}
x^{i}\mapsto \tilde{x}^{i}(t,\mathbf{x}),	\quad t\mapsto\tilde{t}(t,\mathbf{x}).\label{CoTrADM}
\end{align}
In general, operators $\mathcal{O}(g_{\mu\nu},\partial_{\nu})$, invariant under $\mathrm{Diff}(\mathcal{M})$, are constructed by scalar contractions of covariant derivatives $\nabla_{\mu}$ and curvature tensors $R_{\mu\nu\rho\sigma}$. While the action of $\mathrm{Diff}(\mathcal{M})$ on the $D$-dimensional ambient metric $g_{\mu\nu}$ is realized linearly \eqref{Diffeog}, in view of \eqref{LEADM}, the action of $\mathrm{Diff}(\mathcal{M})$ on the ADM variables $N$, $N^{i}$, $\gamma_{ij}$ is non-linearly realized. Thus, only very particular combinations of $\mathrm{Diff}(\mathcal{M})$-invariant operators $\mathcal{O}(N, N^{i},\gamma_{ij}, \partial_i,\partial_t)$ constructed by scalar contractions of the time and space derivatives $\partial_t$ and $\partial_i$ of the ADM variables $N$, $N^{i}$, $\gamma_{ij}$ are allowed.

In contrast to the general coordinate transformations \eqref{CoTrADM}, the coordinate transformations which preserve the foliation include the $d$-dimensional time-dependent spatial  diffeomorphisms and the reparametrizations of time
\begin{align}
x^{i}\mapsto \tilde{x}^{i}(t,\mathbf{x}),	\quad t\mapsto\tilde{t}(t).\label{FDiffCoTra}
\end{align}
Under \eqref{FDiffCoTra}, the ADM fields $N$, $N^{i}$, $\gamma_{ij}$ transform as
\begin{align}
N\mapsto{}\tilde{N}=N\frac{\mathrm{d}t}{\mathrm{d}\tilde{t}},\qquad
N^{i}\mapsto{}\tilde{N}^{i}=\left(N^{j}\frac{\partial \tilde{x}^{i}}{\partial x^j}-\frac{\partial \tilde{x}^{i}}{\partial t}\right)\frac{\mathrm{d}t}{\mathrm{d}\tilde{t}},\qquad
\gamma_{ij}\mapsto{}\tilde{\gamma}_{ij}=\gamma_{k\ell}\frac{\partial x^{k}}{\partial\tilde{x}^{i}}\frac{\partial x^{\ell}}{\partial\tilde{x}^{j}}.
\end{align} 
Combining the action of an infinitesimal diffeomorphism \eqref{Diffeog} on the ambient metric $g_{\mu\nu}$, its decomposition in ADM variables \eqref{Decg} and the decomposition of the generator of infinitesimal diffeomorphisms $\varepsilon^{\mu}=(\varepsilon,\varepsilon^{i})$ with $\varepsilon^{i}(t,\mathbf{x})=\varepsilon^{\mu}e^{i}_{\mu}$ and  $\varepsilon(t,\mathbf{x})=t_{\mu}\varepsilon^{\mu}$, the action of an infinitesimal $\mathrm{Diff}(\mathcal{M})$ on the the ADM fields $\gamma_{ij}$, $N^{i}$ and $N$ is derived as
\begin{align}
\delta_{\varepsilon}N={}&\partial_t\left(\varepsilon N\right)+\mathcal{L}_{\pmb{\varepsilon}}N-NN^{i}\partial_i\varepsilon,\label{DiffLapse}\\
\delta_{\varepsilon}N^{i}={}&
\partial_t\left(\varepsilon N^{i}\right)+\partial_t\varepsilon^{i}+\left(\mathcal{L}_{\pmb{\varepsilon}}N\right)^{i}-\left(N^{i}N^{j}+N^2\gamma^{ij}\right)\partial_{j}\varepsilon,\label{DiffShift}\\
\delta_{\varepsilon}\gamma_{ij}={}&\varepsilon\partial_{t}\gamma_{ij}+\left(\mathcal{L}_{\pmb{\varepsilon}}\gamma\right)_{ij}+2N_{(i}\partial_{j)}\varepsilon.\label{ADMVarDiff}
\end{align}
Here $\mathcal{L}_{\pmb{\varepsilon}}$ denotes the Lie derivative along $\varepsilon^{i}$. The transformation law for the shift vector with covariant index position ${N_i=\gamma_{ij}N^{j}}$ can be obtained by combining the transformation laws \eqref{DiffShift} and \eqref{ADMVarDiff} and reads
\begin{align}
\delta_{\varepsilon}N_{i}=\partial_t\left(\varepsilon N_i\right)+\left(\mathcal{L}_{\pmb{\varepsilon}}N\right)_i+\gamma_{ij}\partial_{\tau}\varepsilon^j+\left(N_{j}N^{j}-N^2\right)\partial_i\varepsilon.\label{DiffShiftCo}
\end{align}
In contrast to the linear transformation \eqref{Diffeog} of the ambient metric $g_{\mu\nu}$, the transformations \eqref{DiffLapse}--\eqref{ADMVarDiff} of the ADM variables under infinitesimal $\mathrm{Diff}({\mathcal M})$ is not linear.
The transformations of the ADM variables under $\mathrm{Diff}_{\mathcal{F}}({\mathcal M})$, for which the time component $\varepsilon$ of the generator $\varepsilon^{\mu}=(\varepsilon,\varepsilon^{i})$ is a function of time only $\varepsilon(t,\mathbf{x})=\varepsilon(t)$, are derived from \eqref{DiffLapse}-\eqref{ADMVarDiff} by neglecting terms involving $\partial_{i}\varepsilon$ and the action of an infinitesimal $\mathrm{Diff}_{\mathcal{F}}({\mathcal M})$ on the ADM variables read
\begin{align}
\delta_{\varepsilon}N={}&\partial_t\left(\varepsilon N\right)+\mathcal{L}_{\pmb{\varepsilon}}N,\label{FDiffLapse}\\
\delta_{\varepsilon}N^{i}={}&
\partial_t\left(\varepsilon N^{i}\right)+\partial_t\varepsilon^{i}+\left(\mathcal{L}_{\pmb{\varepsilon}}N\right)^{i},\label{FDiffShiftUp}\\
\delta_{\varepsilon}\gamma_{ij}={}&\varepsilon\partial_{t}\gamma_{ij}+\left(\mathcal{L}_{\pmb{\varepsilon}}\gamma\right)_{ij}.\label{FDiffMetric}
\end{align}
Likewise, the transformation \eqref{DiffShiftCo} reduces to
\begin{align}
\delta_{\varepsilon}N_{i}=\partial_t\left(\varepsilon N_i\right)+\left(\mathcal{L}_{{\pmb \varepsilon}}N\right)_i+\gamma_{ij}\partial_{\tau}\varepsilon^j.\label{FDiffShift}
\end{align}
Hence, the $\mathrm{Diff}_{\mathcal{F}}({\mathcal M})$ form a subgroup of the $\mathrm{Diff}(\mathcal{M})$ and the absence of terms proportional to $\partial_{i}\varepsilon$ has the effect that the transformations \eqref{FDiffLapse}-\eqref{FDiffShift} act linearly on ADM variables \cite{Horava2009,Rechenberger2013}.

Mathematically, the $\mathrm{Diff}_{\mathcal{F}}({\mathcal{M}})$ are diffeomorphisms which respect the preferred co-dimension-one foliation $\mathcal{F}$ of $D=d+1$ dimensional spacetime $\mathcal{M}$ into spatial $d$-dimensional leaves \cite{Horava2009}. On such a foliation, two classes of functions can be defined: functions that depend on all coordinates $(t,x^{i})$ and functions which are constant on each spatial leave, i.e. which only depend on time $t$. The latter are called ``projectable''. From a canonical perspective with a fundamental dynamical field $\gamma_{ij}$, the shift vector $N^{i}$ might be viewed as the gauge-field associated with the time-dependent spatial diffeomorphisms with infinitesimal generator $\varepsilon^{i}(t,\mathbf{x})$ and and the lapse function $N$ as the gauge-field of the reparametrizations of time with infinitesimal generator $\varepsilon(t)$. It therefore seems natural to restrict $N(\mathrm{x},t)$ to be a function of time only, although both versions  $N(t,\mathbf{x})$ and $N(t)$ are compatible with the $\mathrm{Diff}_{\mathcal{F}}(\mathcal{M})$ symmetry, essentially leading to two variants of HG: 
\begin{itemize}
	\item[i.)]
	\textit{Projectable HG}:\\
	The lapse function only depends on time $N(t)$ and is not considered as dynamical field. By choosing a global time slicing, corresponding to the gauge in which $N(t)=1$, the foliation preserving diffeomorphisms reduce to the time-depended spatial diffeomorphisms.
	
	\item[ii.)]
	\textit{Non-projectable HG}:\\
	The lapse function depends on space and time and $N(t,\mathbf{x})$ is a propagating degree of freedom, i.e. an integration variable in the path integral. Compared to the projectable theory, the main technical challenge is the enlarged set of $\mathrm{Diff}_{\mathcal{F}}(\mathcal{M})$ invariants which involve the acceleration vector \eqref{accV}.
\end{itemize}
Since the two possibilities lead to two different theories with different particle content and different phenomenology, they have to be investigated separately. In particular, the quantization of the non-projectable theory is complicated due to the presence of the fluctuating lapse function leading to non-regular propagators \cite{Barvinsky2016}. In this contribution, I mainly focus on the projectable theory, but highlight at several places important difference to the non-projectable theory. 

\subsection{Projectable HG in $D=2+1$ and $D=3+1$ dimensions}
The action functional of projectable HG in $D=d+1$ dimensions can be formulated in terms of the ADM variables. The natural assignment of the anisotropic scaling dimensions to the ADM variables follows from \eqref{DiffCo} and \eqref{LEADM},
\begin{align}
[\gamma_{ij}]_{\mathrm{S}}=0,\qquad [N^{i}]_{\mathrm{S}}=z-1,\qquad [N]_{\mathrm{S}}=0.\label{ASFields}
\end{align}
Compared to the stringent constraints on the ADM-operators in GR, following from the invariance under $\mathrm{Diff}(\mathcal{M})$, the less restrictive invariance under $\mathrm{Diff}_{\mathcal{F}}({\mathcal{M}})$ allows for a richer structure and consequently for more ADM invariants.
Nevertheless, there are a number of conditions which limit the possible $\mathrm{Diff}_{\mathcal{F}}({\mathcal{M}})$-invariants in the projectable HG:

\begin{enumerate}	
\item \label{p1} Formulated in a manifest $\mathrm{Diff}_{\mathcal{F}}$-invariant way, the shift vector can only arise in combination with a time derivative of the metric $\gamma_{ij}$ in form of the covariant time derivative \eqref{CovDt}. Thus, the invariants in projectable HG can only be constructed by scalar contractions of covariant time derivatives of the metric field $D_t\gamma_{ij}$ (or, equivalently extrinsic curvatures $K_{ij}$), covariant space derivatives $\nabla_{i}$ and spatial curvature tensors $R_{ijkl}$.
\item \label{p2}Invariance under time-reversal and parity only allows invariants with an even number of time or space derivatives. Writing $S_{\mathrm{HG}}=\int\mathrm{d}t\mathrm{d}^dx\mathcal{L}_{\mathrm{HG}}$ and $\mathcal{L}_{\mathrm{HG}}=\sum_{n}c_{(n)}\mathcal{O}_{(n)}(D_t,\nabla_i,\gamma_{ij})$ implies that the operators have the general schematic structure (suppressing the summation index $n$)
\begin{align}
\mathcal{O}(D_t,\nabla_i,\gamma_{ij})=\sqrt{\gamma}\left(D_t\gamma_{ij}\right)^{2k}\left(\nabla_{i}\right)^{2n}\left(R_{ijkl}\right)^{m}.\label{operator}
\end{align}

\item\label{p3} 
For HG to be power counting renormalizable, the action can only include relevant and marginal operators w.r.t. the anisotropic scaling \cite{Horava2009}. Combining the scaling ${[S_{\mathrm{HG}}]_{\mathrm{S}}=0}$ with ${[\int\mathrm{d}t\mathrm{d}^dx]_{\mathrm{S}}=-(d+z)}$ implies $[\mathcal{L}_{\mathrm{HG}}]_{\mathrm{S}}=d+z$. Relevant and marginal operators have scaling $\left[\mathcal{O}_{(j)}(D_t,\nabla_i,\gamma_{ij})\right]_{\mathrm{S}}\leq d+z$.
Combining this with the structure \eqref{operator} yields the constraint
\begin{align}
2(kz+n+m)\leq d+z.\label{constraints}
\end{align}

\item The original motivation of HG to solve the problems with unitarity caused by higher derivative ghosts, requires to restrict the invariants in the action to include only up to second-order time derivatives of the metric. In view of the structure \eqref{operator}, this leaves the two possibilities of $k=1$ and $k=0$. For the kinetic term with $k=1$ and $n=m=0$ to scale marginally under \eqref{AS}, equality in \eqref{constraints} has to be satisfied and implies the \textit{critical scaling} condition
\begin{align}
z=d.\label{CriticalityC}
\end{align}
The operators with $k=0$ correspond to the potential $\mathcal{V}^{d}$, and, for the critical scaling \eqref{CriticalityC}, are restricted by the condition $2(n+m)\leq 2d$.
\end{enumerate}
The action of projectable HG in $D=d+1$ dimensions (in the gauge $N=1$) including all relevant and marginal terms with respect to the critical anisotropic scaling reads
\begin{align}
S_{\mathrm{HG}}={}&\frac{1}{2\,G}\int\mathrm{d}t\mathrm{d}^dx\,\sqrt{\gamma}\left(K_{ij}K^{ij}-\lambda\,K^2-\mathcal{V}^{(d)}\right).\label{PHGAct}
\end{align}
As a consequence of \eqref{CriticalityC}, the structure of the kinetic term is universal, i.e. independent of $d$,
\begin{align}
\sqrt{\gamma}\left(K_{ij}K^{ij}-\lambda\,K^2\right)=\frac{1}{4}\left(D_t\gamma_{ij}\right)\mathcal{G}^{ij,kl}\left(D_t\gamma_{kl}\right).
\end{align}
Here, $\mathcal{G}^{ij,kl}$ is the one-parameter $\lambda$-family of ``generalized DeWitt metrics"
\begin{align}
\mathcal{G}^{ij,kl}:={}&\frac{\sqrt{\gamma}}{2}\left(\gamma^{ik}\gamma^{jl}+\gamma^{il}\gamma^{jk}-2\lambda\gamma^{ij}\gamma^{kl}\right).\label{DWM}
\end{align}
There are two special values of $\lambda$. The first is the ``relativistic'' value $\lambda=1$, which leads to an enhanced symmetry \cite{Horava2009}. The second is the ``conformal'' value $\lambda_{\mathrm{c}}=1/d$, where $\mathcal{G}^{ij,kl}$ is degenerate, which also leads to an enhanced symmetry, namely local
anisotropic Weyl invariance \cite{Horava2009a}. For non-singular values $\lambda\neq \lambda_{\mathrm{c}}$, the inverse is given by
\begin{align}
\mathcal{G}_{ij,kl}={}&\frac{1}{\sqrt{\gamma}}\left(\gamma_{ik}\gamma_{jl}+\gamma_{il}\gamma_{jk}-\frac{2\lambda}{d\lambda-1}\gamma_{ij}\gamma_{kl}\right),\label{InvDeWitt}
\end{align}
For $\lambda<\lambda_{\mathrm{c}}$, \eqref{DWM} is positive definite, for $\lambda>\lambda_{\mathrm{c}}$ indefinite. In the context of GR, this property was found in \cite{Giulini1994} to be directly related to the attractive or repulsive nature of gravity. 

Note the difference of \eqref{PHGAct} to the Einstein-Hilbert action in ADM variables \eqref{EHADM}, where the $\mathrm{Diff}({\mathcal{M}})$ invariance completely fixed the structure of the action, i.e. the relative coefficient between the two terms $K_{ij}K^{ij}$ and $K^2$ in the kinetic terms as well as the coefficient of the potential $R$. In HG, $K_{ij}K^{ij}$, $K^2$ and the terms in $\mathcal{V}^{d}$ are separately invariant under $\mathrm{Diff}_{\mathcal{F}}({\mathcal{M}})$. In particular, $\lambda$ is a free parameter of the theory.

The potential $\mathcal{V}^{(d)}$ of projectable HG is defined in terms of $d$-dimensional curvature invariants, and, according to \eqref{p3}, includes all relevant and marginal operators with respect to the critical anisotropic scaling. In contrast to the kinetic term, the potential is not universal and the number and complexity of invariants in the potential grows with higher $d$. Restricting to $d=2$ and $d=3$, up to total derivatives, the possible curvature invariants read \cite{Sotiriou2009},
\begin{align}
{\mathcal V}^{(d=2)}={}&2\Lambda+\mu R^2,\label{V2}
\\
{\mathcal V}^{(d=3)}={}&2\Lambda-\eta R+\mu_1 R^2+\mu_2\,R_{ij}R^{ij}+\nu_1 R^3+\nu_2 RR_{ij}R^{ij}+\nu_3R^{i}_{\;\;j}R^{j}_{\;\;k}R^{k}_{\;\;i}+\nu_4\nabla_{i}R\nabla^{i}R+\nu_{5}\nabla_{i}R_{jk}\nabla^{i}R^{jk}.\label{V3}
\end{align}
Note that in $d=2$ and $d=3$ all invariants involving the Riemann tensors are absent. In addition, in $d=2$, the linear Einstein-Hilbert term $\sqrt{\gamma}R$ is a total derivative. 
In general, the Riemann tensor in $d$-dimensions has ${d^2(d^2-1)/12}$ independent components. Hence, in $d=2$, there is only one independent component associated with the Ricci scalar
\begin{align}
R_{ijkl}^{(d=2)}=\frac{R}{2}\left(\gamma_{ik}\gamma_{jl}-\gamma_{il}\gamma_{jk}\right).\label{DDI2}
\end{align}
Likewise, in $d=3$, there are only six independent components of the Riemann curvature tensors which are associated with the six components of the Ricci tensor $R_{ij}$. This can be also seen from the fact that in $d=3$, the Weyl tensor $C_{ijkl}\equiv0$ vanishes identically, which allows to express all curvature tensors $R_{ijkl}$ in terms of  $R_{ij}$ and $R$ via
\begin{align}
R_{ijkl}^{(d=3)}=R_{ik}\gamma_{jl}+R_{il}\gamma_{jk}+R_{jk}\gamma_{il}+R_{jl}\gamma_{ik}-\frac{R}{2}\left(\gamma_{ik}\gamma_{jl}-\gamma_{il}\gamma_{jk}\right).
\end{align}
The mass dimensions of the coupling constants follow from $[S_{\mathrm{HG}}]_{\mathrm{M}}=0$, $[\gamma_{ij}]_{\mathrm{M}}=0$,  and $[\partial_{i}]_{\mathrm{M}}=[\partial_t]_{\mathrm{M}}=[N^{i}]_{\mathrm{M}}=1$,
\begin{align}
[G]_{\mathrm{M}}={}&1-d,\qquad [\Lambda]_{\mathrm{M}}={}2,\qquad [\lambda]_{\mathrm{M}}=[\eta]_{\mathrm{M}}={}0,\\
[\mu]_{\mathrm{M}}={}&[\mu_1]_{\mathrm{M}}=[\mu_2]_{\mathrm{M}}=-2,\qquad [\nu_1]_{\mathrm{M}}={}[\nu_2]_{\mathrm{M}}=[\nu_3]_{\mathrm{M}}=[\nu_4]_{\mathrm{M}}=[\nu_5]_{\mathrm{M}}=-4.
\end{align}
A new set of dimensionless couplings $[\tilde{G}]_{\mathrm{M}}=[\tilde{\Lambda}]_{\mathrm{M}}=\left[\tilde{\mu_i}\right]_{\mathrm{M}}=\left[\tilde{\nu_i}\right]_{\mathrm{M}}=0$ is trivially defined by expressing the couplings in units of a common, a priori unspecified, mass scale $M_{*}$,
\begin{align}
\tilde{G}:=\frac{G}{M_{*}^{1-d}},\qquad \tilde{\Lambda}:=\frac{\Lambda}{M_{*}^{2}},\qquad \tilde{\mu}_{i}:=M_{*}^{2}\mu_i,\qquad \tilde{\nu}_i=M_{*}^4\nu_{i}.\label{DimLessCoup}
\end{align}
The parametrization \eqref{DimLessCoup} is useful when discussing phenomenological bounds on HG.

\subsection{Particle spectrum, dispersion relations and phenomenological constraints}
\label{Sec:ParticleSpectrumHG}
The particle spectrum of projectable HG in $d=2$ and $d=3$ is derived along the same lines as for GR by expanding the action around flat space $\bar{\gamma}_{ij}=\delta_{ij}$, $\bar{N}^{i}=0$ to quadratic order in the linear perturbations \footnote{This implies $\Lambda=0$. For a discussion of the cosmological constant in HG, see e.g.~\cite{Appignani:2009dy}.}
\begin{align}
h_{ij}:=\gamma_{ij}-\bar{\gamma}_{ij},\qquad n^{i}:=N^{i}-\bar{N}^{i}.\label{BGFHG}
\end{align}
Inserting the irreducible decomposition of the perturbations
\begin{align}
 n^{i}={}n^{i}_{\mathrm{T}}+\partial^i B,\qquad h_{ij}={}h^{\mathrm{TT}}_{ij}+2\partial_{(i}v^{\mathrm{T}}_{j)}+\left(\delta_{ij}-\frac{\partial_i\partial_j}{\partial^2}\right)\Psi+\frac{\partial_i\partial_j}{\partial^2}E,\label{IrrDecn}
\end{align}
with the three scalars $\Psi$, $E$ and $B$, the differentially constrained transversal vector fields ${\partial_{i}v_{T}^{i}=0}$, ${\partial_{i}n_{T}^{i}=0}$, and the transversal traceless tensor field ${h^{\mathrm{TT}}_{ij}\delta^{ij}=\partial^{i}h^{\mathrm{TT}}_{ij}=0}$ into the quadratic action, ``integrating out'' the non-dynamical modes $v_{i}^{\mathrm{T}}$ and $E$, fixing the gauges $B=0$ and $n^{i}_{\mathrm{T}}=0$, yields after Fourier transformation to momentum space the dispersion relations for the physical propagating degrees of freedom $h^{\mathrm{TT}}_{ij}$ and $\Psi$.
As discussed in the previous section, in $D=2+1$ there are no TT modes $h^{\mathrm{TT}}_{ij}$. However, in contrast to GR, which has no local degrees of freedom in $D=2+1$ dimensions, in HG there is an additional propagating scalar degree of freedom, which is a consequence of the reduced $\mathrm{Diff}_{\mathcal{F}}(\mathcal{M})$ invariance of HG, cf. the discussion in Sec. \ref{Sec:QuantumGR}.
The additional scalar mode persists even for low energies such that there is no smooth limit of HG to GR.

In $d=2$, the additional gravitational scalar has the non-relativistic dispersion relation expressed in terms of the dimensionless couplings \eqref{DimLessCoup}
\begin{align}
\omega^2_{\mathrm{S}}=4\tilde{\mu}\frac{1-\lambda}{1-2\lambda}\frac{k^4}{M_*^2}.\label{Dispd2}
\end{align}
Clearly, the dispersion relation for the additional scalar does not reduce to the linear relativistic form at low energies $k^2/M_{*}^2\ll1$, which again is a consequence of the absence of the relevant linear curvature invariant in the potential \eqref{V2}.
 
In $d=3$, aside from the additional scalar mode, the spectrum encompasses a propagating TT mode. Both have a non-relativistic dispersion relation
\begin{align}
\omega_{\mathrm{TT}}^2={}k^2\left[\eta +\tilde{\mu}_2\frac{k^2}{M_{*}^2}+\tilde{\nu}_5\frac{k^4}{M_{*}^4}\right],\qquad\omega_{\mathrm{S}}^2={}\frac{1-\lambda}{1-3\lambda}k^2\left[-\eta +(8\tilde{\mu}_1+3\tilde{\mu}_2)\frac{k^2}{M_{*}^2}
+(8\tilde{\nu}_4+3\tilde{\nu}_5)\frac{k^4}{M_{*}^4}\right].\label{DRd3TT}
\end{align}

Before discussing experimental constraints on HG, I first briefly review several theoretical restrictions:
\begin{enumerate}
\item
 Despite the critical scaling \eqref{CriticalityC}, which guarantees that the non-relativistic dispersion relations depend only quadratically on the frequency $\omega$, it is essential to make sure that no unitarity violating propagating ghost degrees of freedom enter in HG. Demanding the absence of ghosts leads to the condition $G>0$, which ensures the positivity of the TT kinetic term and the requirement that $\lambda$ must lie in the gaped interval $\lambda<1/d$ or $\lambda>1$, bounded by the points of enhanced symmetry, to ensure the positivity of the scalar kinetic term.

\item 
In contrast to the situation in $D=2+1$, thanks to the presence of the relevant operator $\propto R$ in \eqref{V3}, for low energies $k^2/M_{*}^2\ll1$ both dispersion relations \eqref{DRd3TT} in $D=3+1$ reduce to the linear relativistic relation
\begin{align}
\omega^2_{\mathrm{TT}}=\eta k^2+\mathcal{O}\left(k^2/M_{*}^2\right),\qquad\omega^2_{\mathrm{S}}=-\eta\frac{1-\lambda}{1-3\lambda} k^2+\mathcal{O}\left(k^2/M_{*}^2\right).\label{Dispersion}
\end{align}
However, due to the requirement $(1-3\lambda)/(1-\lambda)>0$, there is no value of $\eta\neq0$ at which both relations \eqref{Dispersion} are simultaneously positive, and for $\eta=0$ the linear relativistic dispersion relation is lost, just as in $D=2+1$. For $\eta>0$, this leads to a tachyonic instability of the scalar mode at low energies $k^2/M_{*}^2\ll1$. An obvious attempt to circumvent this problem is to keep $\eta>0$ and to tune $\lambda$ very close to one, in order to suppress the IR instability of the scalar mode. Unfortunately, this leads to strong coupling for the scalar mode at low energies \cite{Charmousis2009,Blas2009,Blas2011,Koyama2010} invalidating the perturbative treatment which underlies the power counting renormalizability \cite{Papazoglou2010}, see however \cite{Mukohyama2010,Izumi2011,Gumrukcuoglu2012,Cognola:2016gjy,Casalino:2018wnc}. Summarizing without a mechanism by which this IR problem can be avoided, the projectable theory seems to be excluded on phenomenological grounds. 

\item The IR instability problem can be cured in the non-projectable version of HG in which the potential \eqref{V3} involves invariants including the acceleration vector \eqref{accV}, thanks to the propagating lapse function. In order to illustrate the difference to the projectable case, I present the potential and the dispersion relation for the non-projectable theory in $d=2$. In the non-projectable case, the action \eqref{PHGAct} acquires a modified volume element $\mathrm{d}t\mathrm{d}^dx\sqrt{\gamma}\mapsto\mathrm{d}t\mathrm{d}^dxN\sqrt{\gamma}$ and the potential \eqref{V2} for the non-projectable theory in $d=2$ dimensions case is enlarged by additional invariants
\begin{align}
\mathcal{V}^{(2)}_{\mathrm{np}}=2\Lambda-\eta R-\alpha a_{i}a^{i}+\mu R+\rho_1 \Delta R+\rho_2 Ra_{i}a^{i}+\rho_3(a_ia^{i})^2+\rho_4a_ia^{i}\nabla_{j}a^{j}+\rho_5(\nabla_{i}a^{i})^2+\rho_6\nabla_{i}a_j\nabla^{i}a^{j}.\label{Pot2dnp}
\end{align}
Defining the perturbation of the lapse function $\phi:=N-1$ (with the choice $\bar{N}=1$ for the background value of the lapse function), the action expanded around flat background ($\Lambda=0$) up to quadratic order in the linear perturbations lead to the dispersion relation for the single scalar propagating degree of freedom \cite{Barvinsky2016},
\begin{align}
\omega_{\mathrm{S}}^2=\left(\frac{1-\lambda}{1-2\lambda}\right)\frac{\eta^2k^2+(4\alpha\mu+2\eta\rho_1)k^4+[ \rho_1^2-4\mu(\rho_5+\rho_6)]k^6}{\alpha-(\rho_5+\rho_6)k^2}.
\end{align} 
 In particular, among the additional invariants in \eqref{Pot2dnp}, there is a relevant operator proportional to $\alpha N\sqrt{\gamma} a^{i}a_{i}$ which leads to the required modifications of the low energy limit. The freedom in tuning the additional coupling constant $\alpha$ can be used to avoid the IR instability. In \cite{Blas2010} it was found that for $0<\alpha<2$ the instability can be avoided in non-projectable HG. However, as already anticipated in \cite{Horava2009} and supported by different arguments in \cite{Henneaux2010,Blas2011, Barvinsky2016}, the presence of the propagating lapse function $N$ in the non-projectable version leads to essential complications with the quantization, which I briefly comment on in Sec. \ref{Sec:QHG}.
\end{enumerate}
Aside from these theoretical restrictions, there are phenomenological constraints stemming from experimental bounds on Lorentz violation (LV), see e.g. \cite{Colladay1998,Mattingly2005,Jacobson2006,Kostelecky2011,Liberati2013,Eichhorn2019}. In the context of HG, these might be divided into two regimes:
\begin{enumerate}
\item LV in the IR:\\
Despite the suppression of higher order terms in the dispersion relations \eqref{DRd3TT} for low energies $k^2/M_{*}^2\ll1$, HG does not smoothly connect to GR in the IR, but rather to a modified theory of gravity with an additional propagating gravitational scalar degree of freedom. Deviations from GR can be quantified by a variety of experiments and mainly lead to restrictions for the couplings of the relevant operators in the IR. Experimental constraints come from deviations of the observed helium abundance during Big Bang Nucleosynthesis \cite{Carroll2004,Blas2010,Aver2015}, from Post-Newtonian Parameters \cite{Will2006,Blas2011,Blas2011a}, Binary pulsars \cite{Yagi2014}, Black holes \cite{Barausse2011,Barausse2013,Barausse2013a}. The most stringent constraint, however, comes from the recent detection of gravitational waves from the binary neutron star merger event GW170817 \cite{Abbott2017}. The inferred speed of propagation of the TT mode strongly constraints the parameter $|\eta-1|\lesssim10^{-15}$, but the propagation speed of the scalar mode remains largely unconstraint, cf. \cite{EmirGuemruekcueoglu2018}.

\item LV in the UV:	\\
LV effects in the gravitational sector at high energies are not so strongly restricted as in the matter sector provided by the SM particles. In particular, the scale $M_{*}$ might naturally be identified with the LV scale in the gravitational sector. Observations sensitive to the higher-order corrections in the dispersion relations \eqref{DRd3TT} provide a lower bound on $M_{*}$. However, LV effects in the SM are constraint much tighter and a mechanism is needed that prevents LV effects to percolate from the gravitational sector to the matter sector \cite{Liberati2013}. While several such mechanisms have been suggested (see e.g. \cite{Chadha1983,GrootNibbelink2005,Bolokhov2005,Anber2011,Pospelov2012,Kiritsis2013,Bednik2013,Kharuk2016}), it remains an open question whether they can ultimately be realized in HG \cite{Colombo2015,Coates2016}. In case there is a universal LV scale (i.e. in case the LV scale in the matter sector can be identified with the LV scale $M_{*}$ in the gravitational sector), the observation of synchrotron radiation from the crab nebula would provide a lower bound on $M_{*}$ around the Grand Unification Scale $M_{*}>10^{16}$ GeV \cite{Liberati2012}.  
\end{enumerate}
Summarizing, the ``healthy extension'' of the non-projectable model is still phenomenologically viable \cite{Blas2010,EmirGuemruekcueoglu2018}, but stronger constraints on the IR parameter as well as on $M_{*}$ have the potential to rule out the theory. Moreover, regarding the quantum theory, these properties will rely on the IR limit of the RG flow for the couplings of the relevant operators as briefly discussed in Sec.~\ref{Sec:RG2D} for $d=2+1$ dimensional projectable HG.

\section{Quantum Ho\v{r}ava gravity}
\label{Sec:QHG}
So far, all considerations in HG have been purely classical. However, the main motivation for proposing a Lifshitz theory of gravity are its unitarity and perturbative renormalizability, which has been  originally conjectured based on power counting arguments \cite{Horava2009}. While this conjecture has provoked a vast number of articles devoted to specific applications of HG in various scenarios, the question whether HG is indeed perturbatively renormalizable beyond power counting remained open for a long time. It was ultimately answered in the affirmative for the projectable version of HG in \cite{Barvinsky2016}. Furthermore, in order for HG to qualify as a UV-complete theory, also its RG structure must be investigated, which in turn requires explicit loop calculations. In this section, I discuss both aspects.
In order to establish contact with the general formalism in Sec. \ref{Sec:PQDG}, in the remaining sections I use Euclidean signature by Wick rotating $t\mapsto it$, $N^{i}\mapsto -iN^{j}$, which effectively leads to a sign flip of the potential in \eqref{PHGAct}.

\subsection{Non-local gauge-fixing and propagators}
\label{NonLocalGF}
Since HG is a gauge theory with invariance group $\mathrm{Diff}_{\mathcal{F}}\left(\mathcal{M}\right)$, its fluctuation operator \eqref{DefFlucOp} is degenerated and its perturbative quantization requires a gauge-fixing. In contrast to relativistic theories, in Lifshitz theories the situation is more complicated due to the anisotropic scaling between space and time: a standard local gauge-fixing causes the propagators of the theory to behave in an irregular way, ultimately leading to spurious non-local divergences \cite{Barvinsky2016}. Even if, on general grounds, it might be expected that these non-local divergences ultimately cancel order-by order in the perturbative expansion, their presence would greatly complicate the general analysis of renormalizability as well as the intermediate calculations.
Therefore, a new type of non-local gauge-fixing was proposed in \cite{Barvinsky2016}, which leads to regular propagators.

In the background field method, the geometric fields $\gamma_{ij}$ and $N^{i}$ are decomposed according to \eqref{BGFHG}.
As in the general case for relativistic theories \eqref{GBAct}, the gauge-breaking action in HG is quadratic in the the gauge condition $\chi^{i}$, 
 \begin{align}
 S_{\mathrm{gb}}=\frac{\sigma}{2G}\int\mathrm{d}t\mathrm{d}^dx^{i}\sqrt{\gamma}\,\chi^{i}O_{ij}\chi^{j}.\label{GBHGAct}
 \end{align}
Here, $\sigma$ is a gauge parameter. 
Guidance for finding a suitable gauge condition $\chi^{i}$ might be obtained by looking at the spatial part of the relativistic gauges of type \eqref{GCGR}, which expressed in terms of ADM variables \eqref{LEADM}, with the background covariant background derivatives $\bar{D}_t$ and $\bar{\nabla}_{i}$ and the gauge parameter $c_1$, have the general structure
\begin{align}
\chi^{i}[\bar{\gamma},\bar{N};h,n]=\bar{D}_tn^{i}+\left(\bar{\gamma}^{ij}\bar{\gamma}^{k\ell}-c_1\bar{\gamma}^{ik}\bar{\gamma}^{j\ell}\right)\bar{\nabla}_{k}h_{j\ell}.\label{AnisotropicGaugeDeDonder}
\end{align}
A characteristic feature of these ``quasi-relativistic gauge conditions'' is that they artificially render the shift perturbation $n^{i}$ propagating, due to the time derivative $\bar{D}_{t}n^{i}$. However, the gauge condition in the form \eqref{AnisotropicGaugeDeDonder} is not adequate, as it does not scale homogeneously under \eqref{AS}, which can be seen by comparing $[\bar{D}_tn^i]_{\mathrm{S}}=2d-1$ with $[\bar{\gamma}^{ij}\bar{\gamma}^{kl}\bar{\nabla}_{k}h_{jl}]_{\mathrm{S}}=1$. A solution would be to omit the term $\bar{D}_{t}n^i$ in \eqref{AnisotropicGaugeDeDonder}, which however would precisely lead to the aforementioned irregular propagators \cite{Barvinsky2016}. Therefore, keeping the  $\bar{D}_{t}n^{i}$ term, the only option is to increase the scaling dimension of the remaining terms by decorating them with additional spatial derivatives  
\begin{align}
\chi^{i}[\bar{\gamma},\bar{N};h,n]=\bar{D}_tn^{i}+B^{ij}\bar{\gamma}^{k\ell}\left(\bar{\nabla}_{k}h_{j\ell}-c_1\bar{\nabla}_{j}h_{k\ell}\right).\label{AnisotropicGaugeGen}
\end{align}
Here, $B^{ij}(\bar{\gamma};\bar{\nabla})$ is a differential operator of order $2(d-1)$, which aside from $\bar{\nabla}_{i}$ only involves the background metric $\bar{\gamma}_{ij}$.
Without introducing any new dimensional parameter, $S_{\mathrm{gb}}$ should have a marginal anisotropic  scaling ${[S_{\mathrm{gb}}]_{\mathrm{S}}=0}$, which in view of the critical scaling \eqref{CriticalityC} and $[\mathrm{d}t\mathrm{d}^dx^{i}]_{\mathrm{S}}=2d$ implies ${[\chi^{i}O_{ij}\chi^{j}]_{\mathrm{S}}=-2d}$.
Therefore, while \eqref{AnisotropicGaugeGen} with $[B^{ij}]_{\mathrm{S}}=2(d-1)$ ensures a homogeneous scaling $[\chi^{i}]_{\mathrm{S}}=2d-1$, it requires a scaling $[O_{ij}]_{\mathrm{S}}=-2(d-1)$. 
Consequently, if the operator $O_{ij}(\bar{\gamma};\bar{\nabla})$ only includes powers of $\gamma_{ij}$ and $\nabla_{i}$, it must be of the non-local form \footnote{The order of the covariant derivatives in \eqref{OpO} is a matter of choice, as different orders only differ in curvature terms which do not affect the principal part of the fluctuation operator. When lower derivative parts are included in the operator \eqref{OpO}, there might be ``preferred choices'' which simplify the lower derivative parts of the fluctuation operator. In \eqref{OpO}, a symmetric ordering has been chosen. Another natural symmetric choice is e.g. $O_{ij}=-\left(\bar{\nabla}_{i_1}\bar{\nabla}_{i_2}\ldots\bar{\nabla}_{i_{(d-2)/2}}\left(\bar{\Delta}\gamma^{kl}+\xi\bar{\nabla}^{k}\bar{\nabla}^{l}\right)\bar{\nabla}^{i_{(d-2)/2}}\ldots \bar{\nabla}^{i_2}\bar{\nabla}^{i_1}\right)^{-1}.$}
\begin{align}
O_{ij}=(-1)^{d-1}\left(\bar{\Delta}^{(d-1)}\bar{\gamma}^{ij}+\xi\bar{\nabla}^{i}\bar{\Delta}^{(d-2)}\bar{\nabla}^{j}\right)^{-1},\qquad \xi\neq-1.\label{OpO}
\end{align}
For the particularly useful choice $B^{ij}=\left(O^{-1}\right)^{ij}/2\sigma$ and $c_1=\lambda$, the metric and shift fluctuations in the quadratic action of projectable HG decouple, leading to the two-parameter family of $(\xi,\sigma)$ gauge conditions \cite{Barvinsky2016},
\begin{align}
\chi^{i}[\bar{\gamma},\bar{N};h,n]=\bar{D}_tn^{i}+\frac{1}{2\sigma}\left(O^{-1}\right)^{ij}\bar{\gamma}^{k\ell}\left(\bar{\nabla}_{k}h_{j\ell}-\lambda\bar{\nabla}_{j}h_{k\ell}\right).\label{AnisotropicGauge}
\end{align}
The gauge-fixing \eqref{OpO}, \eqref{AnisotropicGauge} leads to the aforementioned regular propagators, discussed in more detail in the following section. Unfortunately, the same gauge-fixing does not seem to work in the non-projectable theory. It leads to irregular terms in the propagators involving the lapse function, which is absent in the projectable theory \cite{Barvinsky2016}.  

\subsection{Regular propagators, superficial degree of divergence and renormalizability}
In the context of Lifshitz theories with anisotropic scaling \eqref{AS}, an important concept is the notion of a \textit{regular} propagator, which also plays a central role in the proof of perturbative renormalizability of HG.
A propagator for two generalized fields $\phi_1$ and $\phi_2$ with anisotropic scaling $[\phi_1]_{\mathrm{S}}=s_1$ and $[\phi_2]_{\mathrm{S}}=s_2$ is of the regular form
\begin{align}
\langle\phi_1,\phi_2\rangle={}&\sum\frac{P(\omega,\mathbf{k})}{D(\omega,\mathbf{k})},\quad D=\prod_{m=1}^{M}\left[A_m\,\omega^2+B_m\,k^{2d}+...\right],\label{RegProp}
\end{align}
iff $P(\omega,k)$ is a polynomial in $\omega$ and $k^{i}$ with leading anisotropic scaling ${[P]_{\mathrm{S}}\leq s_1+s_2+2d(M-1)}$ and $A_m>0$, $B_m>0$ are strictly positive constants. The ellipsis represents terms with subleading scaling dimensions, which generically originate from relevant operators in the action. The scaling properties ensure that the propagator has the right fall-off properties at small distances and time intervals, i.e. scales as $[\langle\phi_1,\phi_2\rangle]_{\mathrm{S}}\leq s_1+s_2-2d$ in the UV limit for high frequencies and momenta in momentum space.

With the choice \eqref{AnisotropicGauge}, the propagators of projectable HG in $D=2+1$ and $D=3+1$ are derived on a flat background $\bar{\gamma}_{ij}=\delta_{ij}$ and $\bar{N}^{i}=0$. Inserting the decomposition \eqref{IrrDecn} for the fluctuations $h_{ij}$ and $n_i$ into the gauge-fixed quadratic action, the gauge-fixed fluctuation operator \eqref{GaugeFixedOp} has a block diagonal form in the scalar, vector and tensor sectors and can be inverted algebraically in momentum space. The propagators for the original $h_{ij}$ and $n_{i}$ fields are recovered by using \eqref{IrrDecn} again. In $D=2+1$ the propagators read \cite{Barvinsky2016},
\begin{align}
\langle h_{ij},h_{kl}\rangle={}&2G\left[\delta_{ik}\delta_{jl}+\delta_{il}\delta_{jl}+\frac{2\lambda}{1-2\lambda}\delta_{ij}\delta_{kl}\right]\mathcal{P}_{\mathrm{S}}(\omega,k),\label{hpropd2}\\
\langle n_{i},n_{j}\rangle={}&4\mu Gk^2\left[\frac{2(1-\lambda)}{(1-2\lambda)}\delta_{ij}-\frac{k_{i}k_{j}}{k^2}\right]\mathcal{P}_{\mathrm{S}}(\omega,k).\label{npropd2}
\end{align}
The tensor combination in \eqref{hpropd2} is just the inverse DeWitt metric \eqref{InvDeWitt} in $d=2$ flat space.
In order to arrive at the final form \eqref{hpropd2} and \eqref{npropd2},  the gauge parameters $(\xi,\sigma)$ have to be chosen such that there is a single pole 
\begin{align}
\mathcal{P}_{\mathrm{S}}(\omega,k)={}&\left[\omega^2+4\mu\frac{1-\lambda}{1-2\lambda}k^4\right]^{-1}\label{poled2},\qquad\sigma=\frac{1-2\lambda}{8\mu(1-\lambda)},\qquad\xi=-\frac{1-2\lambda}{2(1-\lambda)}.
\end{align} 
Clearly, both propagators \eqref{hpropd2} and \eqref{npropd2} are of the regular form \eqref{RegProp}.\footnote{The ghost field propagator $\langle c_{i}^{*},c^{i}\rangle=G\delta_{i}^{j}\mathcal{P}_{\mathrm{S}}(\omega,k)$, which is derived from the gauge-fixing \eqref{AnisotropicGauge} according to the general rule \eqref{Gop}, also acquires the regular form \cite{Barvinsky2016}.}

In $D=3+1$ dimension the analogue procedure leads to the propagators for the $h_{ij}$ and $n_{i}$ fields \cite{Barvinsky2016},
\begin{align}
\langle h_{ij},h_{kl}\rangle={}&2G\left(\delta_{ik}\delta_{jl}+\delta_{il}\delta_{jk}\right)\mathcal{P}_{\mathrm{TT}}-2G\delta_{ij}\delta_{kl}\left[\mathcal{P}_{\mathrm{TT}}-\frac{1-\lambda}{1-3\lambda}\mathcal{P}_{\mathrm{S}}\right]\nonumber\\
&+2G\left(\delta_{ij}\frac{k_kk_l}{k^2}+\delta_{kl}\frac{k_{i}k_{j}}{k^2}\right)\left[\mathcal{P}_{\mathrm{TT}}-\mathcal{P}_{\mathrm{S}}\right]+2G\frac{k_ik_jk_kk_l}{k^4}\left[\frac{7\lambda-5}{1-\lambda}\mathcal{P}_{\mathrm{TT}}+\frac{1-3\lambda}{1-\lambda}\mathcal{P}_{\mathrm{S}}\right],\label{hpropd3}\\
\langle n_{i},n_{j}\rangle={}&G\frac{ \nu_5}{1-\lambda}k^4\left[2(1-\lambda)\delta_{ij}-(1-2\lambda)\frac{k_ik_j}{k^2}\right]\mathcal{P}_{\mathrm{S}},\label{npropd3}
\end{align}
Again, in order to arrive at the final form \eqref{hpropd3} and \eqref{npropd3}, the gauge parameters $(\xi,\sigma)$ have to be chosen in such a way that there are only the two physical poles\footnote{The propagators \eqref{hpropd3}-\eqref{npropd3} with the poles \eqref{Polesd3} are derived by only taking into account those operators in the potential \eqref{V3} which have a marginal anisotropic scaling. If the relevant operators in \eqref{V3} would have been taken into account, they would lead to relevant deformations in the propagators, i.e.~additional terms with lower $k$-dependence. Positive definiteness of $O_{ij}$ requires $\xi>-1$, which is not satisfied for $\lambda>1$ in the gauge \eqref{Polesd3}. This, however, does not seem to lead to difficulties in the perturbative approach, at least as far as it concerns gauge-independent on-shell quantities, such as e.g.~the beta functions of the essential couplings in $2+1$ dimensional HG, discussed in Sec.~\ref{Sec:RG2D}.} 
\begin{align}
\mathcal{P}_{\mathrm{TT}}={}&\left[\omega^2+\nu_5k^6\right]^{-1},\qquad \mathcal{P}_{\mathrm{S}}=\left[\omega^2+\frac{(1-\lambda)(8\nu_4+3\nu_5)}{1-3\lambda}k^6\right]^{-1},\qquad \sigma={}\frac{1}{2\nu_5},\qquad \xi=-\frac{1-2\lambda}{2(1-\lambda)}.\label{Polesd3}
\end{align}
The additional second pole in $D=3+1$ is due to the TT mode, which is absent in $D=2+1$ dimensions. Again, the propagators \eqref{hpropd3} and \eqref{npropd3} are of the regular form \eqref{RegProp}.

The superficial degree of divergence in HG is obtained along the same lines as in \eqref{DoDGR}, but with the anisotropic scaling of loop frequencies and momenta \eqref{ScalingDer}. Provided the propagators are of the regular form, it reads \cite{Barvinsky2016}
\begin{align}
D^{\rm div}_{\mathrm{HG}}=2\,d-d\,T-X-(d-1)\,l_N,\label{DoDHG}
\end{align}
with $T$ and $X$ the number of time derivatives and spatial derivatives acting on external legs and $l_n$ the number of external $n$-legs. The $\mathrm{Diff}_{\mathcal{F}}$ invariance of the counterterms allows to focus on diagrams with $l_n=0$.\footnote{This statement also relies on the $\mathrm{Diff}_{\mathcal{F}}$-invariant structure of the counterterms proven in \cite{Barvinsky2018}, as factors of the shift vector in $\mathrm{Diff}_{\mathcal{F}}$-invariant operators can only occur in form of the covariant time derivative \eqref{CovDt}.}
From \eqref{DoDHG}, it follows that $D^{\mathrm{div}}_{\mathrm{HG}}<0$ with more than two time derivatives or $d$ space derivative on external $h_{ij}$ legs.
If $D^{\mathrm{div}}_{\mathrm{HG}}<0$ would indeed imply the absence of divergences, only local operators with at most two time derivatives or $d$ spatial derivatives acting on $h_{ij}$ would have to be renormalized and HG would be perturbatively renormalizable. There are two complications which prevent to immediately draw this conclusion.
The first is the problem of (overlapping) subdivergences, which is also present in non-relativistic theories, i.e. a diagram might diverge despite $D^{\mathrm{div}}_{\mathrm{HG}}<0$. However, in \cite{Anselmi2007}, it was shown that the combinatorics of the recursive order-by-order subtraction of the Bogoliubov-Parasiuk-Hepp-Zimmermann (BPHZ) scheme \cite{Bogoliubow1957,Hepp1966, Zimmermann1969} works essentially the same as in relativistic theories.

The second problem is similar but inherently related to the non-relativistic nature of the theory. It can be illustrated by considering a generic $L$-loop Feynman integral which is free of subdivergences and has $D^{\mathrm{div}}_{\mathrm{HG}}(\mathcal{I})<0$, 
\begin{align}
\mathcal{I}={}&\int{\mathrm{d}}\omega_{(L)}\,{\rm d}^d\,k_{(L)}\,f\left(\omega_{(L)},k_{(L)}\right),\\
f\left(\omega_{(L)},k_{(L)}\right)={}&\int\prod_{\ell=1}^{L-1} \mathrm{d}\omega_{(\ell)}\mathrm{d}^dk_{(\ell)}\,\tilde{f}\left(\{\omega_{(\ell)}\},\{k_{(\ell)}\};\omega_{(L)},l_{(L)}\right).
\end{align}
The absence of subdivergences implies that the integrations over the $L-1$ loop integrals converge and result in a function $f\left(\omega_{(L)},k_{(L)}\right)$, which, suppressing the dependence on the external momenta, depends only on the $L$th loop frequency $\omega_{(L)}$ and spatial momentum $k_{(L)}$.
The anisotropic scalings $[\omega]_{\mathrm{S}}=d$ and $[k]_{\mathrm{S}}=1$ imply that
${[f]_{\mathrm{S}}=D^{\mathrm{div}}(\mathcal{I})-2d}$. However, in contrast to relativistic theories, in which $f\left(\omega_{(L)},k_{(L)}\right)$ can only depend on the relativistic combination $p^2_{(L)}=\omega_{(L)}^2-k_{(L)}^2$, in Lifshitz theories the anisotropic scaling is less restrictive and $f\left(\omega_{(L)},k_{(L)}\right)$ can acquire different forms, 
such as e.g. 
\begin{align}
f(\omega_{(L)},k_{(L)})
=\begin{cases}
\omega_{(L)}^{-1+n}\,k_{(L)}^{D^{\mathrm{div}}(\mathcal{I})-d(1+n)},\label{DivFreqMom}\\[2mm]
\omega_{(L)}^{-1-n}\,k_{(L)}^{D^{\mathrm{div}}(\mathcal{I})-d(1-n)}.
\end{cases}
\end{align} 
The problem is that, despite the fact that $D^{\mathrm{div}}(\mathcal{I})<0$, the total integral $\mathcal{I}$ might diverge as the individual integrals over the frequency (as in the first case of \eqref{DivFreqMom}) or the spatial momentum (as in the second case of \eqref{DivFreqMom}) diverge. 
In \cite{Barvinsky2016}, it was shown that this problem is absent if the propagators acquire the regular from \eqref{DoDHG}, in which case $D^{\mathrm{div}}_{\mathrm{HG}}(\mathcal{I})<0$ really implies convergence of $\mathcal{I}$. As shown before, all propagators in projectable HG can be brought into the regular form \eqref{RegProp} by the non-local gauge-fixing \eqref{OpO} and \eqref{AnisotropicGauge}. Combined with the $\mathrm{Diff}_{\mathcal{F}}(\mathcal{M})$ invariance of the counterterms shown in \cite{Barvinsky2018}, this completes the proof of perturbative renormalizability of projectable HG \cite{Barvinsky2016}. Unfortunately, the proof does not extend to the non-projectable theory, as not all propagators can be brought into the regular form \eqref{RegProp} for the gauge-fixing \eqref{OpO} and \eqref{AnisotropicGauge} due to the propagating lapse function. This, of course, does not imply that the non-projectable theory is perturbatively non-renormalizable, it simply means that other methods are required to investigate the renormalization structure of the non-projectable theory.

\subsection{Auxiliary field, local formulation and path integral}
 The Euclidean path integral \eqref{PI} for projectable HG acquires the form 
 \begin{align}
 Z_{\mathrm{HG}}=\left(\mathrm{Det}O_{ij}\right)^{1/2}\int\mathcal{D}[N^{i},\gamma_{ij},c^{i},c^{*}_{i}]e^{-S_{\mathrm{tot}}[N^{i},\gamma_{ij},c^{i},c^{*}_{i}]}.\label{PFHGWP}
 \end{align}  
 with the total action, including the HG action \eqref{PHGAct}, the gauge-breaking action \eqref{GBHGAct} and the ghost action $S_{\mathrm{gh}}$, which derives from the gauge condition \eqref{AnisotropicGauge} according to the general definition \eqref{Ghact} with the ghost operator \eqref{Gop},
 \begin{align}
 S_{\mathrm{tot}}=S_{\mathrm{HG}}+S_{\mathrm{gb}}+S_{\mathrm{gh}}.
 \end{align} 
Due to the gauge condition \eqref{AnisotropicGauge} with the non-local operator $O_{ij}$, defined in \eqref{OpO}, the gauge-breaking action $S_{\mathrm{gb}}$ introduces a non-locality in $S_{\mathrm{tot}}$. However, this non-locality only persists in the shift-shift sector of $S_{\mathrm{gb}}$,
\begin{align}
S_{\mathrm{gb}}=\frac{\sigma}{2G}\int\mathrm{d}t\mathrm{d}^dx\sqrt{\bar{\gamma}}\left(\bar{D}_tn^{i}O_{ij}\bar{D}_t n^{j}+\textrm{local terms}\right).
\end{align} 
The non-local part can be rendered local by ``integrating in'' the auxiliary field $\pi_i$ via the Gaussian functional integral, 
\begin{align}
\left(\mathrm{Det}O_{ij}\right)^{1/2}\exp\left[-\int\mathrm{d}t\mathrm{d}^dx\frac{\sigma\sqrt{\bar{\gamma}}}{2G}\bar{D}_tn^{i}O_{ij}\bar{D}_tn^{j}\right]=\int\mathcal{D}[\pi_i]\exp\left[-\int\mathrm{d}t\mathrm{d}^dx\frac{\sqrt{\bar{\gamma}}}{G}\left(\frac{1}{2\sigma}\pi_i\left(O^{-1}\right)^{ij}\pi_j-i\pi_i\bar{D}_tn^{i}\right)\right].\label{StratanovichPi}
\end{align}
The Hubbard-Stratonovich-type transformation \eqref{StratanovichPi} reveals the role of $\pi_i$ as momentum canonically conjugated to $n^{i}$. The field $\pi_i$ also shares similarities with the Nakanishi-Lautrup field used in the BRST formalism to ensure the off-shell nilpotency of the Slavnov operator, see e.g. \cite{Barvinsky2018}.

The field $\pi_{i}$ has mass dimensionality $[\pi_i]_{\mathrm{M}}=1$ and scaling dimensionality $[\pi_i]_{\mathrm{S}}=1$ (for arbitrary $d$).
In \cite{Barvinsky2016}, it was verified that the $\langle\pi_i,\pi_j\rangle$ and $\langle\pi_i,n_j\rangle$ propagators are also of the regular form \eqref{RegProp} and that the presence of the $\pi_i$ field does not affect the regularity of the $h_{ij}$ and $n_i$ propagators. Therefore, within the perturbative quantization, the procedure \eqref{StratanovichPi} is well defined such that the apparent non-locality in the shift sector, induced by the gauge-fixing, does not lead to any problems.
Moreover, \eqref{StratanovichPi} has the effect of absorbing the functional determinant $\left(\mathrm{Det}O_{ij}\right)^{1/2}$ in \eqref{PFHGWP}, such that the partition function acquires the simple form 
\begin{align}
Z_{\mathrm{HG}}=\int\mathcal{D}[N^{i},\pi_{i},\gamma_{ij},c^{i},c^{*}_{i}]e^{-S_{\mathrm{tot}}[N^{i},\pi_i,\gamma_{ij},c^{i},c^{*}_{i}]},\label{PFHG}
\end{align}
with a local action functional  $S_{\mathrm{tot}}[N^{i},\pi_i,\gamma_{ij},c^{i},c^{*}_{i}]$ including the auxiliary $\pi_i$ field.

\section{Explicit calculations and renormalization group flow}
\label{Sec:RG2D}
The proof that projectable HG is perturbatively renormalizable beyond power counting \cite{Barvinsky2016,Barvinsky2018} is an important step towards a unitary quantum theory of gravity. However, in order for this theory to qualify as a fundamental theory, it must be extendable to arbitrarily high energy scales. In other words, perturbative renormalizability does not yet ensure the UV completeness, as the RG flow could drive one or more coupling constants into a Landau pole, leading to divergent interaction strengths at finite energy scales. Another aspect of the RG flow in HG is connected to the IR and the question whether relativistic invariance can effectively be restored dynamically as an emergent symmetry at low energies. The RG analysis and the logarithmic running of the coupling constants requires to calculate the beta functions determined by the UV divergences of the theory. 

Various quantum aspects of Lifshitz theories, in particular in the context of HG, have been considered in \cite{Horava2009a, Horava2009,Orlando2009,Iengo2009,Orlando2010,Giribet2010,Ambjorn2010,Nesterov2011,Baggio2012,Gomes2012,Griffin2012,LopezNacir2012,Griffin2013,Nakayama2012,Rechenberger2013,Contillo2013,Arav2015,Taylor2016,Arav2016,Barvinsky2016,Arav2017,Barvinsky2017a,Keranen2017,Pal2017,MohammadiMozaffar2017,Knorr2019,Angel-Ramelli2019}.
Here, I focus on the calculation of the beta functions in $D=2+1$ dimensional HG. Previous work in this context includes the contributions of Lifshitz scalars to the gravitational beta functions \cite{DOdorico2014,DOdorico2015},
the one-loop beta functions for conformally reduced projectable  
HG in $D=2+1$ \cite{Benedetti2014}, and the renormalization of the cosmological constant in  $D=2+1$ projectable HG \cite{Griffin2017}. In this contribution, I report on the full RG flow of all couplings in projectable HG in $D=2+1$ dimensions which was derived in \cite{Barvinsky2017}. The analogue calculation in $D=3+1$ is technically much more challenging and has not yet been completed. However, recent partial results provide an important first step in this direction \cite{Barvinsky2019}.
 
The Euclidean action for projectable HG in $D=2+1$ dimensions reads\footnote{Note the flipped sign of the $\mu R^2$ term compared to \eqref{PHGAct}.}  
\begin{align}
S_{\mathrm{HG}}^{(d=2)}={}&\frac{1}{2\,G}\int{\rm d}t{\rm d}^2x\,\sqrt{\gamma}\left(K_{ij}K^{ij}-\lambda\,K^2+\mu R^2\right).\label{EuclActPHG2}
\end{align}
The background covariant gauge condition \eqref{AnisotropicGauge} and the non-local operator \eqref{OpO} in $D=2+1$ acquire the form
\begin{align}
\chi^{i}=\bar{D}_tn^{i}+\frac{1}{2\sigma}\left(O^{-1}\right)^{ij}\bar{\gamma}^{k\ell}\left(\bar{\nabla}_{k}h_{j\ell}-\lambda\bar{\nabla}_{j}h_{k\ell}\right),\qquad O_{ij}=-\left(\bar{\Delta}\bar{\gamma}^{ij}+\xi\bar{\nabla}^{i}\bar{\nabla}^{j}\right)^{-1},\qquad \xi\neq-1.\label{OpOd2}
\end{align}
In the background field method the ``quantum fields'' $h_{ij}$, $n^{i}$, $\pi_i$, $c_{i}^{*}$ and $c^{i}$ are integrated out in the path integral, which, within the one-loop approximation, means to perform the functional Gaussian integral \eqref{OneLoop}.
Therefore, only the part of the total action   $S_{\mathrm{tot}}=S_{\mathrm{HG}}+S_{\mathrm{gf}}+S_{\mathrm{gh}}$ quadratic in the perturbations $S_{\mathrm{tot}}^{(2)}$ is required. 
In view of \eqref{Ghact} and \eqref{Gop}, this means that only the ``affine'' parts of the gauge transformations on $h_{ij}$ and $n_{i}$ are required to derive the quadratic part of $S_{\mathrm{gh}}$. In the projectable version of HG in the gauge $N=1$, the $\mathrm{Diff}_{\mathcal{F}}$ reduce to the time-dependent spatial diffeomorphisms (corresponding to $\varepsilon=0$ in \eqref{FDiffShiftUp} and \eqref{FDiffMetric}) and the required gauge transformations in terms of the background covariant time derivative $\bar{D}_t$ and the background covariant spatial derivative $\bar{\nabla}_{i}$ are given by
\begin{align}
\delta_{\varepsilon}h_{ij}=2\bar{\nabla}_{(i}\varepsilon_{j)},\qquad\delta_{\varepsilon} n^{i}=\bar{D}_{t}\varepsilon^i.
\end{align}
The vector-ghost operator $Q^{i}_{j}$ is derived from \eqref{OpOd2} according to the general formula \eqref{Gop} and its quadratic part reads
\begin{align}
\tensor{Q}{^{i}_{j}}=\delta^{i}{}_{j}\bar{D}_t^2  +\frac{1}{4\sigma}&\left\{- 2\delta^{i}{}_{j} \bar{\Delta}^2
+ 2\left[1 + 2 \xi - 2 \lambda (1 + \xi)\right] \bar{\nabla}^{i}\bar{\Delta}\bar{\nabla}_{j}
+\delta^{i}{}_{j}\bar{R} \bar{\Delta} -  (1 - 2 \lambda + 2 \xi) \bar{R} \bar{\nabla}^{i}\bar{\nabla}_{j}\right.\nonumber\\
&\;\;\left.-  2\xi \bar{R}_{;j}\bar{\nabla}^{i}-  2\xi \bar{R}^{;i} \bar{\nabla}_{j} -  2\delta^{i}{}_{j}\bar{R}_{;k}\bar{\nabla}^{k}- \delta^{i}{}_{j} \tensor{\bar{R}}{_{;k}^{k}} -  2\xi \tensor{\bar{R}}{^{;i}_{j}}  \right\}.\label{Ghost}
\end{align}
A virtue of the manifest background covariant treatment in the background field method is that, due to the background $\mathrm{Diff}_{\mathcal{F}}$ invariance, the shift vector $\bar{N}^{i}$ only appears in combination with the time derivative $\partial_t\gamma_{ij}$ in form of the extrinsic curvature, or, equivalently, in form of the covariant time derivative of the metric $D_t\gamma_{ij}=2K_{ij}$. When performing variations of the total action $S_{\mathrm{tot}}=S_{\mathrm{HG}}+S_{\mathrm{gf}}+S_{\mathrm{gh}}$, factors of the shift perturbations $n^{i}$ only arise from the variation of the covariant time derivative, as can be seen from the operator relation,
\begin{align}
[\delta,D_t]=-\mathcal{L}_{\delta\mathbf{N}}.\label{VarId}
\end{align} 
Moreover, a canonical ordering among mixed covariant time derivatives and covariant space derivatives might be chosen in such a way that the covariant time derivatives act first. This requires repeated use of the basic commutator\footnote{The relations \eqref{VarId} and \eqref{baseCom} hold for any $d$, but only in the projectable version of HG. In the non-projectable version the operator version of \eqref{VarId} reads $[\delta,D_t]=-N^{-1}\left(\delta ND_t+\mathcal{L}_{\delta\mathbf{N}}\right)$. Likewise \eqref{baseCom} yields an addition term $a_mD_tT_{i_1\ldots i_s}^{j_1\ldots j_r}$ on the right-hand-side and the covariant spatial derivatives in the definition \eqref{KTensor} must be shifted by the acceleration vector $\nabla_{i}\mapsto\nabla_i+a_i$.}
\begin{align}
[D_t,\nabla_{m}]T_{i_1\ldots i_s}^{j_1\ldots j_r}=\sum_{{j_\ell}}\mathcal{K}^{j_{\ell}}_{mn}T_{i_1\ldots i_s}^{j_1\ldots n \ldots j_r}-\sum_{{i_\ell}}\mathcal{K}^{n}_{mi_{\ell}}T_{i_1\ldots n\ldots i_s}^{j_1\ldots j_r},\label{baseCom}
\end{align}
with the ``anisotropic commutator curvature'' tensor, defined in terms of derivatives of the extrinsic curvature
\begin{align}
\mathcal{K}^{k}_{ij}:=\nabla_{i}K^{k}_{j}+\nabla_{j}K^{k}_{i}-\nabla^{k}K_{ij}.\label{KTensor}
\end{align}
Introducing the auxiliary field $\pi_i$ according to \eqref{StratanovichPi}, making use of \eqref{VarId}, integrating by parts, sorting derivatives with \eqref{baseCom} and reducing curvature tensors by the dimensional-dependent identity \eqref{DDI2}, 
and arranging the fluctuations of the fields $h_{ij}$, $n^{i}$ and $\pi_i$ in a multiplet ${\phi^{A}=(h_{ij},n^{i},\pi_i)^{T}}$, the gauge-fixed fluctuation operator acquires block matrix structure and can be represented in the form
\begin{align}
\tensor{F}{_{AB}}(\bar{D_{t}},\bar{\nabla})=C_{AB}\bar{D}_t^2+D_{AB}^{ijkl}\bar{\nabla}_{i}\bar{\nabla}_{j}\bar{\nabla}_{k}\bar{\nabla}_{l}+T_{AB}\bar{D}_t+W_{AB}^{ij}\bar{\nabla}_{i}\bar{\nabla}_j+\Gamma^{i}_{AB}\bar{\nabla}_i+P_{AB}.\label{FOd2}
\end{align}
The principal part of \eqref{FOd2} is split into a temporal part $C_{AB}$ and a spatial part $D_{AB}^{ijkl}$ for which the derivatives have been made explicit. For brevity, I refrain from presenting the explicit matrices $C_{AB}$, $D_{AB}^{ijkl}$, $T_{AB}$, $W_{AB}^{ij}$, $\Gamma^{i}_{AB}$, and $P_{AB}$, which are functions of the background fields.
The one-loop renormalization requires to calculate the divergent part of the functional traces for the operators \eqref{FOd2} and \eqref{Ghost},
\begin{align}
\Gamma_{1}^{\mathrm{div}}=\frac{1}{2}\left.\mathrm{Tr}\ln F_{AB}\right|^{\mathrm{div}}-\left.\mathrm{Tr}\ln \tensor{Q}{_{i}^{j}}\right|^{\mathrm{div}}.
\end{align} 
In contrast to the relativistic case, standard heat-kernel techniques for the anisotropic case are not available, and in particular, there is no closed algorithm based on a Schwinger-DeWitt representation \eqref{ansatz} for the off-diagonal kernel of \eqref{FOd2}. In addition to the anisotropic character of these operators, they also suffer from further complications. First, the matrices in the principal parts $C_{AB}$ and $D_{AB}$ are degenerate as $n^{i}$ and $\pi^{i}$ enter $F_{AB}$ only with lower derivatives and the ${h-h}$ block of $D_{AB}^{ijkl}$ is a non-minimal fourth order operator.\footnote{The degeneracy is a consequence of the anisotropic scaling: in contrast to $[h_{ij}]_{\mathrm{S}}=0$, the fields $n^{i}$ and $\pi^{i}$ carry non-zero scaling dimension $[\pi_i]_{\mathrm{S}}=[n^{i}]_{\mathrm{S}}$, such that the overall homogeneous scaling $\left[F_{AB}\right]_{\mathrm{S}}=4$ only allows for lower derivatives of $n^{i}$ and $\pi^{i}$.}

Nevertheless, first attempts to deal with anisotropic operators via the heat-kernel technique were suggested in \cite{Nesterov2011, DOdorico2015}. A general algorithm for anisotropic operators, based on the resolvent method, was proposed in \cite{Barvinsky2017a}. In the most general case, however, this algorithm requires the evaluation of a large number of products of nested multi-commutators as well as non-trivial parameter integrals, which is technically challenging.

\begin{figure}[h]
	\includegraphics[width=5.8cm,height=5.8cm]{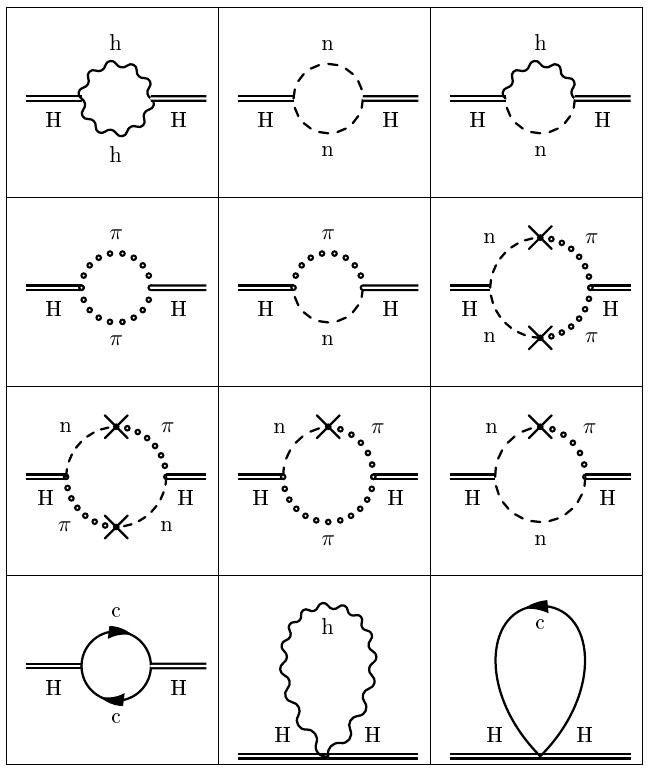}
	\caption{One-loop two-point 1PI diagrams in $D=2+1$ dimensional projectable HG (from \cite{Barvinsky2017}).}
	\label{Fig3}
\end{figure}

Therefore, an alternative way for the calculation of the one-loop divergences might be more suitable, especially since the number of invariants in $D=2+1$ HG is reasonably small and the one-loop calculation via Feynman diagrammatic techniques is still manageable -- in particular when combined with the background field method. After integrating out the ``quantum fields'' $h_{ij}$ and $n^{i}$, $\pi_i$, $c_i^{*}$ and $c^i$ in the path integral, the effective action is a functional of the mean fields, which at the one-loop level can be identified with the background fields. In particular, the divergent part of the effective action is a sum of local operators of the background fields $\bar{\gamma}_{ij}$, $\bar{N}^{i}$ and their time and space derivatives, which, due to the renormalizability of projectable HG, are of the same form as the  manifestly $\mathrm{Diff}_{\mathcal{F}}(\mathcal{M})$-invariant operators already present in the bare action \eqref{EuclActPHG2}. This allows to extract the one-loop renormalization of $G$, $\lambda$ and $\mu$ in a simpler way
by expanding the general background field $\bar{\gamma}_{ij}$ around a flat background  in which $\bar{N}^{i}=0$,
\begin{align}
\bar{\gamma}_{ij}=\delta_{ij}+H_{ij}.\label{Bgam}
\end{align}
Evaluating the bare action \eqref{EuclActPHG2} on the background \eqref{Bgam} and expanding up to quadratic order in $H_{ij}$ yields
\begin{align}
S_{\mathrm{HG}}[\bar{\gamma}_{ij},\bar{N}^{i}]={}&\frac{1}{2G}\int \mathrm{d} t\mathrm{d}^2x\left\{\frac{1}{4} \left(\dot{H}_{ij} \dot{H}^{ij} -   \lambda \dot{H}  \dot{H}\right)+\mu\left[\partial^2 H  \partial^2 H- \partial_{k}\partial_{l}H^{kl} (2\partial^2 H - \partial_{j}\partial_{i}H^{ij})\right] +\mathcal{O}(H^3)\right\},\label{Exp1}
\end{align}
with $\partial^2:=\delta^{\mu\nu}\partial_{\mu}\partial_{\nu}$. The divergent part of the effective action can be expanded in the same way
\begin{align}
\Gamma^{\mathrm{div}}[\bar{\gamma}_{ij},\bar{N}^{i}]=&\int \mathrm{d} t\mathrm{d}^2x\left\{c_{1}^{\mathrm{div}}\dot{H}_{ij} \dot{H}^{ij}+ c_2^{\mathrm{div}}\dot{H}  \dot{H}+c_{3}^{\mathrm{div}}\left[\partial^2 H  \partial^2 H- \mu \partial_{k}\partial_{l}H^{kl} (2\partial^2 H - \partial_{j}\partial_{i}H^{ij})\right] +\mathcal{O}(H^3)\right\}.\label{EffAHG2dH}
\end{align}
In order to access the renormalization of the couplings $G$, $\lambda$ and $\mu$, it is sufficient to calculate the divergent coefficients $c_{1}^{\mathrm{div}}$, $c_{2}^{\mathrm{div}}$, and $c_{3}^{\mathrm{div}}$ of the operators quadratic in $H_{ij}$. The renormalization of $G$ is extracted from $c_{1}^{\mathrm{div}}$, the renormalization of $\lambda/G$ from $c_{2}^{\mathrm{div}}$ and the renormalization of $\mu/G$ by any of the three operators in \eqref{EffAHG2dH}. Disentangling this system, allows to extract the individual renormalization of $G$, $\lambda$ and $\mu$. 
Diagrammatically, the background fields $H_{ij}$ only appear at external legs, while the quantum fields  $h_{ij}$ and $n^{i}$, $\pi_i$, $c_i^{*}$ and $c^i$ propagate in the loops. Hence, according to \eqref{Exp1} and \eqref{EffAHG2dH}, the one-loop renormalization of $G$, $\lambda$ and $\mu$ requires to calculate the divergent part of the 1PI diagrams with two external $H_{ij}$ legs, shown in Fig. \ref{Fig3}.

For the regular gauge \eqref{OpOd2}, with gauge parameters $(\xi,\sigma)$ and pole $\mathcal{P}_{\mathrm{S}}$ as in \eqref{poled2}, the propagators of the quantum fields $h_{ij}$, $n^{i}$ are the same as in \eqref{hpropd2}, \eqref{npropd2}, while those including the $\pi_i$, $c_i^{*}$ and $c^{i}$ fields read \cite{Barvinsky2016},
\begin{align}
\langle\pi_i,n^j\rangle=G\omega\delta_{i}^{j}\mathcal{P}_{\mathrm{S}}(\omega,k),\qquad \langle\pi_i,\pi_j\rangle=\frac{Gk^2}{2}\left[\delta_{ij}+(1-2\lambda)\frac{k_ik_j}{k^2}\right]\mathcal{P}_{\mathrm{S}}(\omega,k),\qquad \langle c_{i}^{*},c^{j}\rangle=G\delta_{i}^{j}\mathcal{P}_{\mathrm{S}}(\omega,k).
\end{align} 
The required three-point and four-point vertices in the gauge $\bar{N}^{i}=0$ are obtained by expanding the background fields in $\mathcal{L}^{(2)}=\delta\phi^{A}F_{AB}\delta\phi^{B}$, with $F_{AB}$ given in \eqref{FOd2}, according to \eqref{Bgam} up to second order in $H_{ij}$. The explicit results for the vertices are rather lengthy and therefore not presented here.
Within dimensional regularization, the divergent part of the one-loop diagrams in Fig. \eqref{Fig3} can be extracted by expanding the propagators in the corresponding integrals around vanishing external frequency and momenta, resulting in a sum of vacuum diagrams from which the logarithmically divergent contributions can easily be extracted by power counting.\footnote{See e.g. \cite{Steinwachs2019} for an application of this method with a particular focus on the combinatorial aspects in the context of relativistic higher-derivative theories.}  
The one-loop beta functions $\beta_{G}$, $\beta_{\lambda}$ and $\beta_{\mu}$, which determine the RG running of the couplings $G$, $\lambda$ and $\mu$, are obtained directly from the logarithmic one-loop divergences, i.e. from the corresponding coefficient of the pole $1/\varepsilon$ in dimension.

\begin{figure}[h]
	\includegraphics[width=7cm,height=6cm]{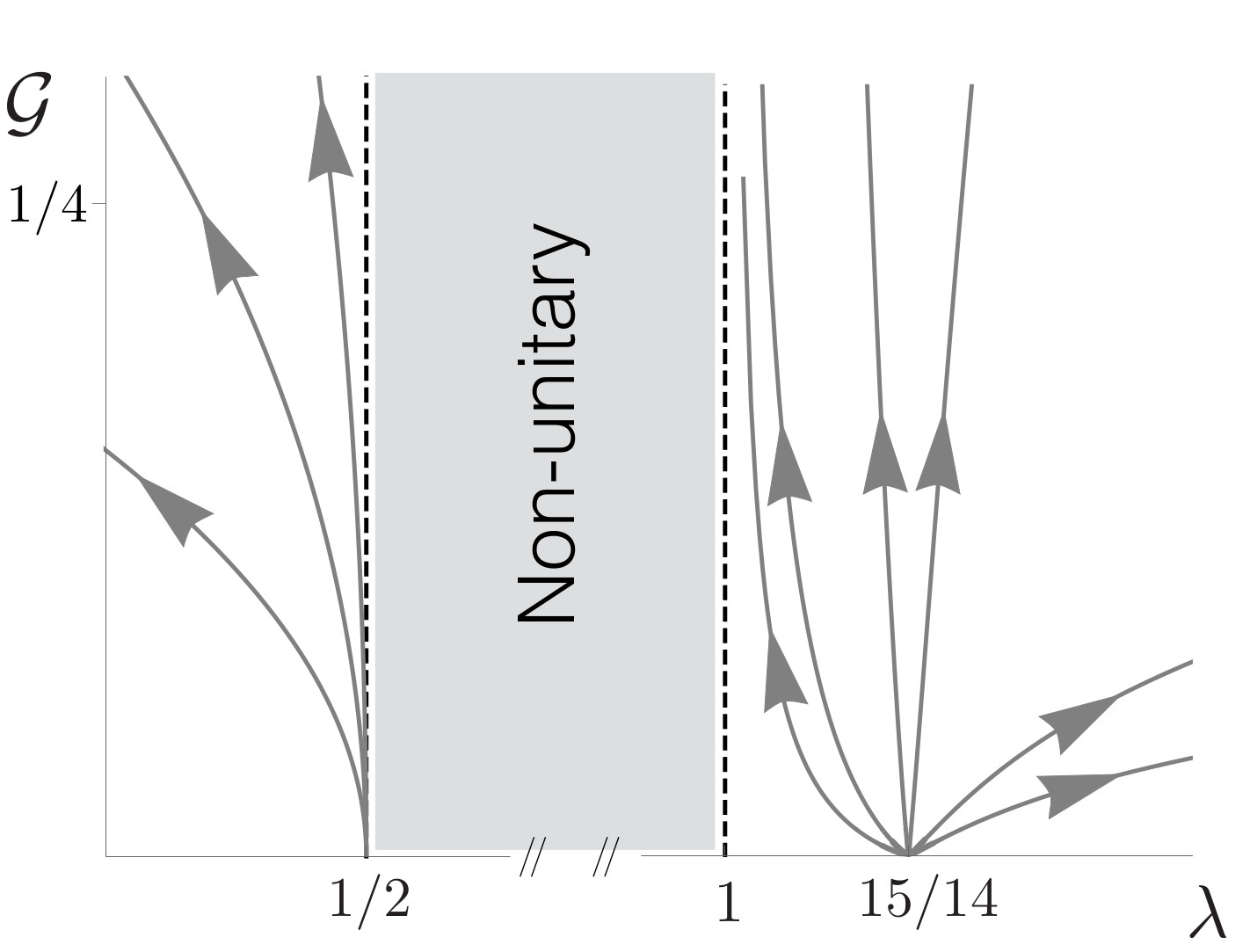}
	\caption{RG flow of essential couplings in $D=2+1$ dimensional HG; arrows point from the UV to the IR (from \cite{Barvinsky2017}).}
	\label{Fig4}
\end{figure}

Finally, in order to discuss the physical implications of the RG flow, it is important to extract the gauge independent physical information from the RG system. In general, the off-shell effective action is parametrization and gauge dependent. On the one hand, a change of the gauge-fixing induces a change $\Gamma^{\mathrm{div}}\mapsto\Gamma^{\mathrm{div}}+\mathcal{E}\delta\Gamma^{\mathrm{div}}$, which is proportional to the equations of motion $\delta\Gamma^{\mathrm{div}}=S_{,i}X^{i}$ with an arbitrary constant $\mathcal{E}$ \cite{DeWitt1967b,Kallosh1973,Kallosh1974}. On the other hand, this change might be compensated by the change $\delta\Gamma^{\mathrm{div}}=(\partial\Gamma^{\mathrm{div}}/\partial G)\delta G+(\partial\Gamma^{\mathrm{div}}/\partial \lambda)\delta \lambda+(\partial\Gamma^{\mathrm{div}}/\partial \mu)\delta \mu$, which is induced by a change in the couplings.
The combinations of couplings for which the corresponding beta function is gauge independent are called essential, all other couplings are called inessential and do not enter physical observables. The problem is therefore to tell apart and disentangle the essential from the inessential couplings. In order to find $X^{i}$ explicitly, one might exploit power counting as $S_{,i}X^i$ must be a local functional with the same scaling as $\Gamma^{\mathrm{div}}$, i.e. in the context of $D=2+1$ projectable HG, $S_{,i}X^i$ can only involve marginal operators with respect to the anisotropic scaling. Since the scaling and the index structure of the $S_{,i}$ are known, this corresponds to a strong constraint on the possible structure of the $X^{i}$. In \cite{Barvinsky2017}, it was found that the unique combination $X^{i}S_{,i}$ which vanishes on-shell is
\begin{align}
\delta\Gamma^{\mathrm{div}}=\mathcal{E}\int\mathrm{d}t\mathrm{d}^2x\left[K_{ij}K^{ij}-\lambda K^2-\mu R^2\right].\label{dGam1}
\end{align}
The variation of $\Gamma^{\mathrm{div}}$ with respect to the couplings reads
\begin{align}
\delta \Gamma^{\mathrm{div}}=\frac{1}{2G}\int\mathrm{d}t\mathrm{d}^2x\sqrt{\gamma}\left[-\frac{\delta G}{G}K_{ij}K^{ij}-\lambda\left(\frac{\delta \lambda}{\lambda}-\frac{\delta G}{G}\right)K^2+\mu\left(\frac{\delta \mu}{\mu}-\frac{\delta G}{G}\right)R^2\right].\label{dGam2}
\end{align} 
Equating \eqref{dGam1} and \eqref{dGam2} yields the desired transformations of the couplings \cite{Barvinsky2017}
,\begin{align}
\delta G=-2G^2\mathcal{E},\qquad\delta\lambda=0,\qquad\delta\mu=-4G\mu\mathcal{E}.
\end{align}
Thus, only $\lambda$ and the combination $\mathcal{G}=G/\sqrt{\mu}$ are essential couplings $(\delta\lambda=\delta\mathcal{G}=0)$ with beta functions \cite{Barvinsky2017},
\begin{align}
		 \beta_{\lambda}=\frac{15-14\lambda}{64\pi}\sqrt{\frac{1-2\lambda}{1-\lambda}}{\mathcal G},\qquad
		\beta_{\mathcal G}=-\frac{(16-33\lambda+18\lambda^2)}{64\pi\,(1-\lambda)^2}\sqrt{\frac{1-\lambda}{1-2\lambda}}{\mathcal G}^2.\label{BF}
\end{align}
The RG flow driven by the beta functions \eqref{BF} is shown in Fig. \ref{Fig4}.
There are two UV fixed points at
\begin{align}
\left(\lambda^{*}_1,\mathcal{G}^{*}_1\right)=\left(1/2,0\right),\qquad\left(\lambda^{*}_2,\mathcal{G}^{*}_2\right)=\left(15/14,0\right).
\end{align}
The first fixed point $\left(\lambda^{*}_1,\mathcal{G}^{*}_1\right)$ lies exactly on the lower boundary of the non-unitary interval $1/2<\lambda<1$ for which the gravitational scalar degree of freedom behaves like a ghost, cf. the discussion in Sec. \ref{Sec:ParticleSpectrumHG}.
For fixed $\mathcal{G}$, the beta function $\beta_{\mathcal{G}}$ develops a divergence in the limit $\lambda\to1/2$. At the same time, however, the limit $\lambda\to1/2$ is accompanied by $\mathcal{G}\to0$, implying that the relevant expansion parameter in this limit is $\tilde{\mathcal{G}}=\mathcal{G}(1-2\lambda)^{-1/2}$. The beta function $\beta_{\tilde{\mathcal{G}}}$ vanishes for $\lambda\to1/2$, which means that there is a one-parameter family of UV fixed points parametrized by the asymptotic value of $\tilde{\mathcal{G}}$. Summarizing, the status of this fixed-point remains inconclusive and higher loop corrections or contributions from matter loops are required to resolve the situation and to decide whether the fixed point is just an artifact of the approximation or has a physical significance. 

In contrast, the second fixed point $\left(\lambda^{*}_2,\mathcal{G}^{*}_2\right)$ is regular, lies in the unitary region $\lambda>1$ and is asymptotically free \cite{Barvinsky2017}.
Although projectable HG in $D=2+1$ dimensions only has the status of a toy model without propagating spin-$2$ particles, it provides the first unitary, perturbatively renormalizable and UV-complete quantum theory of gravitational propagating degrees of freedom. In previous calculations of the one-loop divergences in $D=2+1$ projectable HG, the dynamical content of the metric field was restricted to the conformal mode \cite{Benedetti2014}. In this conformally reduced model, only the fixed point at $(1/2,0)$ has been found. This shows that the formation of the regular fixed point at $(15/14,0)$ requires the full theory \cite{Barvinsky2017}. 

An other interesting feature of the RG flow is that there are RG trajectories which emanate from the regular UV fixed point and asymptotically approach the ``relativistic value'' $\lambda\to1$ in the IR. In addition to the problems with the IR $\lambda\to1$  limit discussed in Sec. \ref{Sec:ParticleSpectrumHG}, the ``gravitational coupling'' $\mathcal{G}$ becomes strongly coupled along these trajectories, requiring a non-perturbative analysis in this regime. Nevertheless, the observed flow towards $\lambda=1$ suggests that the possibility of a dynamical mechanism for an emergent restoration of relativistic symmetry at low energies should be investigated in more detail. First, the phenomenon that a theory, which is asymptotically free in the UV, develops a strong coupling in the IR is well-known. Second, the strong coupling of $\mathcal{G}$ in the IR might just be an artifact of the absence of relevant curvature operators in $D=2+1$. In $D=3+1$ dimensions relevant deformations might be expected to naturally cut off the strong coupling of $\mathcal{G}$. 

All these interesting and encouraging results justify the hope that the RG flow of the more realistic and physically relevant theory in $D=3+1$ dimensions shows similar features. Although, there are no conceptual problems associated with the analogue calculation in $D=3+1$ dimensions, in view of the increased number and complexity of the independent curvature invariants, it is technically much more challenging. A first step towards the RG flow of projectable HG in $D=3+1$ dimensions has been made in \cite{Barvinsky2019a}, were the one-loop beta functions of $G$ and $\lambda$ were derived with Feynman diagrammatic methods in a similar way as in \eqref{Exp1} and \eqref{EffAHG2dH}, by exploiting the gauge invariance of counterterms, allowing to restrict to a flat metric background and to only focus on diagrams with background shift fields at the external legs. However, the gauge-invariant beta functions for the essential coupling constants and the fixed point structure of the theory can only be derived by having access to the renormalization of all couplings, including those in the potential sector. Thus, the complete calculation of the one-loop divergences in $D=3+1$ dimensional projectable HG provides an important task.

\section{Conclusions and outlook}
\label{Sec:CAO}
In this contribution, I have reviewed various attempts to quantize gravity within the framework of perturbative Quantum Field Theory with a particular focus on Ho\v{r}ava Gravity.
I highlighted the merits and difficulties that come along with each of the approaches.
The different approaches to quantum gravity discussed in this contribution might be best characterized by the property that they do \textit{not} share with the other approaches, as shown in TABLE \ref{Table2}.
\begin{table}[h!]
	\begin{tabular}{lll}
		\toprule
		Approach &$\quad$& Property\\
		\midrule
		General Relativity &$\quad$& not renormalizable\\
		Effective Field Theory &$\quad$& not fundamental \\
		Asymptotic Safety &$\quad$& not perturbative\\
		Quadratic Gravity &$\quad$& not unitary or not satisfying mirco-causality\\
		Ho\v{r}ava Gravity &$\quad$& not relativistic\\
		\bottomrule
	\end{tabular}
	\caption{Approaches to quantum gravity characterized by a property they don't have.}
	\label{Table2}
\end{table}

\noindent The status of HG with critical anisotropic scaling can be roughly summarized by dividing the discussion into ``projectable'' vs. ``non-projectable'' and ``phenomenology of the classical theory'' vs. ``properties of the quantum theory''.

From a phenomenological point of view, projectable HG does not seem to qualify as a viable theory, mainly because it suffers from an infrared instability of the additional scalar gravitational mode \cite{Charmousis2009,Blas2009,Blas2011,Koyama2010}. Although other proposals with a more optimistic conclusion on this problem have been made  \cite{Mukohyama2010,Izumi2011,Gumrukcuoglu2012}, they are based on non-perturbative effects which are outside the scope of the weak coupling regime where perturbation theory is applicable.

In contrast, the non-projectable model does not suffer from the infrared instability because additional relevant operators which include powers of the acceleration vector (spatial derivatives of the lapse function) can cure the infrared instability \cite{Blas2010}.
Even if the low energy sector of the non-projectable model is strongly constraint by observational data and a mechanism to avoid a percolation of LV effects from the gravitational sector to the matter sector seems to be needed to avoid conflicts with bounds on LV in the matter sector \cite{Liberati2013}, the non-projectable model is still phenomenologically viable \cite{EmirGuemruekcueoglu2018}.

From a theoretical point of view, regarding the status of HG as consistent quantum theory of gravity, the situation is somewhat opposite to the phenomenological assessment. The projectable theory has been proven to be perturbatively renormalizable (for any dimension $D=d+1$) in the strict sense \cite{Barvinsky2016,Barvinsky2018}. Moreover, the $D=2+1$ dimensional model was shown to be asymptotically free and its RG flow features interesting RG trajectories which emanate from the UV fixed point and asymptote the relativistic value $\lambda=1$ in the IR \cite{Barvinsky2017}. Even if the $D=2+1$ dimensional model must be considered as a toy model without propagating TT modes, it is a unitary, perturbatively renormalizable and UV-complete quantum theory of non-trivial propagating degrees of freedom and captures essential features of HG, which are expected to carry over to the physically relevant $D=3+1$ dimensional case. The situation for the $D=3+1$ dimensional case is not yet conclusively clarified and requires a calculation of the one-loop beta functions. A first step in this direction has ben undertaken in \cite{Barvinsky2019a}, but in order to extract the gauge independent physical information about the running of the essential couplings, the renormalization of all couplings is needed. While there are no new conceptual difficulties, the analogue calculation is technically much more complex as compared to the $D=2+1$ case and requires more efficient methods such as newly developed heat-kernel techniques for anisotropic operators \cite{Nesterov2011,DOdorico2015,Barvinsky2017a}. In any case, the calculation of the one-loop divergences of projectable HG in $D=3+1$ dimensions is certainly a very important result which will give new insights in the structure of the theory.
 
The situation with the quantization of non-projectable model is less clear. Unfortunately, the proof of perturbative renormalizability for the projectable theory \cite{Barvinsky2016,Barvinsky2018} does not extend to the non-projectable theory, mainly because the proof relies on the regular from of all propagators and no gauge-fixing could be found in the non-projectable model which would render all propagators regular. In particular, there seems to be no gauge-fixing which could remove all irregular contributions to the propagator involving the lapse function \cite{Barvinsky2016} -- interpreted in \cite{Barvinsky2016} as a reflection of the instantaneous interaction induced by the lapse function \cite{Blas2011}. 
Therefore new ideas seem to be necessary in order to deal with the perturbative quantization of the non-projectable theory.
 
In summary, HG is an interesting proposal, but, closing with the words of Bryce DeWitt, the theory does not seem yet to have been ``pushed to its logical conclusion'' \cite{DeWitt-Morette}. Further important calculations in $D=3+1$ dimensions are required and might decide upon the fate of Ho\v{r}ava's proposal for a unitary, perturbatively renormalizable and UV-complete quantum theory of gravity.

\section{Acknowledgments}
I thank Benjamin Bahr, Antonio Duarte Pereira and Astrid Eichhorn for inviting me to contribute this article to the Frontiers Research Topic ``Coarse Graining in Quantum Gravity: Bridging the Gap between Microscopic Models and Spacetime-Physics''.
I also would like to thank those of my collaborators with whom I had the pleasure to work in various constellations on different topics that found its way into this contribution, especially Andrei Barvinsky, Diego Blas, Anirudh Gundhi, Lavinia Heisenberg, Mario Herrero-Valea, Alexander Kamenshchik, Claus Kiefer, Dmitry Nesterov, Guillem P\'erez-Nadal, Michael Ruf, Sergey Sibiryakov, Alexei Starobinsky and Matthijs van der Wild.
Part of this work was done during a stay at SISSA (Trieste) in the framework of the ``Young Investigator Training Program 2018'' and it is my pleasure to thank Roberto Percacci and SISSA (Trieste) for the warm hospitality and INFN/ACRI for financial support. I also acknowledge financial support for the article processing charge, funded by the Baden-Wuerttemberg Ministry of Science, Research and Art and the University of Freiburg in the funding programme Open Access Publishing.

\bibliography{ReviewHGV2}

\end{document}